\documentclass[11pt]{article}
\usepackage{amsmath}
\usepackage{amsfonts}
\usepackage{amsthm}
\usepackage{dsfont}
\usepackage{threeparttable}
\usepackage{placeins}
\usepackage{booktabs}
\usepackage{graphicx}
\usepackage{array}
\usepackage{amssymb}
\usepackage{geometry}
\usepackage{bm}
\usepackage{xcolor}
\usepackage{tikz}
\definecolor{figblue}{HTML}{2A78D6}
\definecolor{figorange}{HTML}{EB6834}
\usepackage[onehalfspacing]{setspace}
\usepackage[authoryear,round,longnamesfirst]{natbib}
\usepackage{mathtools}
\usepackage{enumitem}
\usepackage[hang,flushmargin]{footmisc}
\usepackage[colorlinks=true, pdfstartview=FitV, linkcolor=blue, citecolor=blue, urlcolor=blue]{hyperref}
\usepackage{microtype}

\newtheorem{theorem}{Theorem}[section]
\newtheorem{proposition}[theorem]{Proposition}
\newtheorem{lemma}[theorem]{Lemma}
\newtheorem{corollary}[theorem]{Corollary}
\theoremstyle{definition}
\newtheorem{assumption}{Assumption}
\newtheorem{remark}[theorem]{Remark}

\newcommand{\E}{\mathbb{E}}
\newcommand{\Prb}{\mathbb{P}}
\newcommand{\Var}{\operatorname{Var}}
\newcommand{\Cov}{\operatorname{Cov}}
\newcommand{\ind}{\mathds{1}}
\newcommand{\op}{o_p}
\newcommand{\Op}{O_p}
\newcommand{\R}{\mathbb{R}}
\newcommand{\norm}[1]{\lVert #1\rVert}
\newcommand{\trans}{^{\prime}}
\newcommand{\ps}{\psi_\tau}
\newcommand{\dK}{d_{\rm K}}
\renewcommand{\today}{\number\day\space\ifcase\month\or January\or February\or March\or April\or May\or June\or July\or August\or September\or October\or November\or December\fi\space\number\year}

\title{Bias Correction and Inference for Panel Quantile Regression\thanks{This version: \today. We thank Iv\'{a}n Fern\'{a}ndez-Val, Jiaying Gu, Whitney Newey and Yichong Zhang for helpful comments, and Martin Weidner for drawing our attention to his concurrent work with Manuel Arellano and Ayden Higgins. All errors are our own.\medskip}}
\date{}
\author{
	Harold D. Chiang\thanks{Harold D. Chiang: hdchiang@wisc.edu. Department of Economics, University of Wisconsin--Madison, William H. Sewell Social Science Building, 1180 Observatory Drive, Madison, WI 53706, USA.\medskip}
	\quad
	Wei Biao Wu\thanks{Wei Biao Wu: wbwu@uchicago.edu. Department of Statistics, University of Chicago, George Herbert Jones Laboratory, 5747 South Ellis Avenue, Chicago, IL 60637, USA.\medskip}
}

\begin{document}
\maketitle

\begin{abstract}
We develop a large-$n$, large-$T$ theory for the unsmoothed fixed-effects
quantile regression estimator that allows for serially correlated panel
data. An exact rank-score identity for the profiled criterion, with a
convexity argument, delivers the leading incidental parameter bias of the
slope estimator, of order $T^{-1}$, in closed form, paralleling
\citet{HahnNewey04} for smooth nonlinear panel models. When $n$ and $T$
grow at the same rate, the estimator is asymptotically normal but in
general not centred, and conventional intervals are invalid. In stationary
panels whose dependence decays geometrically or polynomially, with $n/T$
bounded, serial correlation changes the structure of this bias: the long-run variance of
the quantile score enters it, and feedback from past errors to current
regressors adds a term that can bias the estimator even at the median. A
split-panel jackknife with contiguous half-panels removes the bias, whereas
one with interleaved half-panels does not in general, and a cluster-robust
variance estimator, clustered by individual, yields valid inference. Under independence over
time, the admissible growth of $n$ relative to $T$, up to $n=o(T^3)$,
depends on the regularity of the conditional error densities, and the
jackknife is valid throughout this range.
\end{abstract}


\vspace{-0.2cm}

\newpage
\section{Introduction}\label{sec:intro}

Quantile regression \citep{koenker1978regression} describes how covariates
affect the whole conditional distribution of an outcome, in applications
ranging from the wage structure
\citep{Buchinsky94,AngristChernozhukovFernandezVal06} to tail risk
\citep{EngleManganelli04,AdrianBrunnermeier16,AdrianBoyarchenkoGiannone19}.
The fixed-effects quantile regression (FEQR) model of \citet{Koenker04}
adds individual intercepts that absorb unobserved, time-invariant
heterogeneity, and quantile-based panel models are used to study, for
example, earnings and consumption dynamics \citep{ArellanoBlundellBonhomme17}
and growth-at-risk across countries \citep{AdrianGrinbergLiangMalikYu22}.
In applications, the number of individuals $n$ usually exceeds the number
of periods $T$: only 6 of the 81 applications of the estimator of
\citet{canay2011simple} surveyed by
\citet{besstremyannaya2019reconsideration} have $n<T$. Panel data with a
long time dimension are, moreover, often serially correlated, through persistent
errors or regressors and through feedback from past errors to current
regressors, and ignoring serial correlation can make standard errors far
too small \citep{BeDuMu2004}.

Because each intercept is estimated from only $T$ observations,
fixed-effects estimators are subject to the incidental parameter problem
\citep{NeymanScott48}, which fixed-$T$ methods solve only in particular
models \citep{Chamberlain80,Lancaster02}. In dynamic linear and smooth
nonlinear panel models, the resulting bias is of order $T^{-1}$, the same
order as the standard deviation when $n$ and $T$ grow at the same rate, so
that conventional confidence intervals are invalid
\citep{HahnKuersteiner02,AlvarezArellano03,HahnNewey04}. Analytical and
jackknife corrections (\citealp{HahnNewey04,DhaeneJochmans15}; see
\citealp{FernandezValWeidner18}, for a review), integrated-likelihood
corrections \citep{ArellanoBonhomme09} and the parametric bootstrap
\citep{HigginsJochmans24} restore valid inference, but all rest on smooth
likelihoods or estimating equations, which quantile regression does not
provide. For the original FEQR estimator in serially independent
panels, a centred normal limit has been established under conditions that
require $n/T\to0$, namely $n^2(\log n)^3/T\to0$ \citep{kato2012asymptotics}
or $n(\log T)^2/T\to0$ \citep{galvao2020unbiased}, and
\citet{galvao2016smoothed} allow $n$ and $T$ of the same order only after smoothing the
criterion. Under serial dependence, the available results for the
unsmoothed estimator also require $n/T\to0$
(\citealp[Section~5]{kato2012asymptotics};
\citealp[Theorem~3]{galvao2020unbiased}). To our
knowledge, apart from the independent work of
\citet{ArellanoHigginsWeidner26} on panels that are independent over time,
discussed below, the distribution of the original estimator has not been
established when $n$ grows as fast as $T$, the typical case in
applications, and the effect of serial dependence on its bias, and hence
on bias correction, has not been characterised.

In this paper, we develop a large-$n$, large-$T$ theory for the unsmoothed
FEQR estimator that covers such panels and allows for serial correlation.
Our theoretical contributions are twofold. First, under independence over
time, $\sqrt{nT}(\widehat\beta-\beta_0)$ is approximated in Kolmogorov
distance by $N(\sqrt{n/T}\,B_n,\Omega_n)$, where the leading bias $B_n/T$
is given in closed form by the conditional error densities and their first
derivatives at zero. This leading bias vanishes if the errors are
independent of the regressors, but not in general under conditional
heteroskedasticity (Appendix~\ref{app:example}). Conventional intervals
are therefore invalid unless $\sqrt{n/T}\,B_n\to0$, as in smooth models.
The admissible growth of $n$ relative to $T$, from bounded $n/T$ up to
$n=o(T^3)$, is governed by the regularity of these densities, not by the
smoothness of the criterion, and the upper end parallels the condition for
the analytical correction of \citet{HahnNewey04}.

Second, and central to applications with a long time dimension, we allow
each individual's observations to form a stationary causal process with
geometrically or polynomially decaying dependence
\citep{Wu05PNAS,Wu07}, with $n/T$ bounded. This extension is not routine,
because serial correlation changes the structure of the bias and not only
the covariance. The long-run variance of the quantile score at level $\tau$ replaces
$\tau(1-\tau)$ in the bias, which also acquires a term from the dependence
between regressors and errors of different periods. This term can make
FEQR biased even at the median under conditional symmetry when past errors
affect current regressors. It is the quantile counterpart of the bias of
the within estimator in dynamic panels \citep{Nickell81,AlvarezArellano03}.
The extension is not routine to prove either. Under independence over
time, the mean of the profiled score follows exactly from a binomial--beta
identity: the periods are independent and identically distributed, so the level at which the other
periods place the cut-off of the rank weights has a beta-mixture law that
does not involve the observation's own regressor. Under serial dependence
the periods are not independent, and the cut-off depends on errors that are
correlated with that regressor, so this identity is no longer available,
and neither is the conditional independence of the regressors given the errors on
which the averaging of the remainders rests. We replace both by coupling
arguments. For each observation, the periods far from it are made
independent of it by replacing the innovations that link them, the order
statistics of these large subsamples are approximated by a linear expansion
whose variance is the long-run variance of the quantile score, and the
nearby periods, whose errors are correlated with the observation's
regressor, produce the new bias term.
Dependence also determines how the bias can be removed: the split-panel
jackknife with contiguous half-panels remains valid, but the one with
interleaved half-panels does not in general, because the odd and even
periods carry a dependence different from that of the full panel. Standard
errors clustered by individual are then valid, whereas those that assume
independence over time are not in general. Appendix~\ref{app:special}
adds two special cases. In a class of panels with possibly unbounded
regressors, which includes Gaussian panels whose regressors and errors
share a common linear filter, the estimator needs no bias correction,
whatever the growth of $n$ relative to $T$
(Proposition~\ref{prop:nobias}). Without a regularity condition on the
joint law of the errors over time, it can have a non-Gaussian limit with
an atom at the true slope (Proposition~\ref{prop:cusp}).
Table~\ref{tab:literature} compares the growth conditions with those in the
literature.

\begin{table}[!t]
\centering
\begin{threeparttable}
\caption{Representative sufficient growth conditions for normal approximations}
\label{tab:literature}
\small
\begin{tabular}{@{}>{\raggedright\arraybackslash}p{0.25\textwidth}>{\raggedright\arraybackslash}p{0.30\textwidth}>{\raggedright\arraybackslash}p{0.39\textwidth}@{}}
\toprule
Reference or result & Temporal setting & Growth condition and conclusion\\
\midrule
\citet[Theorems~1 and~3]{HahnNewey04}
 & Independent; smooth likelihood models
 & $n/T\to\upsilon\in(0,\infty)$; biased normal limit, panel jackknife (scalar parameter)\\
\citet[Theorem~2]{HahnNewey04}
 & Independent; smooth likelihood models
 & $n=o(T^3)$; analytical correction, centred normal limit\\
\addlinespace
\citet[Theorems~3.2 and~5.1]{kato2012asymptotics}
 & Independent, or geometric $\beta$-mixing, bounded pairwise conditional densities
 & $n^2(\log n)^3/T\to0$; centred normal limit\\
\citet[Theorem~1]{galvao2020unbiased}
 & Independent
 & $n(\log T)^2/T\to0$; centred normal limit\\
\citet[Theorem~3]{galvao2020unbiased}
 & Geometric $\beta$-mixing, bounded pairwise conditional densities
 & $n(\log T)^4/T\to0$; centred normal limit\\
\citet{galvao2016smoothed}
 & Smoothed estimator; serial dependence allowed
 & $n$ and $T$ of the same order; biased normal limit and analytical correction\\
\addlinespace
\citet[Theorems~1 and~5]{ArellanoHigginsWeidner26}
 & Independent; unsmoothed estimator
 & $n/T^3\to0$; biased normal limit, jackknife with general subsample sizes
   (centred normal limit)\\
\addlinespace
This paper, Theorems~\ref{thm:main} and~\ref{thm:spj}
 & Independent; unsmoothed estimator
 & From $n/T=O(1)$ to $n=o(T^3)$, by the regularity of the error densities
   (Table~\ref{tab:regimes}); biased normal approximation, split-panel jackknife\\
This paper, Theorem~\ref{thm:dep}
 & Stationary; geometrically or polynomially decaying dependence
   (Assumptions~\ref{ass:causal} and~\ref{ass:dep})
 & $n/T=O(1)$; biased normal approximation with long-run covariance,
   contiguous split-panel jackknife; clustered standard errors
   (Proposition~\ref{prop:cluster})\\
\bottomrule
\end{tabular}
\begin{tablenotes}[flushleft]\footnotesize
\item Each row also requires the identification, density, moment and covariance
conditions of the corresponding result. The table asserts neither necessity
nor an ordering of the assumption sets.
\end{tablenotes}
\end{threeparttable}
\end{table}

These results rest on an argument that never differentiates the criterion.
Once the intercepts are profiled out, each individual's slope score is a
linear rank statistic. Under independence over time, its mean has an exact
representation through a mixture of beta variables, so that the bias
follows from a Taylor expansion of a deterministic function. Averaging the
centred individual remainders, which are independent across individuals,
rather than bounding them one by one, is what permits $n$ and $T$ of the same order. A
convexity argument \citep{Pollard91,HjortPollard93} extends these
calculations, made at deterministic parameter values, to every minimiser
without requiring stochastic equicontinuity of the nonsmooth score
(Remark~\ref{rem:outline}). The argument may be of independent interest.

\citet{ArellanoHigginsWeidner26}, in work developed independently of ours,
study the same estimator, and both papers characterise its incidental
parameter bias through order statistics and the beta distribution. They
study static panels whose observations are independent and identically
distributed over time given the individual heterogeneity, an assumption
that, as they note, rules out serial dependence given that heterogeneity,
including predetermined regressors such as lagged dependent variables. In
this setting, they show that the standard delete-one jackknife does not
remove the leading bias, propose jackknives with general subsample sizes,
whose choice trades bias against variance, cover instrumental-variable
quantile regression and apply their methods to the effect of age on test
scores. Our paper allows for serial dependence, including regressors that
respond to past errors (although not lagged dependent variables, which
condition (b) of Assumption~\ref{ass:dep} excludes). Serial dependence
matters in panels with a long time dimension, since it changes the
structure of the bias, the choice of jackknife and the standard errors. Our
paper also gives a range of
admissible growth rates of $n$ relative to $T$ set by the regularity of the
error densities, inference on the individual effects
(Appendix~\ref{app:alpha}) and the special cases of
Appendix~\ref{app:special}: a class of serially dependent panels in which
no bias correction is needed whatever the growth of $n$, and a model with a
non-Gaussian limit. The two papers are therefore highly complementary.

Our practical contribution is a simple procedure with valid inference in
serially correlated panels. The split-panel jackknife of
\citet{DhaeneJochmans15} removes the leading bias without a density
estimate or bandwidth and, combined with a consistent covariance estimator,
yields valid confidence intervals, from bounded $n/T$ up to $n=o(T^3)$
under independence over time. Under serial dependence, the half-panels
should be contiguous rather than interleaved, and the standard errors
clustered by individual. In our simulations at $\tau=0.9$ with $n=10T$,
nominal 95 per cent intervals cover $0.769$--$0.800$ around FEQR but
$0.890$--$0.933$ around the jackknife in serially independent panels, and,
with clustered standard errors, $0.470$ and $0.102$ around FEQR but $0.875$
and $0.850$ around the contiguous jackknife in two serially dependent
designs. An application to the $q$ theory of investment
\citep{hayashi1982tobin}, using the firm panel of
\citet{schaller1990reexamination}, shows these points at work
(Appendix~\ref{sec:empirical}). Clustering raises the standard errors by
up to about a factor of two. With capital narrowly defined, the
interleaved split misses an estimated upward bias in the effect of output
on investment, in the direction that feedback through the capital stock
suggests. With capital broadly defined, the theory's prediction of how the
bias scales with $T$ indicates that a large jackknife correction reflects a
change in the panel between periods rather than a bias.

This paper also relates to other approaches to distributional heterogeneity
in panels, including quantile effects in nonseparable models
\citep{chernozhukov2013average}, random coefficients
\citep{ArellanoBonhomme12}, location-shift and location-scale models
\citep{canay2011simple,MachadoSantosSilva19}, group-level treatments
\citep{ChetverikovLarsenPalmer16}, quantile factor models
\citep{chen2021quantile}, debiased distribution regression
\citep{ChernozhukovFernandezValWeidner24} and fixed-$T$ identification
\citep{KhanTamer26}. \citet{ChiangGalvaoWei26} allow pervasive common shocks.
More broadly, the paper relates to work on the bias from estimating many
effects, through leave-out estimation \citep{KlineSaggioSolvsten20} and
grouped fixed effects \citep{BonhommeLamadonManresa22}.

The rest of the paper is organised as follows. In
Section~\ref{sec:model}, we set out the model and assumptions. In
Section~\ref{sec:theory}, we develop the theory for the common parameter,
and in Section~\ref{sec:inference}, we turn to bias correction and
inference. We extend the results to serially dependent panels in
Section~\ref{sec:dependence}, present simulation results in
Section~\ref{sec:simulation} and conclude in Section~\ref{sec:conclusion}.
The Appendix contains an example with non-zero bias, an empirical
illustration on Tobin's $q$ and firm investment, the theory for the
individual effects, the proofs and two special cases of serially
correlated panels (Appendix~\ref{app:special}).

\subsection*{Notation}

For random vectors $X$ and $Y$ in $\R^k$, $\mathcal L(X)$ is the
law of $X$ and $\dK\{\mathcal L(X),\mathcal L(Y)\}=\sup_{x\in\R^k}|\Prb(X\leq
x)-\Prb(Y\leq x)|$ the lower-orthant Kolmogorov distance, with
coordinatewise inequalities. We also write $\dK\{\mathcal
L(X),N(\mu,\Sigma)\}$. $\Phi$ and $\phi$ are the standard normal
distribution function and density, $\norm\cdot$ is the Euclidean norm of
a vector and the spectral norm of a matrix, $\norm\cdot_\infty$ the maximum
norm of a vector and the supremum norm of a function, $A\trans$ the
transpose, $\lambda_{\min}$ and $\lambda_{\max}$ the extreme eigenvalues of
a symmetric matrix, $I_m$ the identity matrix and $\ind(\cdot)$ an
indicator. For symmetric matrices, $A\preceq B$ and $B\succeq A$ mean that
$B-A$ is positive semidefinite, and, for a random vector $X$ and $1\leq s<\infty$,
$\norm X_s=(\E\norm X^s)^{1/s}$. Weak convergence is written $\Rightarrow$ and convergence in
probability $\longrightarrow_p$, and limits are taken as $n,T\to\infty$
jointly. $C$ and $c$ are generic finite positive constants, uniform over
individuals, that depend only on $p$, $\tau$, the constants in the
assumptions and, where relevant, the kernel. A subscript records an
additional dependence, and $c_i$, $c_\Gamma$, $C_\Gamma$, $c_V$, $C_V$,
$C_0$, $C_1$, $C_2$, $C_\omega$, $C_h$, $C_\xi$, $C_m$, $C_\delta$, $C_A$, $C_w$,
$c_1$, $c_*$, $c_\upsilon$, $c_{nT}$ and $c_{n,\delta}$ are not generic. We write
$n\asymp T$ if $n/T$ is bounded away from zero and infinity, and
$x_T\sim y_T$ if $x_T/y_T\to1$.

\section{Model, Estimator and Assumptions}\label{sec:model}

\subsection{Model and Estimator}

Let $\{(y_{it},x_{it}\trans)\trans:i=1,\ldots,n,\ t=1,\ldots,T\}$ be panel
data, where $y_{it}\in\R$ is the outcome and $x_{it}\in\R^p$ a vector of
regressors without an intercept, with $p$ fixed. Fix $\tau\in(0,1)$. In the
spirit of \citet{Koenker04}, and following \citet{kato2012asymptotics}, we
consider the quantile regression model with individual fixed effects
\begin{equation}\label{eq:model}
 y_{it}=\alpha_{i0}+x_{it}\trans\beta_0+u_{it},\qquad
 \Prb(u_{it}\leq0\mid x_{it})=\tau,
\end{equation}
so that $\alpha_{i0}+x_{it}\trans\beta_0$ is the conditional
$\tau$-quantile of $y_{it}$ given $x_{it}$. The individual effects, the
common parameter and the errors may all depend on $\tau$, but since $\tau$
is fixed, we suppress this dependence. The individual effects are treated as
fixed parameters, and no restriction is placed on their relation to the
regressors.

We study the original, unpenalised and unsmoothed FEQR estimator
\begin{equation}\label{eq:estimator}
 (\widehat\alpha,\widehat\beta)\in
 \arg\min_{\alpha\in\R^n,\,\beta\in\R^p}
 \frac1{nT}\sum_{i=1}^n\sum_{t=1}^T
 \rho_\tau(y_{it}-\alpha_i-x_{it}\trans\beta),\qquad
 \rho_\tau(v)=v\{\tau-\ind(v\leq0)\},
\end{equation}
where $\widehat\alpha=(\widehat\alpha_1,\ldots,\widehat\alpha_n)\trans$,
and write $\ps(u)=\tau-\ind(u\leq0)$ for the score function, so that
$\rho_\tau(v)=v\,\ps(v)$. The objective is convex and piecewise linear, so
a minimiser exists by the linear-programming formulation of quantile
regression \citep{koenker2005quantile}. It need not be unique, and all
results below apply to every measurable selection.

\subsection{Assumptions}\label{sec:assumptions}

\begin{assumption}[Sampling]\label{ass:sampling}
For each $i$, the pairs $(u_{it},x_{it})$, $t=1,\ldots,T$, are independent
and identically distributed (i.i.d.), and different individuals are
independent. The distributions may differ across $i$, and
\eqref{eq:model} holds for every $i$.
\end{assumption}

\begin{assumption}[Regressors and error densities]\label{ass:density}
There is a constant $M$ with $\norm{x_{it}}\leq M$ almost surely. Write
$F_i(u\mid x)=\Prb(u_{i1}\leq u\mid x_{i1}=x)$ and
$F_i(u)=\E F_i(u\mid x_{i1})$. The conditional error density $f_i(u\mid x)$
exists and is bounded uniformly over $i$, $u$ and $x$, and $f_i(u)=\E
f_i(u\mid x_{i1})$ denotes the marginal error density.
\end{assumption}

The next assumption comes in four versions of increasing strength. Suprema
over $x$ may be taken over the support of the regressor law, for suitable
versions of the conditional densities.

\begin{assumption}[Density regularity]\label{ass:smooth}
There is $\epsilon_0>0$ such that one of the following holds.
\begin{enumerate}[leftmargin=3.2em]
\item[(W)] On $(-\epsilon_0,\epsilon_0)$ the conditional densities
$f_i(\cdot\mid x)$ are continuously differentiable, $f_i'$ is uniformly
bounded, and the first derivatives are uniformly equicontinuous:
\[
 \omega_f(a):=
 \sup_{\substack{i,x;\ |u|,|v|<\epsilon_0\\ |u-v|\leq a}}
 |f_i'(u\mid x)-f_i'(v\mid x)|\longrightarrow0
 \quad \text {as }a\downarrow 0.
\]
Moreover, the marginal error density at zero satisfies
$a_i:=f_i(0)=\E f_i(0\mid x_{i1})\geq c>0$ for all $i$.
\item[(H$_\eta$)] For some $\eta\in(0,1]$, condition (W) holds and
$\omega_f(a)\leq Ca^\eta$.
\item[(I)] Condition (W) holds, the conditional densities are twice
continuously differentiable on $(-\epsilon_0,\epsilon_0)$, and
$\sup_{i,x,|u|<\epsilon_0}|f_i''(u\mid x)|<\infty$.
\item[(II)] Condition (I) holds, the conditional densities are three
times continuously differentiable on $(-\epsilon_0,\epsilon_0)$, and
$\sup_{i,x,|u|<\epsilon_0}|f_i'''(u\mid x)|<\infty$.
\end{enumerate}
\end{assumption}

Condition (I) implies (H$_1$). Throughout, $\omega$ denotes a fixed
non-decreasing function on $[0,1]$ with $\omega(a)\downarrow0$ as
$a\downarrow0$ and $\omega(a)\geq C_\omega\{\omega_f(C_\omega a)+a\}$ for a
constant $C_\omega$ large enough for the finitely many uses below. Under
(H$_\eta$), one may take $\omega(a)=Ca^\eta$, and under (I) and (II),
$\omega(a)=Ca$.

Define the density-weighted regressor mean, the centred regressor, and
the Jacobian and variance matrices
\begin{equation}\label{eq:matrices}
\begin{gathered}
 \gamma_i=\frac{\E[f_i(0\mid x_{i1})x_{i1}]}{a_i},
 \qquad
 z_{it}=x_{it}-\gamma_i,
 \\
 \Gamma_n=\frac1n\sum_{i=1}^n
     \E[f_i(0\mid x_{i1})z_{i1}z_{i1}\trans],
 \qquad
 V_n=\frac1n\sum_{i=1}^n\E[z_{i1}z_{i1}\trans].
\end{gathered}
\end{equation}

\begin{assumption}[Non-degeneracy]\label{ass:nondeg}
There are constants $0<c_\Gamma\leq C_\Gamma$ and $0<c_V\leq C_V$ such that,
for all sufficiently large $n$,
$c_\Gamma\leq\lambda_{\min}(\Gamma_n)\leq\lambda_{\max}(\Gamma_n)\leq
C_\Gamma$ and $c_V\leq\lambda_{\min}(V_n)\leq\lambda_{\max}(V_n)\leq C_V$.
\end{assumption}

\begin{assumption}[Panel growth]\label{ass:growth}
$n,T\to\infty$ and, according to the version of
Assumption~\ref{ass:smooth} in force,
\[
 \text{(W): } n/T=O(1);\qquad
 \text{(H$_\eta$): } n=o(T^{1+\eta});\qquad
 \text{(I): } n=o(T^{2});\qquad
 \text{(II): } n=o(T^{3}).
\]
\end{assumption}

We call a matching pair of versions of Assumptions~\ref{ass:smooth}
and~\ref{ass:growth} a \emph{regime}, labelled W, H$_\eta$, I and II, and
write ``Assumptions~\ref{ass:sampling}--\ref{ass:growth}'' for
Assumptions~\ref{ass:sampling}, \ref{ass:density} and \ref{ass:nondeg}
together with one regime. Table~\ref{tab:regimes} summarises the regimes.
Its third column gives the order of a deterministic sequence $\epsilon_T$,
which bounds the error in the mean of the profiled slope score once its
leading $T^{-1}$ bias is subtracted, uniformly in $i$ and in small
displacements of the slope (Lemma~\ref{lem:beta}). Here the profiled slope
score is the score for $\beta$ after the intercepts are profiled out
(Remark~\ref{rem:outline}). Regime I coincides with H$_1$ as far as orders
are concerned. We list it separately because (I) is the familiar
twice-differentiability condition. The order $\epsilon_T=O(T^{-2})$ of regime II also holds if (I) is
imposed and $f_i''$ is uniformly Lipschitz. In each regime,
\begin{equation}\label{eq:growth}
 \sqrt{nT}\,\epsilon_T\longrightarrow0,\qquad
 L_{nT}:=\log(nT)=O(\log T).
\end{equation}

\begin{table}[!ht]
\centering
\begin{threeparttable}
\caption{Density regularity, mean-score error and admissible panel growth}
\label{tab:regimes}
\begin{tabular}{llll}
\toprule
Regime & Assumption~\ref{ass:smooth} & Mean error $\epsilon_T$ & Assumption~\ref{ass:growth}\\
\midrule
W & (W) & $o(T^{-1})$ & $n/T=O(1)$\\
H$_\eta$ & (H$_\eta$) & $O(T^{-1-\eta/2})$ & $n=o(T^{1+\eta})$\\
I & (I), or (H$_1$) & $O(T^{-3/2})$ & $n=o(T^2)$\\
II & (II) & $O(T^{-2})$ & $n=o(T^3)$\\
\bottomrule
\end{tabular}
\end{threeparttable}
\end{table}

We now discuss the assumptions. Assumption~\ref{ass:sampling} requires
independence across individuals and over time, as in
\citet{kato2012asymptotics} and in the independent-data results of
\citet{galvao2020unbiased}. It allows, however, heterogeneous distributions
across individuals and arbitrary contemporaneous dependence between $u_{it}$
and $x_{it}$, so that conditional heteroskedasticity is permitted. No
moments of the errors are assumed. Independence over time is used in two
essential places: the exact binomial--beta rank calculation of
Lemma~\ref{lem:beta} and the conditional independence of the regressors
given the errors in Lemma~\ref{lem:remainder}. Section~\ref{sec:dependence}
replaces both arguments for serially dependent panels. Bounded regressors
are a standard simplifying condition in this literature \citep{Koenker04}.

Assumption~\ref{ass:smooth} is the key condition. In smooth nonlinear panel
models, the order of the admissible bias expansion is tied to the number of
derivatives of the criterion. Here, by contrast, only the regularity of the
conditional error densities enters, and under (W) neither a second
derivative nor a quantitative modulus of continuity is required. A mere
bound on $f_i'$ without its uniform continuity, however, does not give the
uniform $o(T^{-1})$ remainder of regime W. Assumption~\ref{ass:nondeg}
requires $\Gamma_n$ and $V_n$ to be uniformly well conditioned but not to
converge. The upper bounds follow from Assumption~\ref{ass:density}, and the
lower bound on $V_n$ follows from that on $\Gamma_n$ and the density bound.
Finally, no separate identification condition for consistency is needed,
because the local convexity argument of Appendix~\ref{app:slope} localises
every slope minimiser.

\subsection{Bias and Covariance}\label{sec:bias-cov}

Put $v_\tau=\tau(1-\tau)$ and define
\begin{equation}\label{eq:bias}
 c_i=\E[f_i'(0\mid x_{i1})z_{i1}],\qquad
 b_n=-\frac{v_\tau}{2n}\sum_{i=1}^n\frac{c_i}{a_i^2},\qquad
 B_n=\Gamma_n^{-1}b_n,\qquad
 \Omega_n=v_\tau\Gamma_n^{-1}V_n\Gamma_n^{-1}.
\end{equation}
The vectors $c_i$ and $b_n$ are uniformly bounded. Under
Assumption~\ref{ass:nondeg}, so is $B_n$, and the eigenvalues of $\Omega_n$
are bounded above and away from zero, although neither $B_n$ nor $\Omega_n$
needs to converge. Since $\E[f_i(0\mid x_{i1})z_{i1}]=0$, $\Gamma_n$ equals
the Jacobian $n^{-1}\sum_i\E[f_i(0\mid x_{i1})z_{i1}x_{i1}\trans]$. Being
symmetric, it appears as $\Gamma_n^{-1}$ in both outer factors of
$\Omega_n$.
Section~\ref{sec:theory} shows that $B_n/T$ is the leading incidental
parameter bias of $\widehat\beta$ and $\Omega_n$ the covariance of its
leading stochastic term at the $\sqrt{nT}$ scale. The
ratio $c_i/a_i^2$ is the derivative, with respect to the quantile level
$v$ at $v=\tau$, of the mean of $x_{i1}$ given that $u_{i1}$ equals its
$v$-quantile, and $b_n/T$ is a half-variance-times-curvature term, as in
a second-order delta method (Remark~\ref{rem:outline}).

\section{Asymptotic Theory}\label{sec:theory}

This section states the main results for the common parameter,
including the cases in which the bias vanishes. The individual
effects are treated in Appendix~\ref{app:alpha}.

\begin{theorem}[Gaussian approximation for the slope estimator]\label{thm:main}
Under Assumptions~\ref{ass:sampling}--\ref{ass:growth}, the slope
component $\widehat\beta$ of every measurable minimiser in
\eqref{eq:estimator} satisfies
\begin{equation}\label{eq:expansion}
 \widehat\beta-\beta_0
 =\Gamma_n^{-1}\frac1{nT}\sum_{i=1}^n\sum_{t=1}^Tz_{it}\ps(u_{it})
     +\frac{B_n}{T}+\op\bigl((nT)^{-1/2}\bigr)
\end{equation}
and
\begin{equation}\label{eq:centered}
 \sup_{x\in\R^p}
 \left|\Prb\left\{\sqrt{nT}
    \left(\widehat\beta-\beta_0-\frac{B_n}{T}\right)\leq x\right\}
              -\Prb\{N(0,\Omega_n)\leq x\}\right|\longrightarrow0.
\end{equation}
Equivalently, for the uncorrected estimator,
\begin{equation}\label{eq:shifted}
 \dK\left\{\mathcal L\bigl(\sqrt{nT}(\widehat\beta-\beta_0)\bigr),
       N\bigl(\sqrt{n/T}\,B_n,\Omega_n\bigr)\right\}\longrightarrow0.
\end{equation}
In particular, already under regime W, if $n=T$, the law of
$n(\widehat\beta-\beta_0)$ is approximated in Kolmogorov distance by
$N(B_n,\Omega_n)$.
\end{theorem}

Theorem~\ref{thm:main} shows that the uncorrected estimator is approximately
normal with mean $\beta_0+B_n/T$. Equation \eqref{eq:centered} replaces a
central limit theorem with a fixed limit by a Kolmogorov-distance approximation with the finite-$n$
covariance $\Omega_n$. When the relevant limits exist, the familiar form
follows.

\begin{corollary}[Limit distribution]\label{cor:limit}
If, in addition, $\Omega_n\to\Omega$, $B_n\to B$ and
$n/T\to\upsilon\in[0,\infty)$, then
$\sqrt{nT}(\widehat\beta-\beta_0)\Rightarrow N(\sqrt\upsilon\,B,\Omega)$.
\end{corollary}

No correction is needed when the normalised bias vanishes, which happens
in two leading cases.

\begin{corollary}[Centred cases]\label{cor:centred}
Under Assumptions~\ref{ass:sampling}--\ref{ass:growth}, if
\begin{equation}\label{eq:vanishing-bias}
 \sqrt{n/T}\,B_n\longrightarrow0,
\end{equation}
then $\dK\{\mathcal L(\sqrt{nT}(\widehat\beta-\beta_0)),N(0,\Omega_n)\}
\to0$. Condition \eqref{eq:vanishing-bias} holds with $B_n=0$ identically
if
\begin{equation}\label{eq:zerobias}
 \E[f_i'(0\mid x_{i1})(x_{i1}-\gamma_i)]=0\quad\text{for every }i,
\end{equation}
and \eqref{eq:zerobias} holds (i) if $u_{i1}$ is independent of $x_{i1}$
for every $i$, and then $m_i(v;0)$ of Remark~\ref{rem:outline} is constant in $v$ and the profiled
score at the true slope has mean exactly zero for every $T$; and (ii) if $\tau=1/2$ and
every conditional error density is symmetric about zero, conditional
heteroskedasticity being allowed.
\end{corollary}

Case (i) is in line with \citet{GrahamHahnPowell09}, who show that, for $T=2$,
i.i.d.\ asymmetric Laplace errors and regressors independent of the
errors, the FEQR estimator equals the least absolute deviations
estimator of the differenced model, so that there is no incidental
parameter problem. When $n/T\to\infty$, mere convergence $B_n\to0$ is not
enough, and zero leading bias does not follow from $\tau=1/2$ in a
conditionally asymmetric model.

\begin{remark}[Comparison with smooth panel models]\label{rem:HN}
\citet{HahnNewey04} show that, in nonlinear panel models, the
fixed-effects estimator is asymptotically biased when $T$ grows at the same
rate as $n$, so that asymptotic confidence intervals are incorrect.
Corollary~\ref{cor:limit} is the counterpart for the nonsmooth FEQR
criterion. When $n/T\to\upsilon>0$, the normalised estimator converges to a
normal law with mean $\sqrt\upsilon\,B$, because the incidental parameter
bias $B_n/T$ is of the same order as the standard deviation. In addition,
Theorem~\ref{thm:main} gives $B_n$ in closed form and shows that the
validity of the expansion is governed by the regularity of the conditional
error densities rather than by the differentiability of the criterion.
Conventional intervals centred at $\widehat\beta$ are therefore invalid
whenever $\sqrt{n/T}\,B_n$ does not vanish, and Section~\ref{sec:inference}
provides bias-corrected alternatives. \hfill $\lozenge$
\end{remark}

\begin{remark}[Outline of the argument]\label{rem:outline}
We replace the higher-order expansions of smooth models by three devices:
an exact formula for the mean of the profiled score (from which the bias
follows), an averaging step across individuals and a convexity argument. The
formal statements are in Appendix~\ref{app:lemmas}.

\emph{Rank scores and their exact mean.} For a deterministic
$d=\beta-\beta_0$, put $R_{it}(d)=u_{it}-x_{it}\trans d$,
$\ell=\lfloor T\tau\rfloor$ and $r=T\tau-\ell$, and let the
rank weight $w_{it}(d)$ be one for the $\ell$ smallest of
$R_{i1}(d),\ldots,R_{iT}(d)$, $r$ for the next and zero otherwise. For
fixed $d$, the optimal intercept is a sample $\tau$-quantile, and profiling
it out gives the profiled criterion
$\mathbb M_{iT}(d)=T^{-1}\sum_tR_{it}(d)\{\tau-w_{it}(d)\}$ and the
profiled slope score
$S_{iT}(d)=T^{-1}\sum_tx_{it}\{\tau-w_{it}(d)\}$, a linear rank statistic.
Since the pairs $(x_{it},R_{it}(d))$ are i.i.d.\ over $t$, the
binomial--beta identity gives the mean of each rank weight. Let $F_{i,d}$ be
the distribution function of $R_{i1}(d)$. For an observation at level
$V=F_{i,d}(R_{i1}(d))=v$ of its distribution, the rank weight has mean
$\Prb(Q_T\geq v)$, where $Q_T$ is $\operatorname{Beta}(\ell,T-\ell)$ with
probability $1-r$ and $\operatorname{Beta}(\ell+1,T-\ell-1)$ with
probability $r$, so that $\E Q_T=\tau$ and $\Var(Q_T)=v_\tau/T+O(T^{-2})$.
With $m_i(v;d)=\E(x_{i1}\mid V=v)$ and $G_i(v;d)=\int_0^vm_i(s;d)\,ds$,
this gives the exact identity of Lemma~\ref{lem:beta},
\[
 \E S_{iT}(d)=\tau\E x_{i1}-\E G_i(Q_T;d),
 \qquad
 S_i(d)=\tau\E x_{i1}-G_i(\tau;d),
\]
where $S_i(d)$ is the population score. In words, the mean score is the
population score evaluated not at $\tau$ but at the random level $Q_T$, on
the rank scale, at which the other $T-1$ observations place the cut-off of
the rank weights, and averaged over $Q_T$.

\emph{The bias.} A second-order Taylor expansion of the deterministic
function $G_i(\cdot;d)$ at $\tau$ then gives
$\E S_{iT}(d)=S_i(d)-v_\tau\partial_vm_i(\tau;d)/(2T)+O(\epsilon_T)$, with
$\partial_vm_i(\tau;0)=c_i/a_i^2$. The $T^{-1}$ term is a
half-variance-times-curvature term, and the order of $\epsilon_T$ is set by
the regularity of the conditional densities (Table~\ref{tab:regimes}). If
$u_{i1}$ is independent of $x_{i1}$, then $G_i(\cdot;0)$ is linear and
$\E S_{iT}(0)=0$ exactly. Averaged over individuals, the mean score is
approximately $-\Gamma_nd+b_n/T$. Its root, $B_n/T$, is the displaced centre
of $\widehat\beta-\beta_0$.

\emph{Averaging.} The remainder $S_{iT}(0)-T^{-1}\sum_tz_{it}\ps(u_{it})$
involves only the $O_p(T^{1/2})$ observations whose errors lie between zero
and the sample quantile. Its fluctuations are of order $T^{-3/4}$, centred
and independent across individuals, so their average contributes only
$O_p(T^{-1/4})$ at the $\sqrt{nT}$ scale, whatever the value of $n/T$
(Lemma~\ref{lem:remainder}). This stochastic averaging step is what permits
$n\asymp T$. Growth is restricted only by the mean of the remainder: a
deterministic error of order $T^{-s}$ left after removing $B_n/T$
contributes $\sqrt{nT}\,T^{-s}=T^{(q+1)/2-s}$ when $n=T^q$, which gives the
ranges in Assumption~\ref{ass:growth} (Appendix~\ref{app:growth}).

\emph{Convexity.} Because the profiled criterion is convex, a quadratic
approximation at fixed $d=B_n/T+\theta/\sqrt{nT}$ holds uniformly on compacts
(Lemma~\ref{lem:local}). A bound on the distance between minimisers
\citep{HjortPollard93} then transfers it to every minimiser, so that the
score never needs to be evaluated at the random $\widehat\beta-\beta_0$.
\hfill $\lozenge$
\end{remark}

\begin{remark}[Comparison with existing FEQR theory]\label{rem:KGM}
\citet[Theorem~3.2]{kato2012asymptotics} require $n^2(\log n)^3/T\to0$.
They bound the individual score remainders uniformly, without averaging
across individuals, which leaves an error of order
$\sqrt{nT}(\log n/T)^{3/4}=\{n^2(\log n)^3/T\}^{1/4}$ at the $\sqrt{nT}$
scale. This error does not settle the distribution when $n$ and $T$ are of
comparable size. \citet{galvao2020unbiased} establish unbiased asymptotic
normality under $n(\log T)^2/T\to0$, and Theorem~\ref{thm:main} shows why a
condition of this type is needed for a centred limit. The bias contributes
$\sqrt{n/T}\,B_n$ at the $\sqrt{nT}$ scale, so that at $T=n$ the
uncorrected estimator need not have a mean-zero limit
(Appendix~\ref{app:example}). For $n=\lfloor T^q\rfloor$ with $q>1$, the
sufficient ranges are $q<1+\eta$ in regime H$_\eta$, $q<2$ in regime I and
$q<3$ in regime II. If $n/T\to\infty$, \eqref{eq:expansion} gives
$T(\widehat\beta-\beta_0)-B_n\to_p0$. A diverging normalised bias is thus
compatible with \eqref{eq:centered}, and the deterministic centring is
essential. Finally, the proof yields a Berry--Esseen rate $(nT)^{-1/2}$ for
the linear term in \eqref{eq:expansion}, but only an $o_p(1)$ coupling error
for the estimator. For this reason, \eqref{eq:centered} asserts convergence
to zero and not a rate. \hfill $\lozenge$
\end{remark}

\section{Bias Correction and Inference}\label{sec:inference}

In this section, we develop the split-panel jackknife, which we recommend
as the bias-corrected estimator of the common parameter, together with a
covariance estimator for feasible inference. Appendix~\ref{app:alpha} gives
feasible intervals and simultaneous bands for the individual effects, and
Remark~\ref{rem:abc} briefly discusses an analytical correction.

\subsection{Split-Panel Jackknife}\label{sec:spj}

Split the time indices deterministically into two disjoint sets
$\mathcal T_1$ and $\mathcal T_2$ of sizes $T_1$ and $T_2=T-T_1$, with
$T_j/T\to\lambda_j\in(0,1)$. Fit \eqref{eq:estimator} on each set with
all $n$ individuals, denote the resulting slope estimates by
$\widehat\beta^{(1)}$ and $\widehat\beta^{(2)}$ and, following
\citet{DhaeneJochmans15}, define the split-panel jackknife
(SPJ) estimator
\begin{equation}\label{eq:SPJ}
 \widehat\beta_{\mathrm{SPJ}}=2\widehat\beta
 -\left\{\frac{T_1}{T}\widehat\beta^{(1)}
              +\frac{T_2}{T}\widehat\beta^{(2)}\right\},
\end{equation}
which for $T_1=T_2=T/2$ is the half-panel jackknife
$2\widehat\beta-(\widehat\beta^{(1)}+\widehat\beta^{(2)})/2$.

\begin{theorem}[Split-panel jackknife]\label{thm:spj}
Under Assumptions~\ref{ass:sampling}--\ref{ass:growth}, in every regime,
\begin{equation}\label{eq:spj-expansion}
\begin{gathered}
 \widehat\beta_{\mathrm{SPJ}}-\beta_0
 =\Gamma_n^{-1}\frac1{nT}\sum_{i=1}^n\sum_{t=1}^Tz_{it}\ps(u_{it})
             +\op\bigl((nT)^{-1/2}\bigr),
\\
 \dK\left\{\mathcal L\bigl(\sqrt{nT}
       (\widehat\beta_{\mathrm{SPJ}}-\beta_0)\bigr),
                   N(0,\Omega_n)\right\}\longrightarrow0.
\end{gathered}
\end{equation}
\end{theorem}

\begin{remark}[Relation to the delete-one panel jackknife]\label{rem:jackknife}
The panel jackknife of \citet{HahnNewey04} is a delete-one jackknife
\citep[see][]{DhaeneJochmans15}. It combines the full-sample estimator with
the average of the $T$ leave-one-period-out estimators, with weights $T$ and
$-(T-1)$. For the nonsmooth estimator, however, Theorem~\ref{thm:main}
provides only an $\op((nT)^{-1/2})$ remainder for each fit, and a weight of
order $T$ would require much sharper control of the differences between the
remainders of the $T$ fits. The split-panel jackknife avoids this problem
because its weights are bounded, so that the three remainders stay
negligible. Its bias cancellation, moreover, uses only the $T^{-1}$ form of
the leading bias, which the three fits share because observations are
identically distributed over time. The three fits are correlated, and this
correlation reproduces the full-sample first-order covariance. Fractional
ranks in the subsamples affect only the residual mean errors, which vanish
in every regime.

Extending the range beyond $n=o(T^3)$ by a higher-order jackknife would face
an additional difficulty. By \eqref{eq:beta-exact},
$\Var(Q_T)=v_\tau/T+\{r(1-r)-v_\tau\}/T^2+O(T^{-3})$ depends on the
fractional rank $r$, which does not vary monotonically with $T$. A
second-order bias expansion is therefore not a power series in $T^{-1}$ with
$T$-free coefficients, and different splits can have different fractional
ranks. Consistently with this, \citet{ArellanoHigginsWeidner26} show that
the delete-one jackknife does not remove the leading bias. In the present
notation, the fractional rank changes between $T$ and $T-1$ periods, and
weights of order $T$ turn the difference between the $T^{-2}$ terms of the
two biases into a term of order $T^{-1}$. \hfill $\lozenge$
\end{remark}

\subsection{Covariance Estimation}\label{sec:covariance}

Let $\widehat u_{it}=y_{it}-\widehat\alpha_i-x_{it}\trans\widehat\beta$
be the full-sample residuals.

\begin{assumption}[Kernel and bandwidth]\label{ass:kernel}
The kernel $K:\R\to[0,\infty)$ is compactly supported, twice continuously
differentiable and integrates to one, and $h=C_hT^{-1/8}$ for a fixed
constant $C_h>0$.
\end{assumption}

Define
\begin{align*}
 \widehat a_i&=\frac1{Th}\sum_{t=1}^TK(\widehat u_{it}/h),&
 \widehat\gamma_i&=
   \frac{\sum_tK(\widehat u_{it}/h)x_{it}}
        {\sum_tK(\widehat u_{it}/h)},&
 \widehat z_{it}&=x_{it}-\widehat\gamma_i,\\
 \widehat\Gamma_n&=\frac1{nTh}\sum_{i=1}^n\sum_{t=1}^T
       K(\widehat u_{it}/h)\widehat z_{it}\widehat z_{it}\trans,&
 \widehat V_n&=\frac1{nT}\sum_{i=1}^n\sum_{t=1}^T
       \widehat z_{it}\widehat z_{it}\trans,&
 \widehat\Omega_n&=v_\tau\widehat\Gamma_n^{-1}
                         \widehat V_n\widehat\Gamma_n^{-1},
\end{align*}
with arbitrary values on the events of a zero denominator or a singular
matrix, whose probabilities tend to zero under the conditions of
Proposition~\ref{prop:cov}.

\begin{proposition}[Covariance estimation]\label{prop:cov}
Under Assumptions~\ref{ass:sampling}--\ref{ass:kernel},
\begin{equation}\label{eq:cov-consistency}
 \norm{\widehat\Gamma_n-\Gamma_n}
 +\norm{\widehat V_n-V_n}
 +\norm{\widehat\Omega_n-\Omega_n}\longrightarrow_p0.
\end{equation}
\end{proposition}

This is consistency for a sequence of matrices, not convergence to a fixed
matrix. The proof needs only $h\to0$, $L_{nT}/(Th)\to0$ and
$h^{-2}\sqrt{L_{nT}/T}\to0$. We choose the rate $T^{-1/8}$ because it
also yields the uniform accuracy \eqref{eq:alpha-density-log} of the
$\widehat a_i$ that the bands for the individual effects require
(Appendix~\ref{sec:alpha-inference}).

\begin{corollary}[Confidence intervals for the common parameter]\label{cor:slope-CI}
Under Assumptions~\ref{ass:sampling}--\ref{ass:kernel}, for any fixed
non-zero $v\in\R^p$,
$\sqrt{nT}\,v\trans(\widehat\beta_{\rm SPJ}-\beta_0)/
\sqrt{v\trans\widehat\Omega_n v}\Rightarrow N(0,1)$, so that for
$\delta\in(0,1)$ the interval
\begin{equation}\label{eq:slope-CI}
 v\trans\widehat\beta_{\rm SPJ}\ \pm\
\Phi^{-1}(1-\delta/2)\sqrt{\frac{v\trans\widehat\Omega_n v}{nT}}
\end{equation}
has asymptotic coverage $1-\delta$ for $v\trans\beta_0$.
\end{corollary}

\begin{remark}[Analytical bias correction]\label{rem:abc}
The other remedy of \citet{HahnNewey04}, subtracting a plug-in estimate of
$B_n/T$ from $\widehat\beta$, is also possible here. It requires, however,
stronger conditions than the jackknife. The constant $B_n$ involves the
derivatives of the conditional error densities at zero, and estimating these
derivatives nonparametrically needs an additional bandwidth and, beyond
bounded $n/T$, a rate of convergence that we do not establish. The jackknife
needs neither, and its point estimate involves no bandwidth. This is why we
recommend it. \hfill $\lozenge$
\end{remark}

\section{Serially Dependent Panels}\label{sec:dependence}

Assumption~\ref{ass:sampling} rules out serial correlation in the errors,
serially dependent regressors and feedback from past errors to current
regressors, all of which are common in panels with a long time dimension.
\citet[Section~5]{kato2012asymptotics} extend their results to observations
that are stationary and $\beta$-mixing over time for each individual and
independent across individuals. In this section, we drop independence over
time and work in the framework of \citet{Wu05PNAS,Wu07}. Each individual's
observations form a stationary causal process, and dependence is measured by
coupling, that is, by the effect of replacing either the whole past by an
independent copy, as in the proof of Proposition~5 of \citet{Wu07}, or a
single innovation, as in the functional dependence measure of
\citet{Wu05PNAS}. The dependence on the past may decay at a geometric rate,
as under the geometric moment contraction of \citet{WuShao04}, or at a
polynomial rate. \citet{Wu07} treats M-estimation, including quantile
regression, with nonstochastic regressors. Here, by contrast, regressors and
errors are generated jointly, so that regressors may respond to past
errors, within the limits of condition (b) of Assumption~\ref{ass:dep}
below. Except in Remark~\ref{rem:dep-growth}, $n/T$ is bounded throughout the section. As we show, dependence changes the
bias and the covariance, determines which split-panel jackknife remains
valid and calls for clustering by individual
(Section~\ref{sec:dep-inference}). Appendix~\ref{app:special} treats two
special cases, in which $n/T$ need not be bounded: a class of panels with
possibly unbounded regressors in which the estimator needs no bias
correction, and a model showing that condition (b) of
Assumption~\ref{ass:dep} below cannot be dropped.

\subsection{Framework}\label{sec:dep-framework}

For each $i$, let $(\eta_{it})_{t\in\mathbb Z}$ be i.i.d.\
innovations, of any dimension, with the sequences independent across
individuals, and put $\mathcal F_{it}=(\ldots,\eta_{i,t-1},\eta_{it})$.
Let $(\eta_{it}')_{t\in\mathbb Z}$ be an independent copy of
$(\eta_{it})_{t\in\mathbb Z}$, independent of all other innovations, and,
for $t\geq0$, let
$\mathcal F_{it}^*=(\ldots,\eta'_{i,-1},\eta'_{i0},\eta_{i1},\ldots,\eta_{it})$,
in which the whole past up to time zero is replaced, and let
$\mathcal F_{it}^{\langle0\rangle}=(\ldots,\eta_{i,-1},\eta'_{i0},\eta_{i1},\ldots,\eta_{it})$,
in which only $\eta_{i0}$ is replaced.

\begin{assumption}[Causal panels]\label{ass:causal}
For each $i$ there is a measurable function $H_i$ such that
$W_{it}:=(x_{it}\trans,u_{it})\trans=H_i(\mathcal F_{it})$ for
$t\in\mathbb Z$, \eqref{eq:model} holds for every $i$ and $t$, and
$\norm{x_{it}}\leq M$ almost surely.
\end{assumption}

Under Assumption~\ref{ass:causal}, each $(W_{it})_t$ is strictly
stationary, so the quantities of
Section~\ref{sec:model} are defined with $x_{i0}$ and $u_{i0}$ in place of
$x_{i1}$ and $u_{i1}$. Write $W_{it}^*=H_i(\mathcal F_{it}^*)$,
which has the law of $W_{it}$ and is independent of $\mathcal F_{i0}$, and
$W_{it}^{\langle0\rangle}=H_i(\mathcal F_{it}^{\langle0\rangle})$. In the
following assumption, $\bar m_T=\log T$ under (G) and
$\bar m_T=T^{1/4}$ under (P).

\begin{assumption}[Short-range dependence]\label{ass:dep}
There are constants $\varsigma\in(0,1]$, $C$ and $c>0$ such that the
following hold, where no constant, including $\varrho$, $\nu$ and $C'$
below, depends on $i$, and $\epsilon_0$ is that of
Assumption~\ref{ass:smooth}:
\begin{enumerate}[label=(\alph*),leftmargin=2.4em]
\item one of the following holds:
\begin{itemize}[leftmargin=2.4em]
\item[(G)] for some $\varrho\in(0,1)$,
$\E\norm{W_{it}-W_{it}^*}^\varsigma\leq C\varrho^t$ for all $t\geq1$;
\item[(P)] for some $\nu>9(1+\varsigma)$,
$\E\norm{W_{ik}-W_{ik}^{\langle0\rangle}}^\varsigma\leq Ck^{-\nu-1}$
for all $k\geq1$;
\end{itemize}
\item for every finite $J\subset\mathbb Z\setminus\{0\}$, the conditional
law of $u_{i0}$ given $x_{i0}$ and $(W_{is})_{s\in J}$ has a density
$f_{i,J}$ on $(-\epsilon_0,\epsilon_0)$ with
$|f_{i,J}|+|\partial_uf_{i,J}|\leq C$;
\item $\lambda_{\min}(\Lambda_n)\geq c$ for all sufficiently large
$n$, with $\Lambda_n$ defined in \eqref{eq:dep-bias} below;
\item for every $C_A<\infty$ there is $C'<\infty$ such that, for
every $T\geq3$, every set $\mathcal A\subseteq\{1,\ldots,T\}$ that omits at most
$C_A\bar m_T$ elements of $\{1,\ldots,T\}$ and every set
$J\subset\mathbb Z\setminus\mathcal A$ of at most $C_A\bar m_T$ periods, the counts
$N_{i,\mathcal A}(v,d)=\sum_{s\in\mathcal A}\ind(u_{is}-x_{is}\trans d\leq v)$ have
conditional probabilities
$\pi_{i,\mathcal A,J}(j;v,d)=\Prb\{N_{i,\mathcal A}(v,d)=j\mid(W_{is})_{s\in J}\}$ that
satisfy, outside an event of probability at most $C'T^{-2}$,
\[
 \sup_{|v|,\norm d<\epsilon_0}\max_{j\in\mathbb Z}
 |\pi_{i,\mathcal A,J}(j;v,d)-\pi_{i,\mathcal A,J}(j+1;v,d)|
 \leq\frac{C'(\log T)^{C'}}T.
\]
\end{enumerate}
\end{assumption}

Condition (a)(G) is the geometric moment contraction: the process forgets
its past at a geometric rate. For iterated random functions
$W_{it}=\varpi_i(W_{i,t-1},\eta_{it})$, it holds uniformly in $i$ under
Conditions~1 and~2 of \citet[Theorem~2]{WuShao04}. This is the case, for
instance, when the maps $\varpi_i(\cdot,\eta)$ contract on average, in the
sense that $\E\log\operatorname{Lip}_i(\eta_{i0})<0$, with
$\operatorname{Lip}_i(\eta)$ the Lipschitz constant of $\varpi_i(\cdot,\eta)$,
and satisfy mild moment conditions \citep{DiaconisFreedman99}, as in
threshold and ARCH-type models.
Condition (a)(G) also holds for linear processes with geometrically decaying
coefficients and innovations with a finite moment of some positive order.
Since $\varsigma$ may be small, the errors need not have a finite mean.

Condition (a)(P) bounds the functional dependence measure, that is, the
effect of a single innovation on the observation $k$ periods later. It
implies $\E\norm{W_{it}-W_{it}^*}^\varsigma\leq Ct^{-\nu}$
(Lemma~\ref{lem:dep-poly}) and holds for linear processes
$W_{it}=\sum_{j\geq0}\Pi_{ij}\eta_{i,t-j}$ with
$\E\norm{\eta_{i0}}^\varsigma\leq C$ and
$\norm{\Pi_{ij}}\leq Cj^{-(\nu+1)/\varsigma}$ for $j\geq1$, whose
coefficients may decay too slowly for (G). The proofs compare an observation
with those more than $m_T$ periods away, where $m_T$ is of order $\log T$
under (G) and $T^\mu$, $\mu<1/4$, under (P). The threshold $9(1+\varsigma)$
is sufficient, not sharp.
Condition (a) measures dependence by coupling rather than by mixing. If
$\varsigma=1$ and the errors are integrable, it implies, by Jensen's
inequality, that $W_{it}-\E W_{it}$ is an adapted $L^1$-mixingale, with
coefficients at lag $k$ of order $\varrho^k$ under (G) and, by
Lemma~\ref{lem:dep-poly}, $k^{-\nu}$ under (P) (see also
\citealp[Remark~2.2]{Hill25}, for the implication from functional
dependence to the mixingale property). A
strong mixing sequence with a finite moment of order greater than one is
also an $L^1$-mixingale, but condition (a) covers processes that are not
strong mixing, and hence lie outside the $\beta$-mixing framework of
\citet{kato2012asymptotics} and \citet{galvao2020unbiased}, such as
regressors that follow a first-order autoregression with a coefficient in
$(0,1/2]$ and Bernoulli innovations, in which the current value determines
the whole past (\citealp[Remark~2.4]{Hill25}, and the references there).

Condition (b) extends the density bounds of Assumption~\ref{ass:density}
and condition (W), which are its case $J=\emptyset$, to
conditioning on finitely many other periods.
It cannot be dropped: Proposition~\ref{prop:cusp} in
Appendix~\ref{app:special} gives a 1-dependent model whose one-period
error density is constant near zero but in which the conditional law of
$u_{i0}$ given $u_{i1}$ has an atom, and the estimator has a non-Gaussian
limit when $n/T$ tends to a positive constant.
Condition (b) also limits the feedback from errors to regressors. It allows
regressors that respond to past errors provided that, given its own
regressor and the observations of finitely many other periods, the current
error keeps a bounded and smooth conditional density near zero, as it can
when an independent shock with a smooth density also enters the regressors,
which is the form of feedback used in design PRE in
Section~\ref{sec:sim-dep}. Condition (b) fails when the current error is
determined by its own regressor and the regressors of other periods, as
with a lagged dependent variable, for which the conditional law of
$u_{i0}$ given $x_{i0}$ and $x_{i1}=y_{i0}$ is degenerate.
Condition (c) bounds the long-run covariance matrix $\Lambda_n$ from below.
Condition (d) requires the probabilities of the counts of errors below a
level to change by $O(T^{-1})$, up to a logarithmic factor, between
neighbouring integers, as they do for a binomial count. The exceptional
event allows for extreme values of the conditioning observations, which
matter when the errors are unbounded and persistent, as in a Gaussian
autoregression. Condition (d) holds for
independent observations, after reducing $\epsilon_0$ if necessary, and
rules out, for instance, errors that all come
in tied pairs, which would distort the spacings near the sample quantile at
the scale $T^{-1}$ on which the bias depends (Lemma~\ref{lem:dep-mean}).
Ties among a positive
fraction of the errors, as in Proposition~\ref{prop:cusp}, are compatible
with (d) but excluded by (b).

\subsection{Bias and Gaussian Approximation}\label{sec:dep-bias}

With $\ps$ the score function of Section~\ref{sec:model}, define
\begin{equation}\label{eq:dep-bias}
\begin{gathered}
 \Omega_{\psi,i}=\sum_{h\in\mathbb Z}\E\{\ps(u_{i0})\ps(u_{ih})\},\qquad
 \zeta_i=\sum_{h\neq0}\E\{z_{i0}\ps(u_{ih})\mid u_{i0}=0\},\\
 \Lambda_n=\frac1n\sum_{i=1}^n\sum_{h\in\mathbb Z}
     \E\{z_{i0}z_{ih}\trans\ps(u_{i0})\ps(u_{ih})\},\\
 b_n^{\rm dep}=-\frac1n\sum_{i=1}^n
   \left(\frac{\Omega_{\psi,i}}2\frac{c_i}{a_i^2}+\zeta_i\right),\qquad
 B_n^{\rm dep}=\Gamma_n^{-1}b_n^{\rm dep},\qquad
 \Omega_n^{\rm dep}=\Gamma_n^{-1}\Lambda_n\Gamma_n^{-1},
\end{gathered}
\end{equation}
where
$\E\{z_{i0}\ps(u_{ih})\mid u_{i0}=0\}=a_i^{-1}\E[z_{i0}\ps(u_{ih})f_{i,\{h\}}(0\mid x_{i0},W_{ih})]$
and $c_i$, $a_i$, $z_{it}$ and $\Gamma_n$ are as in
Section~\ref{sec:model}. Lemma~\ref{lem:dep-decay} in
Appendix~\ref{app:dependence} shows that the summands decay geometrically
or polynomially in $|h|$, uniformly in $i$, so the series converge
absolutely and, under Assumption~\ref{ass:nondeg}, all quantities in
\eqref{eq:dep-bias} are uniformly bounded.

The bias has two parts. In the first, the long-run variance
$\Omega_{\psi,i}$ replaces $v_\tau$ in the half-variance-times-curvature term
of Remark~\ref{rem:outline}. This is natural, since $\Omega_{\psi,i}/T$ is,
to first order, the variance of the estimated intercept on the rank scale.
The second part is new. The rank weight of an observation whose error is
near zero depends, through the sample quantile, on the errors of the other
periods. If these errors are correlated with the observation's own
regressor, so is its rank weight, and $\zeta_i/T$ is the resulting
covariance.

\emph{Why a different proof is needed.} Under independence over time,
Lemma~\ref{lem:beta} gives the mean score exactly: the pairs
$(x_{it},u_{it})$ are i.i.d.\ over $t$, so the level at which the other
$T-1$ observations place the cut-off of the rank weights has the
beta-mixture law of $Q_T$, whatever the observation's own regressor, and
the bias follows from a Taylor expansion of a deterministic function at
$\tau$ (Remark~\ref{rem:outline}). Under serial dependence neither step
survives. The periods are not independent, so the cut-off no longer has a
beta-mixture law, and it is not independent of the observation's
regressor, because the errors of nearby periods are correlated with that
regressor; this dependence is what produces $\zeta_i$. The averaging step
of Lemma~\ref{lem:remainder}, which conditions on the errors and uses the
independence of the regressors across periods, fails for the same reason.
Appendix~\ref{app:dependence} therefore proceeds differently. For each
period $t$, it splits the other periods into those within $m_T$ periods of $t$
(Section~\ref{sec:dep-framework}) and the rest, and replaces the innovations that link the distant
periods to period $t$ by independent copies (Lemma~\ref{lem:dep-moments}).
After this coupling, $(x_{it},u_{it})$ is independent of the distant
periods, whose order statistics are approximated, up to a negligible error,
by a linear function of the average of their quantile scores, so that
their spread is governed by $\Omega_{\psi,i}$ rather than $v_\tau$
(Lemma~\ref{lem:dep-quantile}). The distant periods give the first part of
the bias, and the nearby periods, which change the rank weight of period $t$
only when $u_{it}$ is close to the cut-off, give $\zeta_i$
(Lemma~\ref{lem:dep-mean}). Moment bounds derived from the coupling
replace conditional independence in the averaging step
(Lemma~\ref{lem:dep-remainder}). The coupling errors are summed over all
$T$ periods and must remain $o(T^{-1})$, which is what the conditions on
the decay of dependence in Assumption~\ref{ass:dep} secure.

A single function captures both parts. For small $d$, let $q_i(d)$ be the
$\tau$-quantile of $u_{i0}-x_{i0}\trans d$ and
$e_{it}(d)=u_{it}-x_{it}\trans d-q_i(d)$. Let
$A_i(d)=\E f_i\{q_i(d)+x_{i0}\trans d\mid x_{i0}\}$ be the density of
$e_{i0}(d)$ at zero and
$\Omega_{\psi,i}(d)=\sum_{h\in\mathbb Z}\E[\ps\{e_{i0}(d)\}\ps\{e_{ih}(d)\}]$
the long-run variance of the quantile score at the displaced slope, and put
$\kappa_i(d)=\Omega_{\psi,i}(d)/A_i(d)$. Then $A_i(0)=a_i$ and
$\Omega_{\psi,i}(0)=\Omega_{\psi,i}$.

\begin{proposition}[The bias as a gradient]\label{prop:kappa}
Under Assumptions~\ref{ass:density}, \ref{ass:causal} and
\ref{ass:dep}(a), (b), with condition (W) of Assumption~\ref{ass:smooth},
each $\kappa_i$ is differentiable at zero and
\[
 -\frac{\Omega_{\psi,i}}2\frac{c_i}{a_i^2}-\zeta_i=\frac12\nabla\kappa_i(0),
 \qquad\text{so that}\qquad
 b_n^{\rm dep}=\frac1{2n}\sum_{i=1}^n\nabla\kappa_i(0).
\]
\end{proposition}

Proposition~\ref{prop:kappa} thus expresses $b_n^{\rm dep}$ as half the
average gradient of $\kappa_i$, the long-run variance of the quantile score
divided by the density, both evaluated at a displaced slope. Under
independence over time, $\Omega_{\psi,i}(d)=v_\tau$, and the gradient reduces
to the summand of $b_n$ in \eqref{eq:bias}. Under dependence,
$\nabla\Omega_{\psi,i}(0)=-2a_i\zeta_i$ supplies the cross-period term.

\begin{corollary}[Special cases]\label{cor:dep-special}
Under the conditions of Proposition~\ref{prop:kappa} and
Assumption~\ref{ass:nondeg}, the following hold.
\begin{enumerate}[label=(\roman*),leftmargin=2.4em]
\item If the $W_{it}$ are i.i.d.\ over $t$, then $\Omega_{\psi,i}=v_\tau$,
$\zeta_i=0$ and $\Lambda_n=v_\tau V_n$, so that $B_n^{\rm dep}=B_n$ and
$\Omega_n^{\rm dep}=\Omega_n$.
\item If the processes $(x_{it})_t$ and $(u_{it})_t$ are independent,
then $c_i=\zeta_i=0$, so that $B_n^{\rm dep}=0$, and the profiled score of
Remark~\ref{rem:outline} has mean zero at the true slope for every $T$. In general,
$\Omega_n^{\rm dep}\neq\Omega_n$.
\item If
$\Prb\{u_{it}\leq0\mid x_{it},(W_{is})_{s<t}\}=\tau$, so that
$\ps(u_{it})$ is a martingale difference, then $\Omega_{\psi,i}=v_\tau$,
$\Lambda_n=v_\tau V_n$ and
$\zeta_i=\sum_{h\geq1}\E\{z_{ih}\ps(u_{i0})\mid u_{ih}=0\}$.
\end{enumerate}
\end{corollary}

Case (iii) is of particular interest. With martingale-difference scores,
$\Omega_n^{\rm dep}=\Omega_n$, but $\zeta_i\neq0$ in general when past errors
affect current regressors. The estimator is then biased even when $c_i=0$.
This is the quantile counterpart of the bias of the within estimator in
dynamic panels \citep{Nickell81}.

\begin{theorem}[Serially dependent panels]\label{thm:dep}
Suppose that Assumptions~\ref{ass:density}, \ref{ass:nondeg},
\ref{ass:causal} and \ref{ass:dep} hold, together with condition (W) of
Assumption~\ref{ass:smooth}, and that $n,T\to\infty$ with $n/T=O(1)$.
Then the slope component of every measurable minimiser in
\eqref{eq:estimator} satisfies
\begin{equation}\label{eq:dep-expansion}
 \widehat\beta-\beta_0
 =\Gamma_n^{-1}\frac1{nT}\sum_{i=1}^n\sum_{t=1}^Tz_{it}\ps(u_{it})
   +\frac{B_n^{\rm dep}}T+\op\bigl((nT)^{-1/2}\bigr)
\end{equation}
and
\begin{equation}\label{eq:dep-shifted}
 \dK\left\{\mathcal L\bigl(\sqrt{nT}(\widehat\beta-\beta_0)\bigr),
   N\bigl(\sqrt{n/T}\,B_n^{\rm dep},\Omega_n^{\rm dep}\bigr)\right\}
   \longrightarrow0.
\end{equation}
If the split-panel jackknife \eqref{eq:SPJ} uses the contiguous sets
$\mathcal T_1=\{1,\ldots,T_1\}$ and $\mathcal T_2=\{T_1+1,\ldots,T\}$
with $T_1/T\to\lambda_1\in(0,1)$, then
$\dK\{\mathcal L(\sqrt{nT}(\widehat\beta_{\rm SPJ}-\beta_0)),
N(0,\Omega_n^{\rm dep})\}\to0$.
\end{theorem}

Theorem~\ref{thm:dep} shows that, with $n/T$ bounded, the estimator remains
approximately normal with a bias of order $T^{-1}$ and that the contiguous
jackknife removes this bias, as in Theorems~\ref{thm:main}
and~\ref{thm:spj}. Only the bias and the covariance change, and the bias can
be larger or smaller than under independence.

\begin{remark}[Faster growth of $n$]\label{rem:dep-growth}
Theorem~\ref{thm:dep} requires bounded $n/T$ only because
Lemma~\ref{lem:dep-mean} expands the mean of the profiled score with an
$o(T^{-1})$ error. Suppose instead that, for some $\epsilon_T=o(T^{-1})$,
$\E S_{iT}(d)=S_i(d)+b_i(d)/T+O(\epsilon_T)$ uniformly in $i$ and in $d$
near zero, with bounded and uniformly Lipschitz functions $b_i$ that do not
depend on $T$. Then
$b_i(0)=\nabla\kappa_i(0)/2$ by Lemma~\ref{lem:dep-mean} and
Proposition~\ref{prop:kappa}, and the proof of Theorem~\ref{thm:dep} gives
its expansion and Gaussian approximation whenever
$\sqrt{nT}(\epsilon_T+T^{-2})\to0$, and its conclusion for the contiguous
jackknife if this also holds with $\epsilon_{T_1}$ and $\epsilon_{T_2}$ in
place of $\epsilon_T$. The expansion of the mean can be
verified for various classes of processes, but the arguments are lengthy
and are omitted. Appendix~\ref{sec:dep-nobias} gives a class of panels,
with possibly unbounded regressors, in which the mean of the profiled score
is an exact multiple of the population score, so that it vanishes at the
true slope for every $T$, and in which a separate argument
removes any restriction on the growth of $n$. \hfill $\lozenge$
\end{remark}

\begin{remark}[Choice of the split]\label{rem:dep-split}
The contiguous sets are stationary panels of lengths $T_1$ and $T_2$ with
the same constant $B_n^{\rm dep}$ as the full panel (Figure~\ref{fig:split}),
so the bias cancels as in Theorem~\ref{thm:spj}. The interleaved sets of odd
and even periods are also stationary panels, but they inherit the
dependence of the subsampled process $(W_{i,2t})_t$, which satisfies
Assumption~\ref{ass:causal} and conditions (a) and (b) of
Assumption~\ref{ass:dep} whenever $(W_{it})_t$ does. Their constant
$B_n^{{\rm dep},(2)}$ is given by \eqref{eq:dep-bias} with $h$ replaced by
$2h$. If the subsampled processes also satisfy conditions (c) and (d) of
Assumption~\ref{ass:dep}, the jackknife based on them is approximately
$N(2\sqrt{n/T}(B_n^{\rm dep}-B_n^{{\rm dep},(2)}),\Omega_n^{\rm dep})$ at the
$\sqrt{nT}$ scale. It is therefore not centred unless
$\sqrt{n/T}\,\norm{B_n^{\rm dep}-B_n^{{\rm dep},(2)}}\to0$, and the difference
can be large. In design AR of Section~\ref{sec:sim-dep} at $\tau=0.9$,
$B_n^{\rm dep}=-2.896$ and $B_n^{{\rm dep},(2)}=-1.943$, so the interleaved
jackknife retains a $T\times$bias of $-1.906$. In design PRE, the subsampled
process is i.i.d.\ over $t$, so that $B_n^{{\rm dep},(2)}=B_n$; at
$\tau=0.5$, where $B_n=0$, the interleaved jackknife doubles the bias of the
uncorrected estimator. Finally, the analytical correction of
Remark~\ref{rem:abc} targets $B_n$ rather than $B_n^{\rm dep}$, and so
removes the bias only if $\sqrt{n/T}\,\norm{B_n-B_n^{\rm dep}}\to0$.
\hfill $\lozenge$
\end{remark}

\begin{figure}[!htb]
\centering
\begin{tikzpicture}[x=0.5cm,y=1cm,
  h1/.style={circle,fill=figblue,draw=white,line width=0.7pt,inner sep=0pt,minimum size=6.4pt},
  h2/.style={rectangle,fill=figorange,draw=white,line width=0.7pt,inner sep=0pt,minimum size=5.8pt}]
 \node[anchor=east,font=\small] at (0.3,0) {(a) Contiguous};
 \node[anchor=east,font=\small] at (0.3,-1.6) {(b) Interleaved};
 \foreach \t in {1,...,12} {
   \ifnum\t<7 \node[h1] (a\t) at (\t,0) {}; \else \node[h2] (a\t) at (\t,0) {}; \fi
   \ifodd\t \node[h1] (b\t) at (\t,-1.6) {}; \else \node[h2] (b\t) at (\t,-1.6) {}; \fi }
 \foreach \t [evaluate=\t as \u using int(\t+1)] in {1,...,5}
   \draw[figblue,line width=0.9pt] (a\t) to[bend left=70] (a\u);
 \foreach \t [evaluate=\t as \u using int(\t+1)] in {7,...,11}
   \draw[figorange,line width=0.9pt] (a\t) to[bend right=70] (a\u);
 \foreach \t [evaluate=\t as \u using int(\t+2)] in {1,3,5,7,9}
   \draw[figblue,line width=0.9pt] (b\t) to[bend left=45] (b\u);
 \foreach \t [evaluate=\t as \u using int(\t+2)] in {2,4,6,8,10}
   \draw[figorange,line width=0.9pt] (b\t) to[bend right=45] (b\u);
 \node[anchor=west,font=\small,align=left] at (12.7,0)
   {neighbours in a half are one period\\ apart: constant $B_n^{\rm dep}$};
 \node[anchor=west,font=\small,align=left] at (12.7,-1.6)
   {neighbours in a half are two periods\\ apart: constant $B_n^{{\rm dep},(2)}$};
 \node[font=\small] at (6.5,-2.45) {periods $t=1,\ldots,T$ of one individual};
\end{tikzpicture}
\caption{Contiguous and interleaved half-panels.
Circles mark the periods of the first half-panel and squares those of the
second, and arcs join periods that are adjacent within a half-panel. Contiguous halves keep the
dependence of $(W_{it})_t$ and hence the constant $B_n^{\rm dep}$ of the
full panel, so that the jackknife removes the bias. Interleaved halves
carry the dependence of $(W_{i,2t})_t$, whose constant
$B_n^{{\rm dep},(2)}$ differs in general (Remark~\ref{rem:dep-split}).}
\label{fig:split}
\end{figure}

\subsection{Inference}\label{sec:dep-inference}

The estimator $\widehat\Omega_n$ of Section~\ref{sec:covariance} uses
$v_\tau\widehat V_n$ in place of $\Lambda_n$. It is therefore inconsistent
unless $\Lambda_n-v_\tau V_n\to0$, as happens for martingale-difference
scores (Corollary~\ref{cor:dep-special}(iii)). Clustering by individual
\citep{arellano1987computing,Hansen07} instead estimates the long-run
covariance without a lag truncation.
With the residuals and the
estimators of Section~\ref{sec:covariance}, let
\begin{equation}\label{eq:cluster}
 \widehat s_i=\sum_{t=1}^T\widehat z_{it}\ps(\widehat u_{it}),\qquad
 \widehat\Lambda_n^{\rm CR}=\frac1{nT}\sum_{i=1}^n\widehat s_i\widehat s_i\trans,
 \qquad
 \widehat\Omega_n^{\rm CR}=\widehat\Gamma_n^{-1}\widehat\Lambda_n^{\rm CR}
                            \widehat\Gamma_n^{-1}.
\end{equation}

\begin{proposition}[Clustered covariance estimation]\label{prop:cluster}
Under the conditions of Theorem~\ref{thm:dep} and
Assumption~\ref{ass:kernel},
$\norm{\widehat\Lambda_n^{\rm CR}-\Lambda_n}+\norm{\widehat\Omega_n^{\rm CR}
-\Omega_n^{\rm dep}}\to_p0$. Consequently, for fixed non-zero $v\in\R^p$
and $\delta\in(0,1)$, the interval \eqref{eq:slope-CI}, with
$\widehat\Omega_n^{\rm CR}$ in place of $\widehat\Omega_n$ and with
$\widehat\beta_{\rm SPJ}$ computed from the contiguous split, has asymptotic
coverage $1-\delta$ for $v\trans\beta_0$.
\end{proposition}

Under Assumptions~\ref{ass:sampling}, \ref{ass:density}
and~\ref{ass:nondeg} and condition (W), the observations satisfy
Assumptions~\ref{ass:causal} and~\ref{ass:dep}, after reducing $\epsilon_0$
if necessary. With bounded $n/T$, the clustered estimator can therefore be
used whether or not the data are serially dependent.

\begin{remark}[Bartlett HAC alternative]\label{rem:hac}
A pooled heteroskedasticity and autocorrelation consistent
(HAC) estimator weights the within-individual autocovariances
$\widehat\Lambda_n(j)=(nT)^{-1}\sum_{i=1}^n\sum_{t=j+1}^T\widehat s_{it}\widehat s_{i,t-j}\trans$
of the fitted scores $\widehat s_{it}=\widehat z_{it}\ps(\widehat u_{it})$
with the Bartlett kernel \citep{NeweyWest87,Andrews91}:
\[
 \widehat\Lambda_n^{\rm HAC}=\widehat\Lambda_n(0)
   +\sum_{j=1}^{b_T}\Bigl(1-\frac j{b_T+1}\Bigr)
   \{\widehat\Lambda_n(j)+\widehat\Lambda_n(j)\trans\}.
\]
Under the conditions of Proposition~\ref{prop:cluster}, it is consistent
for $\Lambda_n$ if $b_T\to\infty$ and $b_T=o(T^{1/8})$
(Appendix~\ref{app:hac}). With positively autocorrelated scores, however,
the Bartlett weights make it too small by an amount that decreases no faster
than $1/b_T$, while its variance grows with $b_T$, so that it needs a large
$T$. The clustered estimator, by contrast, needs no truncation, and its
precision is governed by $n$. HAC estimators converge faster under weak
dependence \citep{Hansen07}, but this matters only when $n$ is small and $T$
is large.
\hfill $\lozenge$
\end{remark}

\section{Simulation Studies}\label{sec:simulation}

In this section, we use simulated data to examine the finite-sample
performance of the uncorrected estimator, the split-panel jackknife and the
proposed standard errors for the common parameter, first in independent and
then in serially dependent panels. Simulation results for the individual
effects are in Appendix~\ref{sec:alpha-sim}.

\phantomsection\label{sec:sim-design}
Following \citet{kato2012asymptotics} and
\citet{ChiangGalvaoWei26}, we generate data from the location-scale model
\[
 y_{it}=\alpha_i+\beta x_{it}+u_{it},\qquad
 u_{it}=(1+\vartheta x_{it})(\varepsilon_{it}-q_\tau),\qquad
 x_{it}=\omega_{it}+0.3\,\alpha_i,
\]
where $\alpha_i\sim U(0,1)$, $\omega_{it}\sim\chi^2_3$ and
$\varepsilon_{it}\sim N(0,1)$ are mutually independent and i.i.d., and
$q_\tau=\Phi^{-1}(\tau)$. We set $\beta=\vartheta=1$ throughout. Since
$1+\vartheta x_{it}>0$, the conditional $\tau$-quantile of $u_{it}$ given
$x_{it}$ is zero, so that $\beta_0=1$ at every $\tau$. As in those papers, the
regressor is unbounded, which goes beyond the bounded regressors of
Assumption~\ref{ass:density}. We vary
$\tau\in\{0.5,0.75,0.9\}$, $T\in\{50,100,200\}$ and $n\in\{T,10T\}$ across
designs, with 2000 Monte Carlo replications for each. The uncorrected estimator FE is computed with the sparse
interior-point routine of \texttt{quantreg} \citep{quantreg,KoenkerNg03}, and
SPJ uses the first and last $T/2$ periods. Both $\widehat\Omega_n$ and
$\widehat a_i$ use the triweight kernel with bandwidth $h=\widehat sT^{-1/8}$,
where $\widehat s$ is the interquartile range of the pooled residuals divided
by $1.349$; this data-dependent scale takes the place of the fixed constant
$C_h$ of Assumption~\ref{ass:kernel}.

\begin{table}[!t]
\centering
\begin{threeparttable}
\caption{Bias, dispersion, standard errors and coverage of 95 per cent intervals for the common parameter}
\label{tab:sim-slope}
\small
\setlength{\tabcolsep}{8pt}
\begin{tabular}{rrrrrrrrrr}
\toprule
 & & & \multicolumn{2}{c}{$T\times$bias} & \multicolumn{2}{c}{sd$/\sigma_{nT}$} & & \multicolumn{2}{c}{Coverage}\\
\cmidrule(lr){4-5}\cmidrule(lr){6-7}\cmidrule(lr){9-10}
$T$ & $n$ & $r_{nT}$ & FE & SPJ & FE & SPJ & se$/\sigma_{nT}$ & FE & SPJ\\
\midrule
\multicolumn{10}{l}{$\tau=0.9$: $B_n=-1.534$, $\Omega_n=18.76$}\\
50 & 50 & 0.35 & $-$1.51 & 0.64 & 1.03 & 1.21 & 1.00 & 0.920 & 0.888\\
 & 500 & 1.12 & $-$1.66 & 0.38 & 1.03 & 1.19 & 1.02 & 0.769 & 0.890\\
100 & 100 & 0.35 & $-$1.57 & 0.01 & 1.01 & 1.14 & 1.00 & 0.926 & 0.912\\
 & 1000 & 1.12 & $-$1.54 & 0.06 & 1.00 & 1.12 & 1.00 & 0.800 & 0.913\\
200 & 200 & 0.35 & $-$1.50 & 0.06 & 1.00 & 1.08 & 0.99 & 0.933 & 0.931\\
 & 2000 & 1.12 & $-$1.59 & $-$0.01 & 0.99 & 1.07 & 1.00 & 0.788 & 0.933\\
\midrule
\multicolumn{10}{l}{$\tau=0.75$: $B_n=-0.513$, $\Omega_n=11.92$}\\
50 & 50 & 0.15 & $-$0.55 & 0.04 & 1.03 & 1.15 & 1.01 & 0.932 & 0.914\\
 & 500 & 0.47 & $-$0.60 & $-$0.04 & 1.00 & 1.09 & 1.02 & 0.920 & 0.933\\
100 & 100 & 0.15 & $-$0.46 & 0.08 & 0.98 & 1.07 & 1.01 & 0.953 & 0.939\\
 & 1000 & 0.47 & $-$0.55 & 0.01 & 1.00 & 1.06 & 1.02 & 0.925 & 0.940\\
200 & 200 & 0.15 & $-$0.50 & 0.01 & 1.01 & 1.07 & 1.01 & 0.946 & 0.934\\
 & 2000 & 0.47 & $-$0.54 & $-$0.01 & 1.01 & 1.06 & 1.01 & 0.920 & 0.943\\
\midrule
\multicolumn{10}{l}{$\tau=0.5$: $B_n=0$, $\Omega_n=10.08$}\\
50 & 50 & 0.00 & $-$0.10 & $-$0.11 & 1.00 & 1.11 & 1.02 & 0.942 & 0.916\\
 & 500 & 0.00 & 0.01 & 0.01 & 1.01 & 1.09 & 1.03 & 0.953 & 0.929\\
100 & 100 & 0.00 & 0.04 & 0.03 & 1.01 & 1.10 & 1.02 & 0.948 & 0.930\\
 & 1000 & 0.00 & 0.01 & 0.01 & 1.01 & 1.08 & 1.02 & 0.947 & 0.933\\
200 & 200 & 0.00 & $-$0.04 & $-$0.03 & 1.03 & 1.09 & 1.01 & 0.949 & 0.933\\
 & 2000 & 0.00 & 0.00 & $-$0.01 & 1.00 & 1.05 & 1.01 & 0.955 & 0.941\\
\bottomrule
\end{tabular}
\begin{tablenotes}[flushleft]\footnotesize
\item \emph{Notes:} FE is the uncorrected estimator $\widehat\beta$ of
\eqref{eq:estimator}, and SPJ is $\widehat\beta_{\rm SPJ}$ of \eqref{eq:SPJ}
with $T_1=T_2=T/2$. $\sigma_{nT}=\{\Omega_n/(nT)\}^{1/2}$ is the
first-order standard deviation of both, and
$r_{nT}=\sqrt{n/T}\,|B_n|/\Omega_n^{1/2}$ is the absolute first-order bias
$|B_n|/T$ of FE in units of $\sigma_{nT}$. $B_n$ and $\Omega_n$ of
\eqref{eq:bias} are computed by quadrature, with the average over $i$
replaced by the expectation over $\alpha_i$. $T\times$bias is $T$ times
the Monte Carlo mean of the estimation error, sd its Monte Carlo
standard deviation and se the Monte Carlo mean of
$\{\widehat\Omega_n/(nT)\}^{1/2}$. Coverage is that of the nominal 95
per cent intervals of $\pm1.96$ standard errors around FE and SPJ. With 2000 replications per
design, the Monte Carlo standard errors are at most $0.100$ (FE) and
$0.118$ (SPJ) for $T\times$bias, and that of a coverage rate near $0.95$ is
$0.0049$.
\end{tablenotes}
\end{threeparttable}
\end{table}

\phantomsection\label{sec:sim-slope}
Table~\ref{tab:sim-slope} summarises the results, which we highlight in
four points. First, in every design, the $T\times$bias of FE is close to
$B_n$, which is zero at $\tau=0.5$ (Corollary~\ref{cor:centred}), and its
standard deviation is $0.98$--$1.03$ times
$\sigma_{nT}=\{\Omega_n/(nT)\}^{1/2}$, as Theorem~\ref{thm:main} predicts.
Second, the conventional interval therefore undercovers when the bias is
large relative to the standard deviation. At $\tau=0.9$, its asymptotic
coverage $\Phi(1.96-r_{nT})-\Phi(-1.96-r_{nT})$, with
$r_{nT}=\sqrt{n/T}\,|B_n|/\Omega_n^{1/2}$, is $0.936$ for $n=T$ and $0.799$
for $n=10T$ (Corollary~\ref{cor:limit}), and for $T\geq100$ the simulated
coverage is $0.926$--$0.933$ and $0.788$--$0.800$, respectively. Third, SPJ
removes the bias for $T\geq100$. At $T=50$ and $\tau=0.9$, however, it
over-corrects, because the $T^{-1}$ form of the bias on which it relies
(Remark~\ref{rem:jackknife}) is not yet accurate in half-panels of 25
periods. Fourth, the standard error is accurate in every design
(Proposition~\ref{prop:cov}). For $T\geq100$, the interval
\eqref{eq:slope-CI} around SPJ covers $0.912$--$0.943$, falling short of
$0.95$ because SPJ is more dispersed than FE in finite samples, with a
standard deviation of $1.05$--$1.14$ times $\sigma_{nT}$.

In summary, the jackknife pays off when the normalised bias $r_{nT}$ is
large. At $n=10T$ and $\tau=0.9$, where $r_{nT}=1.12$, its interval covers
$0.890$--$0.933$, against $0.769$--$0.800$ for the conventional interval.
When $r_{nT}\leq0.35$, as for $n=T$, the conventional interval is at least
as accurate.

\phantomsection\label{sec:sim-dep}
We use two designs to illustrate Section~\ref{sec:dependence}. In design AR,
the model is as above, except that $\varepsilon_{it}$ is a stationary
Gaussian first-order autoregression with coefficient $0.5$ and unit
variance. Since the $\varepsilon_{it}$ are independent of the regressors,
$\zeta_i=0$, and the bias is that of the independent design multiplied by
$\Omega_{\psi,i}/v_\tau$. In design PRE, the $\varepsilon_{it}$ are i.i.d.\
$N(0,1)$, $x_{it}=0.3\,\alpha_i+\varepsilon_{i,t-1}+\omega_{it}$ with
$\omega_{it}$ i.i.d.\ $N(0,1)$, and
$u_{it}=\exp(x_{it}/2)(\varepsilon_{it}-q_\tau)$, so that the regressor
responds to the previous shock. The scores are martingale differences, but
this feedback gives $B_n^{\rm dep}=-0.370$ at $\tau=0.5$, where $B_n=0$. We
take $\tau\in\{0.5,0.9\}$ and
$(T,n)\in\{(50,50),(50,500),(100,100),(200,200)\}$, with 2000 replications
(1000 at $T=200$) and a burn-in of 50 periods. SPJ-c uses the first and last
$T/2$ periods, and SPJ-i the odd and even periods. The standard errors are
those of Propositions~\ref{prop:cov} (iid) and~\ref{prop:cluster}
(cluster-robust, CR).

\begin{table}[!t]
\centering
\begin{threeparttable}
\caption{Serially dependent panels: bias, dispersion, standard errors and coverage of 95 per cent intervals for the common parameter}
\label{tab:sim-dep}
\footnotesize
\setlength{\tabcolsep}{3.6pt}
\begin{tabular}{rrrrrrrrrrrrr}
\toprule
 & & & \multicolumn{3}{c}{$T\times$bias} & \multicolumn{2}{c}{sd$/\sigma^{\rm dep}_{nT}$} & \multicolumn{2}{c}{se$/\sigma^{\rm dep}_{nT}$} & \multicolumn{3}{c}{Coverage}\\
\cmidrule(lr){4-6}\cmidrule(lr){7-8}\cmidrule(lr){9-10}\cmidrule(lr){11-13}
$T$ & $n$ & $r^{\rm dep}_{nT}$ & FE & SPJ-c & SPJ-i & FE & SPJ-c & iid & CR & FE, CR & SPJ-c, iid & SPJ-c, CR\\
\midrule
\multicolumn{13}{l}{Design AR, $\tau=0.9$: $B_n^{\rm dep}=-2.896$ ($B_n=-1.534$), $\Omega_n^{\rm dep}=21.35$ ($\Omega_n=18.76$)}\\
50 & 50 & 0.63 & $-$3.06 & 0.10 & $-$1.43 & 1.00 & 1.21 & 0.93 & 0.94 & 0.863 & 0.871 & 0.872\\
 & 500 & 1.98 & $-$2.84 & 0.28 & $-$1.30 & 1.01 & 1.22 & 0.94 & 0.97 & 0.470 & 0.866 & 0.875\\
100 & 100 & 0.63 & $-$2.88 & 0.02 & $-$1.75 & 1.00 & 1.16 & 0.93 & 0.97 & 0.889 & 0.880 & 0.894\\
200 & 200 & 0.63 & $-$2.98 & $-$0.13 & $-$1.95 & 0.99 & 1.10 & 0.93 & 0.98 & 0.890 & 0.898 & 0.914\\
\midrule
\multicolumn{13}{l}{Design AR, $\tau=0.5$: $B_n^{\rm dep}=B_n=0$, $\Omega_n^{\rm dep}=12.14$ ($\Omega_n=10.08$)}\\
50 & 50 & 0.00 & $-$0.10 & $-$0.12 & $-$0.09 & 1.01 & 1.14 & 0.92 & 0.99 & 0.940 & 0.886 & 0.912\\
 & 500 & 0.00 & $-$0.02 & $-$0.02 & $-$0.03 & 1.01 & 1.13 & 0.92 & 1.01 & 0.941 & 0.895 & 0.916\\
100 & 100 & 0.00 & 0.01 & $-$0.01 & 0.02 & 0.99 & 1.07 & 0.92 & 1.01 & 0.952 & 0.904 & 0.928\\
200 & 200 & 0.00 & 0.03 & 0.00 & 0.05 & 1.00 & 1.07 & 0.92 & 1.01 & 0.947 & 0.913 & 0.936\\
\midrule
\multicolumn{13}{l}{Design PRE, $\tau=0.9$: $B_n^{\rm dep}=-1.490$ ($B_n=-1.295$), $\Omega_n^{\rm dep}=\Omega_n=1.541$}\\
50 & 50 & 1.20 & $-$1.46 & 0.21 & $-$0.39 & 1.11 & 1.35 & 1.16 & 1.08 & 0.789 & 0.900 & 0.869\\
 & 500 & 3.79 & $-$1.42 & 0.22 & $-$0.38 & 1.16 & 1.40 & 1.16 & 1.10 & 0.102 & 0.872 & 0.850\\
100 & 100 & 1.20 & $-$1.41 & 0.00 & $-$0.52 & 1.07 & 1.24 & 1.09 & 1.05 & 0.791 & 0.914 & 0.899\\
200 & 200 & 1.20 & $-$1.48 & $-$0.08 & $-$0.55 & 1.08 & 1.19 & 1.04 & 1.02 & 0.765 & 0.911 & 0.899\\
\midrule
\multicolumn{13}{l}{Design PRE, $\tau=0.5$: $B_n^{\rm dep}=-0.370$ ($B_n=0$), $\Omega_n^{\rm dep}=\Omega_n=0.829$}\\
50 & 50 & 0.41 & $-$0.43 & 0.05 & $-$0.85 & 1.09 & 1.23 & 1.20 & 1.20 & 0.945 & 0.951 & 0.942\\
 & 500 & 1.29 & $-$0.42 & 0.08 & $-$0.84 & 1.09 & 1.19 & 1.20 & 1.22 & 0.808 & 0.947 & 0.949\\
100 & 100 & 0.41 & $-$0.39 & 0.05 & $-$0.81 & 1.05 & 1.14 & 1.16 & 1.18 & 0.954 & 0.951 & 0.949\\
200 & 200 & 0.41 & $-$0.37 & 0.05 & $-$0.74 & 0.98 & 1.04 & 1.13 & 1.16 & 0.965 & 0.966 & 0.969\\
\bottomrule
\end{tabular}
\begin{tablenotes}[flushleft]\footnotesize
\item \emph{Notes:} FE is the uncorrected estimator, and SPJ-c and SPJ-i
are the half-panel jackknives with contiguous and with interleaved
halves. $\sigma^{\rm dep}_{nT}=\{\Omega_n^{\rm dep}/(nT)\}^{1/2}$ and
$r^{\rm dep}_{nT}=\sqrt{n/T}\,|B_n^{\rm dep}|/(\Omega_n^{\rm dep})^{1/2}$, with
$B_n^{\rm dep}$ and $\Omega_n^{\rm dep}$ of \eqref{eq:dep-bias} computed by
quadrature. The values $B_n$ and $\Omega_n$ in parentheses are those of
\eqref{eq:bias}, which assume independence over time. $T\times$bias, sd
and se are as in Table~\ref{tab:sim-slope}, with iid and CR the
standard errors of Propositions~\ref{prop:cov} and~\ref{prop:cluster}.
Coverage is that of the nominal 95 per cent intervals of $\pm1.96$
standard errors around FE, with CR, and around SPJ-c, with iid and with
CR. The Monte Carlo standard errors of $T\times$bias are at most $0.15$
(FE) and $0.16$ (SPJ-c and SPJ-i) in design AR and $0.05$ in design PRE,
and that of a coverage rate near $0.95$ is
$0.0049$ with 2000 and $0.0069$ with 1000 replications.
\end{tablenotes}
\end{threeparttable}
\end{table}

Table~\ref{tab:sim-dep} summarises the results. First, the $T\times$bias of
FE is close to $B_n^{\rm dep}$ rather than to $B_n$, and in design PRE, FE is
biased even at the median. Second, SPJ-c removes the bias, whereas SPJ-i does
not: in the two cases of Remark~\ref{rem:dep-split}, its $T\times$bias
approaches the predicted values, $-1.906$ and $-0.740$, as $T$ grows. Third,
in design AR, the iid standard error understates the dispersion, and the
clustered one largely corrects this. Fourth, with clustered standard errors,
the interval around SPJ-c covers $0.912$--$0.969$ at $\tau=0.5$ and
$0.850$--$0.914$ at $\tau=0.9$, where, as in Table~\ref{tab:sim-slope}, the
excess dispersion of the jackknife keeps it below $0.95$. By contrast, the
interval around FE covers only $0.470$ (AR) and $0.102$ (PRE) at $\tau=0.9$
and $n=10T$. These results are consistent with Theorem~\ref{thm:dep} and
Proposition~\ref{prop:cluster}.

\FloatBarrier
\section{Conclusion}\label{sec:conclusion}

In this paper, we develop a large-$n$, large-$T$ theory for the unsmoothed
FEQR estimator that allows for serially correlated panel data. Its leading
bias is of order $T^{-1}$ and available in closed form, and the split-panel
jackknife, combined with a consistent covariance estimator, gives valid
inference, under independence over time, over a range of growth rates set
by the regularity of the error densities. Serial dependence changes the structure of this bias: the
long-run variance of the quantile score enters it, and feedback from past
errors to current regressors adds a term of its own. The jackknife then
remains valid with contiguous half-panels but not, in general, with
interleaved ones, and inference needs standard errors that are robust to
serial dependence, such as those clustered by individual. In a class of
serially dependent panels with possibly unbounded regressors, no bias
correction is needed, whatever the growth of $n$ relative to $T$, and
without a regularity condition on the joint law of the errors over time,
the limit can be non-Gaussian (Appendix~\ref{app:special}). Simulation studies
support these results, and an application to firm investment illustrates
how the scaling of the bias with $T$ can help to separate an incidental
parameter bias from a change in the panel over time.

Three extensions remain open. First, under serial dependence, the theory
remains to be developed for faster growth of $n$ outside the class of
Proposition~\ref{prop:nobias} (Remark~\ref{rem:dep-growth}) and for panels
whose law changes over time, which the application suggests can matter in
practice. Second, with two-way fixed
effects, treated for smooth models by \citet{FernandezValWeidner16}, the
residuals net of the estimated time effects are no longer i.i.d.\ over time.
Third, in dynamic panels, studied by \citet{HahnKuersteiner11} and, for
quantile regression, by \citet{galvao2011quantile}, the lagged outcomes are
unbounded unless the process is restricted, and they violate condition (b)
of Assumption~\ref{ass:dep}.

\newpage
\appendix
\section*{Appendix}
\addcontentsline{toc}{section}{Appendix}

\section{An Example with Non-Zero Bias}\label{app:example}

Let all individuals share the regressor--error law
\[
 x\sim\operatorname{Bernoulli}(1/2),\quad
 u=(1+x)(\varepsilon-q_\tau),\quad
 \varepsilon\sim N(0,1)\ \text{independent of }x,
\]
where $q_\tau=\Phi^{-1}(\tau)$. This design satisfies \eqref{eq:model} with $\beta_0=0$ although $u$ and $x$ are
dependent through the scale. We drop the indices $i$ and $n$, so that
$a$, $\gamma$, $c$, $b$, $\Gamma$, $V$, $B$ and $\Omega$ denote the
common values of $a_i$, $\gamma_i$, $c_i$, $b_n$, $\Gamma_n$, $V_n$,
$B_n$ and $\Omega_n$. The conditional density is $f(u\mid x)=\phi\{u/(1+x)+q_\tau\}/(1+x)$, so that
$f(0\mid x)=\phi(q_\tau)/(1+x)$, $f'(0\mid x)=-q_\tau\phi(q_\tau)/(1+x)^2$
and
\[
\begin{gathered}
 a=\tfrac34\phi(q_\tau),\quad \gamma=\tfrac13,\quad
 \Gamma=\frac{\phi(q_\tau)}6,\quad V=\frac5{18},\quad
 c=\frac{q_\tau\phi(q_\tau)}{12},\\
 B=-\frac{4\tau(1-\tau)q_\tau}{9\phi(q_\tau)^2},\qquad
 \Omega=\frac{10\tau(1-\tau)}{\phi(q_\tau)^2}.
\end{gathered}
\]
At $\tau=0.75$, $B\simeq-0.556609$ and $\Omega\simeq18.567675$, so at
$T=n$ Corollary~\ref{cor:limit} gives
$n(\widehat\beta-\beta_0)\Rightarrow N(-0.556609,18.567675)$, whereas, by
Theorem~\ref{thm:spj}, $n(\widehat\beta_{\rm SPJ}-\beta_0)\Rightarrow
N(0,18.567675)$. The example thus shows that, under conditional heteroskedasticity, the
mean-zero limit theorems available when $T$ dominates $n$ need not
extend to $n\asymp T$. The normal density satisfies (II), so for
$n=\lfloor T^q\rfloor$ with $1<q<3$,
$T^{(q+1)/2}(\widehat\beta-\beta_0-B/T)\Rightarrow N(0,\Omega)$ and
$T^{(q+1)/2}(\widehat\beta_{\mathrm{SPJ}}-\beta_0)\Rightarrow N(0,\Omega)$,
while the normalised bias of the uncorrected estimator grows like
$T^{(q-1)/2}$.

\begin{table}[!htb]
\centering
\begin{threeparttable}
\caption{Deterministic check of the bias coefficient and the curvature ($\tau=0.75$)}
\label{tab:check}
\begin{tabular}{rrrr}
\toprule
$T$ & $T\E S_{iT}(0)$ & $T\beta_T$
    & $-\partial_d\E S_{iT}(0)$\\
\midrule
25  & $-0.030435$ & $-0.598119$ & $0.050851$\\
50  & $-0.030222$ & $-0.581827$ & $0.051923$\\
100 & $-0.029444$ & $-0.561173$ & $0.052458$\\
200 & $-0.029464$ & $-0.558912$ & $0.052711$\\
400 & $-0.029472$ & $-0.557765$ & $0.052837$\\
800 & $-0.029476$ & $-0.557188$ & $0.052900$\\
\midrule
Limit & $-0.029480$ & $-0.556609$ & $0.052963$\\
\bottomrule
\end{tabular}
\begin{tablenotes}[flushleft]\footnotesize
\item \emph{Notes:} $S_{iT}$ is the profiled slope score \eqref{eq:profile-score} and $\beta_T$
the root of $\E S_{iT}(\cdot)=0$. The limits of the last three columns are
$b=\Gamma B$, $B$ and $\Gamma$.
\end{tablenotes}
\end{threeparttable}
\end{table}

Table~\ref{tab:check} checks the coefficients by one-dimensional
quadrature of the exact representation \eqref{eq:beta-exact} of Lemma~\ref{lem:beta}. Here $\beta_T$ is the root of the mean profiled
score, the pseudo-true slope for fixed $T$ (recall $\beta_0=0$), so
$T\beta_T$ should approach $B$. These are
population quantities, not Monte Carlo estimates. The third column
confirms the sign and scale of $B$, the fourth the curvature $\Gamma$. The
irregularity of the second and third columns between $T=25$, $50$ and
$100$ comes from the term of $\Var(Q_T)$ in \eqref{eq:beta-exact} that
depends on the fractional part of $T\tau$.

\section{Empirical Illustration: Tobin's \texorpdfstring{$q$}{q} and Firm Investment}\label{sec:empirical}

Under quadratic adjustment costs, the $q$ theory of investment
\citep{tobin1969general,hayashi1982tobin} makes a firm's investment rate
linear in its marginal $q$. Following the firm-level estimates of
\citet{salinger1983tax}, this relation has been estimated on firm
panels with firm effects, with average $q$ in place of marginal $q$ and
often with output or cash flow as an additional regressor
\citep[among others]{fazzari1988financing,schaller1990reexamination,blundell1992investment}.
These estimates describe the conditional mean, whereas
\citet{galvao2015efficient} estimate quantile regressions with fixed effects for investment.

We use the panel of \citet{schaller1990reexamination}: US firms from
Compustat observed in every year from 1951 to 1985, with capital either
broadly defined, including inventories, or narrowly defined, excluding
them. Our within estimates of his standard model, which regresses the
investment rate $(I/K)_{it}$ on Tobin's $q_{it}$ with firm effects, reproduce
his Tables~II and~III. We add his output--capital ratio
$S_{it}=Y_{it}/\{(1-\delta)K_{i,t-1}\}$
\citep[eq.~(10)]{schaller1990reexamination}, where $\delta$ is the depreciation rate, and estimate
\[
 (I/K)_{it}=\alpha_i+\beta_1q_{it}+\beta_2S_{it}+u_{it}
\]
on a balanced panel of $n=187$ firms over 1952--1985, so that $T=34$ and
$n/T=5.5$.

Two features of this panel call for the theory of
Section~\ref{sec:dependence}. First, the errors are serially correlated:
Schaller reports first-order serial correlation of the residuals of 0.20
and 0.38 for the two definitions. Second, the output--capital ratio is
predetermined rather than strictly exogenous. Its denominator $K_{i,t-1}$
accumulates past investment, so that an unusually high investment rate in
one year raises the capital stock and lowers $S_{is}$ in later years $s$.
Such feedback from past errors to current regressors enters the bias
through $\zeta_i$ in \eqref{eq:dep-bias}, as in
Corollary~\ref{cor:dep-special}(iii). With $S_{it}$ as the only regressor
and martingale-difference scores, it tends to make $\zeta_i$ negative and
so to add a positive term to the bias. Serial dependence can thus affect the bias of
FEQR here, and not only its variance.

We compare the within estimator with FEQR and two split-panel jackknives:
with contiguous halves, 1952--1968 and 1969--1985, which
Theorem~\ref{thm:dep} shows to be centred under serial dependence, and
with interleaved halves, the odd and even years, which are not in general
(Remark~\ref{rem:dep-split}). We report the standard errors of
Proposition~\ref{prop:cov} and the cluster-robust (CR) ones of
Proposition~\ref{prop:cluster}, both common to FEQR and the jackknives,
and, for the within estimator, heteroskedasticity-robust and
cluster-robust ones \citep{arellano1987computing}. As in the within
regressions, output is treated as contemporaneously exogenous.

\begin{table}[!htb]
\centering
\begin{threeparttable}
\caption{Coefficients in the investment equation ($\times10^3$)}
\label{tab:empirical}
\footnotesize
\begin{tabular}{lrrrrr}
\toprule
$\tau$ & 0.1 & 0.25 & 0.5 & 0.75 & 0.9\\
\midrule
\multicolumn{6}{l}{\emph{Broad definition, coefficient on $q$; within LS 6.03 (0.54) [0.98]}}\\
FEQR & 5.56 & 5.05 & 5.41 & 6.27 & 8.81\\
SPJ, interleaved & 5.43 & 5.04 & 5.70 & 6.89 & 9.36\\
SPJ, contiguous & 4.85 & 3.96 & 4.70 & 5.50 & 8.50\\
Standard error & (0.60) & (0.54) & (0.47) & (1.03) & (1.40)\\
CR standard error & [0.81] & [0.89] & [0.84] & [1.46] & [1.56]\\[2pt]
\multicolumn{6}{l}{\emph{Broad definition, coefficient on $S$; within LS 20.62 (2.48) [2.71]}}\\
FEQR & 12.45 & 18.55 & 21.62 & 24.78 & 31.43\\
SPJ, interleaved & 8.28 & 18.05 & 20.90 & 22.41 & 30.97\\
SPJ, contiguous & 1.26 & 10.77 & 10.97 & 11.14 & 27.38\\
Standard error & (4.11) & (2.37) & (2.31) & (3.45) & (4.52)\\
CR standard error & [3.64] & [3.25] & [3.57] & [4.37] & [4.71]\\[2pt]
\multicolumn{6}{l}{\emph{Narrow definition, coefficient on $q$; within LS 3.02 (0.35) [0.55]}}\\
FEQR & 2.42 & 2.57 & 2.88 & 3.51 & 3.43\\
SPJ, interleaved & 2.35 & 2.54 & 2.90 & 3.61 & 3.43\\
SPJ, contiguous & 2.19 & 2.41 & 2.52 & 3.24 & 2.93\\
Standard error & (0.23) & (0.26) & (0.26) & (0.33) & (0.45)\\
CR standard error & [0.34] & [0.44] & [0.47] & [0.46] & [0.45]\\[2pt]
\multicolumn{6}{l}{\emph{Narrow definition, coefficient on $S$; within LS 5.96 (0.71) [1.07]}}\\
FEQR & 4.26 & 6.10 & 7.76 & 9.68 & 11.20\\
SPJ, interleaved & 4.17 & 5.95 & 7.25 & 8.91 & 11.09\\
SPJ, contiguous & $-$0.65 & 1.91 & 4.25 & 7.17 & 9.84\\
Standard error & (0.71) & (0.55) & (0.77) & (1.05) & (1.71)\\
CR standard error & [1.18] & [1.08] & [1.28] & [1.58] & [1.74]\\
\bottomrule
\end{tabular}
\begin{tablenotes}[flushleft]\footnotesize
\item \emph{Notes:} 187 firms, 1952--1985. FEQR is
\eqref{eq:estimator}, and SPJ is \eqref{eq:SPJ} with odd and even years
(interleaved) or with 1952--1968 and 1969--1985 (contiguous) as the two
halves. Standard errors in parentheses are from
Proposition~\ref{prop:cov} and in brackets from
Proposition~\ref{prop:cluster} (CR), and both are common to FEQR and SPJ.
For within LS, they are heteroskedasticity-robust (parentheses) and
cluster-robust by firm (brackets).
\end{tablenotes}
\end{threeparttable}
\end{table}

\begin{figure}[!htb]
\centering
\includegraphics[width=\textwidth]{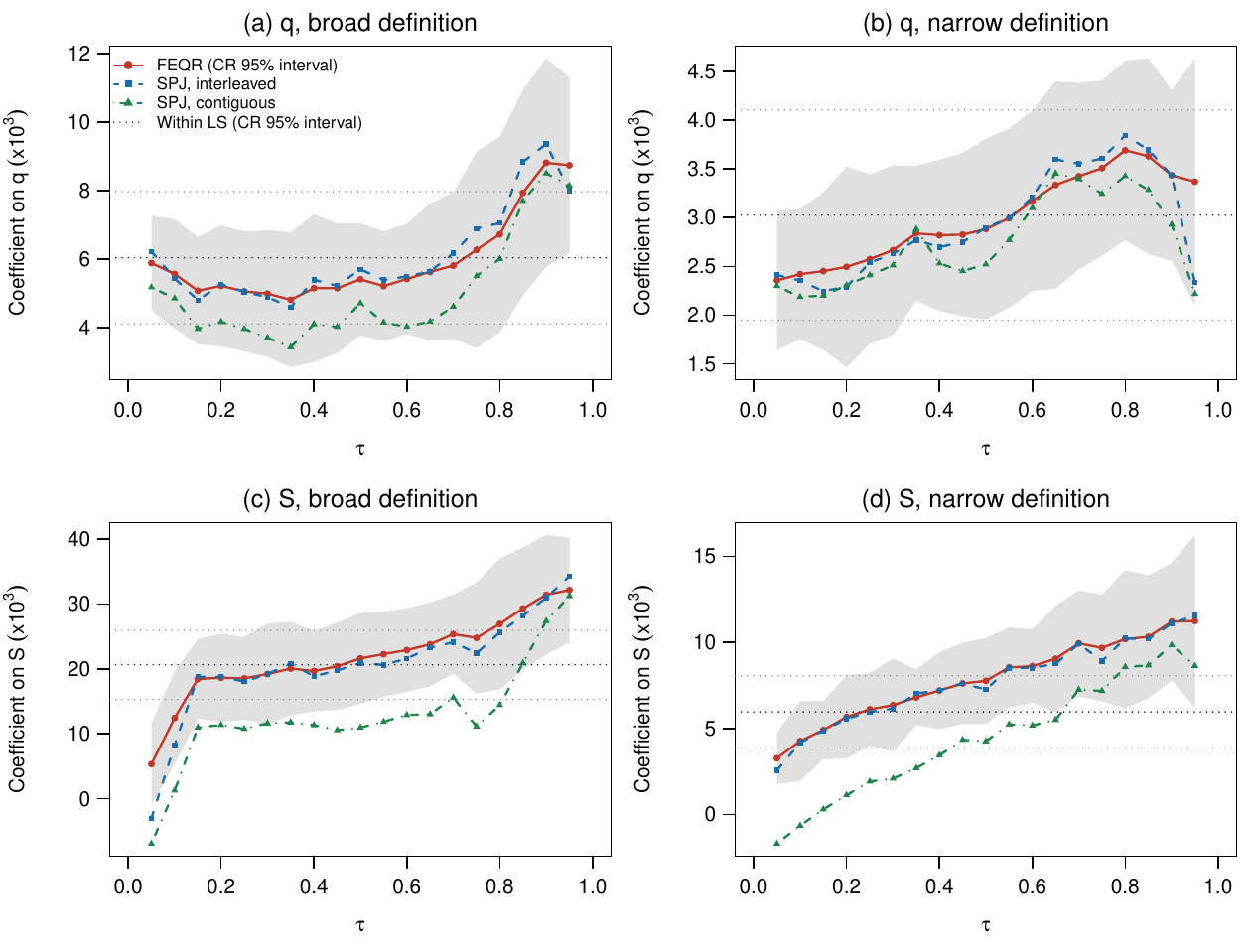}
\caption{Coefficients by quantile: FEQR with nominal 95 per cent intervals
($\pm1.96$ CR standard errors), SPJ with interleaved and contiguous halves,
and the within estimate with its nominal 95 per cent CR interval.}
\label{fig:empirical}
\end{figure}

Table~\ref{tab:empirical} and Figure~\ref{fig:empirical} give three
findings that do not depend on the choice of jackknife; coefficients are
multiplied by $10^3$ throughout. First, serial dependence matters for
inference: over $\tau=0.10,0.15,\ldots,0.90$, the CR standard errors of the
four coefficients are 0.89 to 2.06 times those of
Proposition~\ref{prop:cov}, with medians of 1.37 to 1.67. Second, the
effect of output is heterogeneous. The slope on $S$ rises across quantiles
from 12.45 to 31.43 for the broad and from 4.26 to 11.20 for the narrow
definition, and the CR 95 per cent intervals at $\tau=0.1$ and $\tau=0.9$
do not overlap: they are $[5.32,19.59]$ and $[22.19,40.66]$ for the broad
and $[1.95,6.57]$ and $[7.79,14.61]$ for the narrow definition. The
jackknife estimates are also higher at $\tau=0.9$ than at $\tau=0.1$. The
single within slopes thus understate the responses at high investment rates
and, for the broad definition, overstate them at low ones. Third, the
jackknife corrections of the slope on $q$ are moderate: over the same
quantiles, they are at most 0.59 (interleaved) and 1.70 (contiguous) CR
standard errors.

For the slope on output, by contrast, the two jackknives disagree. The
interleaved jackknife moves FEQR by at most 1.15 CR standard errors,
whereas the contiguous one lowers the slope at the median by about half,
by 2.99 and 2.75 CR standard errors for the two definitions. Two
explanations fit this pattern. The first is a bias from serial dependence
that the interleaved halves largely miss. Under the conditions of
Theorem~\ref{thm:dep}, the interleaved jackknife is centred at
$2(B_n^{\rm dep}-B_n^{{\rm dep},(2)})/T$ rather than at zero
(Remark~\ref{rem:dep-split}), so that a small interleaved correction does
not indicate a small bias: in design AR of Section~\ref{sec:sim-dep} at
$\tau=0.9$, the interleaved jackknife removes only about a third of the
bias. The second is a change in the law of the panel between 1952--1968
and 1969--1985, for instance through firm effects that shift between the
periods; \citet[footnote~20]{schaller1990reexamination} finds the slope on
$q$ unstable between 1951--1967 and 1968--1985. The full-panel fit is then
distorted by the change, but the contiguous half-panel fits are not, so
that the three fits no longer share the $T^{-1}$ form of the bias on which
the jackknife rests (Remark~\ref{rem:jackknife}). The contiguous jackknife
then doubles the distortion, whereas the interleaved halves, which span
both periods, share it with the full panel.

Theorem~\ref{thm:dep} implies a check that can help to separate the two
explanations. Under its conditions, each period is itself a stationary
panel, with the same constant $B_n^{\rm dep}$ and half as many years. The
contiguous correction $\widehat\beta_{\rm SPJ}-\widehat\beta$, whose
leading term is $-B_n^{\rm dep}/T$, should therefore roughly double when it
is computed within each period, with the halves 1952--1959 and 1960--1968,
and 1969--1976 and 1977--1985. A change between the periods, by contrast,
enters the full-panel correction but not the within-period ones. The check
presumes that the dependence dies out within a few years, so that the
$T^{-1}$ form of the bias is accurate in half-periods of eight or nine
years; with more persistent dependence, the within-period corrections fall
short of twice the full-panel one even in a stationary panel. The check is
also rough at $\tau=0.1$ and 0.9, where such a half-period contains about
one observation per firm beyond the quantile.
Table~\ref{tab:empirical-periods} reports it.

\begin{table}[!htb]
\centering
\begin{threeparttable}
\caption{Estimates by period and the scaling check ($\times10^3$)}
\label{tab:empirical-periods}
\footnotesize
\setlength{\tabcolsep}{4pt}
\begin{tabular}{lrrrrr}
\toprule
$\tau$ & 0.1 & 0.25 & 0.5 & 0.75 & 0.9\\
\midrule
\multicolumn{6}{l}{\emph{Broad definition, coefficient on $q$}}\\
FEQR, 1952--68 & 6.26 [1.04] & 5.67 [1.16] & 5.66 [1.05] & 5.56 [1.38] & 6.03 [1.10]\\
FEQR, 1969--85 & 6.29 [1.19] & 6.62 [1.08] & 6.56 [1.41] & 8.50 [1.53] & 12.20 [2.20]\\
Twice the 1952--85 correction & $-$1.43 & $-$2.19 & $-$1.40 & $-$1.53 & $-$0.62\\
Correction within 1952--68 & 0.54 & $-$0.18 & $-$2.32 & $-$1.21 & $-$1.34\\
Correction within 1969--85 & $-$7.64 & $-$7.18 & $-$5.99 & $-$4.25 & $-$0.72\\[2pt]
\multicolumn{6}{l}{\emph{Broad definition, coefficient on $S$}}\\
FEQR, 1952--68 & 18.31 [3.97] & 24.47 [3.73] & 30.54 [3.40] & 34.94 [7.08] & 33.49 [4.45]\\
FEQR, 1969--85 & 28.98 [6.17] & 28.17 [5.37] & 34.02 [7.01] & 41.90 [6.20] & 37.45 [8.10]\\
Twice the 1952--85 correction & $-$22.38 & $-$15.55 & $-$21.31 & $-$27.28 & $-$8.09\\
Correction within 1952--68 & 1.36 & $-$3.07 & $-$2.80 & $-$4.93 & $-$1.49\\
Correction within 1969--85 & 0.46 & $-$1.42 & $-$2.21 & $-$1.24 & 3.62\\[2pt]
\multicolumn{6}{l}{\emph{Narrow definition, coefficient on $q$}}\\
FEQR, 1952--68 & 2.96 [0.59] & 3.13 [0.51] & 3.67 [0.53] & 3.44 [0.69] & 2.97 [0.87]\\
FEQR, 1969--85 & 2.34 [0.45] & 2.35 [0.54] & 2.82 [0.55] & 4.10 [0.98] & 4.91 [1.49]\\
Twice the 1952--85 correction & $-$0.47 & $-$0.33 & $-$0.72 & $-$0.53 & $-$1.01\\
Correction within 1952--68 & 1.02 & 0.92 & 0.61 & 0.40 & $-$0.47\\
Correction within 1969--85 & $-$1.39 & $-$3.09 & $-$3.13 & $-$1.46 & $-$2.95\\[2pt]
\multicolumn{6}{l}{\emph{Narrow definition, coefficient on $S$}}\\
FEQR, 1952--68 & 3.65 [2.86] & 6.10 [1.79] & 8.09 [2.57] & 11.31 [2.14] & 12.14 [4.70]\\
FEQR, 1969--85 & 14.71 [1.34] & 14.48 [1.36] & 14.44 [1.69] & 13.04 [1.56] & 12.97 [1.83]\\
Twice the 1952--85 correction & $-$9.83 & $-$8.38 & $-$7.01 & $-$5.00 & $-$2.72\\
Correction within 1952--68 & $-$4.44 & $-$7.03 & $-$7.30 & $-$5.81 & $-$5.17\\
Correction within 1969--85 & $-$2.24 & $-$3.48 & $-$3.29 & $-$4.89 & $-$6.56\\
\bottomrule
\end{tabular}
\begin{tablenotes}[flushleft]\footnotesize
\item \emph{Notes:} 187 firms. FEQR is \eqref{eq:estimator} on each period,
with CR standard errors (Proposition~\ref{prop:cluster}) in brackets. A
correction is $\widehat\beta_{\rm SPJ}-\widehat\beta$ for the contiguous
jackknife \eqref{eq:SPJ}, with the halves 1952--1959 and 1960--1968 within
1952--1968, and 1969--1976 and 1977--1985 within 1969--1985; the 1952--1985
correction is that of Table~\ref{tab:empirical}. Under the conditions of
Theorem~\ref{thm:dep}, each within-period correction is about twice the
1952--1985 correction.
\end{tablenotes}
\end{threeparttable}
\end{table}

For the slope on output, the check separates the two definitions. With the
broad definition, the within-period corrections are far smaller than
predicted: averaged over the five quantiles, they are $-2.19$ and $-0.16$,
against $-18.92$, and the largest is 0.82 CR standard errors of the period
fit. Unless the dependence is very persistent, the large full-panel
correction thus reflects a change between the periods rather than an
incidental parameter bias, which is then small within each period. For
this definition, the period estimates, which need little correction, are
the relevant ones. They lie above the full-panel FEQR at every quantile and
are higher at $\tau=0.9$ than at $\tau=0.1$, 33.49 against 18.31 in
1952--1968 and 37.45 against 28.98 in 1969--1985, although they are much
less precise.

With the narrow definition, the check is broadly met: all ten
within-period corrections are negative, and they average $-5.95$ and
$-4.09$, against $-6.59$ predicted. Since a change between the periods
does not enter these corrections, they are consistent with an incidental
parameter bias of FEQR that persists within both periods and that the
interleaved jackknife largely misses: in all three samples, the
interleaved jackknife moves FEQR by at most 0.58 CR standard errors. A
change within 1969--1985, which the corrections for $q$ below suggest, may
also affect the 1969--1985 corrections, but not those for 1952--1968. Half
the average of the within-period corrections, which estimates
$-B_n^{\rm dep}/34$ under the conditions of Theorem~\ref{thm:dep} in each
period, ranges from $-1.67$ to $-2.93$ across the five quantiles, or from
1.41 to 2.44 CR standard errors of the full-panel fit. FEQR thus appears to
overstate the narrow output slope, in the direction that the stylised
feedback argument above predicts. At $\tau=0.1$ and 0.25, however, the
period estimates also differ markedly, 3.65 and 6.10 in 1952--1968 against
14.71 and 14.48 in 1969--1985, so that a change between the periods
contributes to the full-panel correction at these quantiles; in
1969--1985, the narrow output slope varies little across quantiles, from
12.97 to 14.71.

For the slope on $q$, the within-period corrections in 1969--1985 are
several times larger than predicted, $-5.16$ and $-2.40$ on average for
the two definitions against $-1.43$ and $-0.61$, whereas those in
1952--1968 average $-0.90$ and $0.49$. This points to a change in the $q$
relation within 1969--1985.

Adding year effects, which our theory does not cover, leaves the main
findings for the slope on output in place. FEQR still rises across
quantiles, from 12.50 to 29.26 and from 4.69 to 8.54. The contiguous
jackknife again lowers it, by up to 3.69 and 5.18 CR standard errors at
the five quantiles of Table~\ref{tab:empirical}, whereas the interleaved
one moves it by at most 1.20. For the broad definition, the within-period
corrections are again far smaller than twice the full-panel ones, $-0.21$
and $-3.82$ against $-14.86$ on average, so that the change between the
periods is not a common shift in investment rates, which year effects
would absorb. For the narrow definition, they are about half the predicted
size, $-3.06$ and $-4.17$ against $-7.45$, so that the check is less
clear-cut there.

The application thus illustrates what the theory for serially dependent
panels adds in practice. Clustered standard errors are needed, and they
are up to about twice as large as those that assume independence. The
choice of split matters: the interleaved jackknife, which is valid under
independence, can miss much of a bias from serial dependence, and for the
narrow output slope, the within-period corrections indicate such a bias of
1.4 to 2.4 CR standard errors. And the scaling of the bias with $T$, which
Theorem~\ref{thm:dep} makes explicit, gives a check of the stationarity on
which the contiguous jackknife rests, which a comparison of two jackknives
alone cannot provide. If the dependence is short-lived, the large
contiguous correction of the broad output slope reflects a change between
the periods, whereas that of the narrow output slope mainly reflects a
bias, except at the lowest quantiles. Throughout, the output slope is
higher at $\tau=0.9$ than at $\tau=0.1$: in the full panel for FEQR and
both jackknives, with and without year effects, and, for the broad
definition, within each period.

\section{Proofs for Section~\ref{sec:theory}}\label{app:slope}

As in the main text, $C$ and $c$ denote generic finite positive constants
that are uniform over individuals; this convention is used throughout the
proofs. This appendix first sets out the profiled criterion, the rank-score
representation and the three key lemmas summarised in
Remark~\ref{rem:outline}, and then proves them, Theorem~\ref{thm:main} and
Corollaries~\ref{cor:limit} and~\ref{cor:centred}; Appendix~\ref{app:growth}
discusses the growth conditions.

\subsection{Profiling, Rank Scores and the Key Lemmas}\label{app:lemmas}

\paragraph{Exact profiling.}
Write $d=\beta-\beta_0$, $R_{it}(d)=u_{it}-x_{it}\trans d$, and let
\[
 \mathbb M_{iT}(d)=\min_{\alpha\in\R}\frac1T
      \sum_{t=1}^T\rho_\tau(R_{it}(d)-\alpha)
\]
be the profiled criterion, a convex and $M$-Lipschitz function;
minimising $\sum_i\mathbb M_{iT}$ over $d$ is equivalent to minimising
\eqref{eq:estimator} over $\beta$. Let $\ell=\lfloor T\tau\rfloor$,
$r=T\tau-\ell$, let $R_{i(1)}(d)\leq\cdots\leq R_{i(T)}(d)$ be the order
statistics, with ties, if any, broken by a fixed rule, $x_{i[j]}(d)$ the
regressor paired with $R_{i(j)}(d)$ (its concomitant),
$\overline R_i(d)=T^{-1}\sum_tR_{it}(d)$ and
$\overline x_i=T^{-1}\sum_tx_{it}$. Then, for every realisation, the
minimum over $\alpha$ is attained at $\alpha=R_{i(\ell+1)}(d)$ and
\begin{equation}\label{eq:profile}
 \mathbb M_{iT}(d)=\tau\overline R_i(d)
  -\frac1T\left\{\sum_{j=1}^{\ell}R_{i(j)}(d)
                  +rR_{i(\ell+1)}(d)\right\},
\end{equation}
and the profiled slope score
\begin{equation}\label{eq:profile-score}
 S_{iT}(d)=\tau\overline x_i
 -\frac1T\left\{\sum_{j=1}^{\ell}x_{i[j]}(d)
                    +r x_{i[\ell+1]}(d)\right\}
\end{equation}
satisfies the subgradient inequality
\[
 \mathbb M_{iT}(d')\geq\mathbb M_{iT}(d)-(d'-d)\trans S_{iT}(d)
 \qquad\text{for all }d'\in\R^p,
\]
so that $-S_{iT}(d)$ is a subgradient of $\mathbb M_{iT}$ at $d$; where
$\mathbb M_{iT}$ is differentiable, $S_{iT}(d)$ is its negative gradient.
Indeed, the rank weights $w_{it}(d)$ of Remark~\ref{rem:outline} lie in
$[0,1]$ and sum to $T\tau$ exactly, and $\rho_\tau(v)\geq v(\tau-w)$ for
every $w\in[0,1]$, so
$\mathbb M_{iT}(d')\geq T^{-1}\sum_tR_{it}(d')\{\tau-w_{it}(d)\}$, with
equality at $d'=d$ by \eqref{eq:profile}; the right-hand side is affine
in $d'$ with gradient $-S_{iT}(d)$. Both formulas hold also when $T\tau$ is an integer, so that $r=0$ and the
minimising $\alpha$ need not be unique; no convention for choosing it is
needed. The
score is a linear rank statistic whose mean can be computed exactly. Let
$F_{i,d}$ be the distribution function of $R_{i1}(d)$,
\[
 f_{i,d}(s)=\E f_i(s+x_{i1}\trans d\mid x_{i1}),\qquad
 g_{i,d}(s)=\E[x_{i1}f_i(s+x_{i1}\trans d\mid x_{i1})],\qquad
 q_i(d)=F_{i,d}^{-1}(\tau),
\]
and, for $v$ near $\tau$,
\[
 m_i(v;d)=\frac{g_{i,d}(F_{i,d}^{-1}(v))}{f_{i,d}(F_{i,d}^{-1}(v))},\qquad
 G_i(v;d)=\int_0^v m_i(w;d)\,dw.
\]
Thus $m_i(v;d)=\E\{x_{i1}\mid F_{i,d}(R_{i1}(d))=v\}$; outside a fixed
neighbourhood of $\tau$ we use a bounded measurable version of this
conditional mean. Since the regressors and the conditional densities are
bounded (Assumption~\ref{ass:density}), and $f_i'$ is bounded on
$(-\epsilon_0,\epsilon_0)$ and $a_i\geq c$ under
Assumption~\ref{ass:smooth}, there are constants
$\delta_0,\delta_1>0$, not depending on $i$, such that, for
$\norm d\leq\delta_1$, $f_{i,d}\geq c/2$ on an interval around zero that
$F_{i,d}$ maps onto a set containing $[\tau-\delta_0,\tau+\delta_0]$, with
all density arguments in $(-\epsilon_0,\epsilon_0)$; below, small $d$ and
$v$ near $\tau$ refer to these neighbourhoods. Let
$S_i(d)=\E[x_{i1}\{\tau-\ind(R_{i1}(d)\leq q_i(d))\}]$ be the population
score.

\begin{lemma}[Exact mean of the profiled score]\label{lem:beta}
Let $T$ be large enough that $1\leq\ell$ and $\ell+1\leq T-1$, and let
$Q_T$ be $\operatorname{Beta}(\ell,T-\ell)$ with probability $1-r$ and
$\operatorname{Beta}(\ell+1,T-\ell-1)$ with probability $r$. Under Assumptions~\ref{ass:sampling}, \ref{ass:density} and any version of
Assumption~\ref{ass:smooth},
\begin{equation}\label{eq:beta-exact}
 \E S_{iT}(d)=\tau\E x_{i1}-\E G_i(Q_T;d),\qquad
 \E Q_T=\tau,\qquad
 \Var(Q_T)=\frac{v_\tau}{T+1}+\frac{r(1-r)}{T(T+1)},
\end{equation}
and, uniformly over $i$ and deterministic $d$ in a common neighbourhood
of zero,
\begin{equation}\label{eq:meanexp}
 \E S_{iT}(d)=S_i(d)
       -\frac{v_\tau}{2T}\partial_v m_i(\tau;d)
       +O(\epsilon_T),
\end{equation}
with $\epsilon_T$ of the order given in Table~\ref{tab:regimes} for that
version. At $d=0$,
\begin{equation}\label{eq:scorebias}
 S_i(0)=0,\qquad m_i(\tau;0)=\gamma_i,\qquad
 \partial_vm_i(\tau;0)=c_i/a_i^2.
\end{equation}
\end{lemma}

\paragraph{Averaging the centred remainder.}
At a small deterministic displacement $d$, let
$e_{it}(d)=u_{it}-x_{it}\trans d-q_i(d)$ and
\[
 \xi_{it}(d)=\{x_{it}-m_i(\tau;d)\}\ps(e_{it}(d)),\qquad
 \mathcal R_{iT}(d)=S_{iT}(d)-\frac1T\sum_{t=1}^T\xi_{it}(d);
\]
in particular
$\mathcal R_{iT}(0)=S_{iT}(0)-T^{-1}\sum_tz_{it}\ps(u_{it})$. The shifted
errors satisfy $\Prb(e_{it}(d)\leq0)=\tau$, hence $\E\xi_{it}(d)=S_i(d)$.

\begin{lemma}[Individual remainder and averaging]\label{lem:remainder}
Under Assumptions~\ref{ass:sampling}, \ref{ass:density} and any version of
Assumption~\ref{ass:smooth}, $\E\norm{\mathcal R_{iT}(d)}^2\leq CT^{-3/2}$ uniformly in $i$ and
small deterministic $d$, and, with $\epsilon_T$ as in
Lemma~\ref{lem:beta},
\begin{equation}\label{eq:R-bounds}
 \E\mathcal R_{iT}(0)=-\frac{v_\tau}{2T}\frac{c_i}{a_i^2}+O(\epsilon_T).
\end{equation}
Consequently,
\begin{equation}\label{eq:averaged-score}
 \frac1n\sum_{i=1}^n S_{iT}(0)
 =\frac1{nT}\sum_{i=1}^n\sum_{t=1}^Tz_{it}\ps(u_{it})+\frac{b_n}{T}
       +\Op(n^{-1/2}T^{-3/4})+O(\epsilon_T),
\end{equation}
and the last two terms are $\op((nT)^{-1/2})$ under
Assumption~\ref{ass:growth}.
\end{lemma}

\paragraph{The local criterion at the displaced centre.}
Define the score at the displaced centre and the localised criterion
\[
 \Delta^c_{nT}=\frac{\sqrt{nT}}n\sum_{i=1}^n S_{iT}(B_n/T),
 \qquad
 H^c_{nT}(\theta)=T\sum_{i=1}^n
 \{\mathbb M_{iT}(B_n/T+\theta/\sqrt{nT})-\mathbb M_{iT}(B_n/T)\},
\]
whose minimisers are exactly the points
$\widehat\theta_{nT}=\sqrt{nT}(\widehat\beta-\beta_0-B_n/T)$, with
$\widehat\beta$ the slope component of a minimiser in
\eqref{eq:estimator}. Lemmas
\ref{lem:beta} and \ref{lem:remainder}, together with Lemmas
\ref{lem:bias-modulus} and \ref{lem:shifted-score} below, yield
\begin{equation}\label{eq:center-score-main}
 \Delta^c_{nT}
 =\frac1{\sqrt{nT}}\sum_{i=1}^n\sum_{t=1}^Tz_{it}\ps(u_{it})+\op(1)=O_p(1)
\end{equation}
and the following approximation, whose proof is in
Appendix~\ref{app:proof-local}.

\begin{lemma}[Local quadratic approximation]\label{lem:local}
Under Assumptions~\ref{ass:sampling}--\ref{ass:growth}, uniformly on
compact subsets of $\R^p$,
\begin{equation}\label{eq:LAQ}
 H^c_{nT}(\theta)=-\theta\trans\Delta^c_{nT}
            +\tfrac12 \theta\trans\Gamma_n\theta+\op(1).
\end{equation}
\end{lemma}

\subsection{Proof of Lemma~\ref{lem:beta}}

Because $d$ is deterministic and $F_{i,d}$ is continuous, ties among
$R_{i1}(d),\ldots,R_{iT}(d)$ have probability zero, so the rule for
breaking ties in Appendix~\ref{app:lemmas} plays no role in this proof.
Conditionally on $x_{i1}$ and
$F_{i,d}(R_{i1}(d))=v$, the other $T-1$ residuals are i.i.d.\ and each lies
below $R_{i1}(d)$ with probability $v$, so the rank of $R_{i1}(d)$ minus one
is $\operatorname{Bin}(T-1,v)$. Since
$\Prb\{\operatorname{Bin}(T-1,v)\leq k-1\}=\Prb\{\operatorname{Beta}(k,T-k)\geq v\}$
for $1\leq k\leq T-1$, its expected rank weight is
\[
 \Prb\{\operatorname{Bin}(T-1,v)\leq\ell-1\}
       +r\,\Prb\{\operatorname{Bin}(T-1,v)=\ell\}
       =\Prb(Q_T\geq v).
\]
Exchangeability over time and Fubini's theorem give the first identity in
\eqref{eq:beta-exact}. If $A=\ell$ with probability $1-r$ and $A=\ell+1$
with probability $r$, then $\E A=T\tau$ and $\Var(A)=r(1-r)$. The
identities $\E(Q_T\mid A)=A/T$ and $\E(Q_T^2\mid A)=A(A+1)/\{T(T+1)\}$ give
the mean and the variance.

Put $H_T=Q_T-\tau$. The beta mixture has exponentially small tails outside
a fixed neighbourhood of $\tau$. By the rising-factorial formula for beta
moments, its fourth central moment is $\E H_T^4=3v_\tau^2/T^2+O(T^{-3})$
(see \eqref{eq:fourth-beta-moment} below), and Lyapunov's inequality then
bounds the lower absolute moments:
\[
 \E H_T^4=O(T^{-2}),\qquad
 \E|H_T|^{2+\eta}\leq(\E H_T^4)^{(2+\eta)/4}=O(T^{-1-\eta/2}),
 \quad 0<\eta\leq1.
\]
The density-ratio formula for $m_i$, its derivative, and the uniformly
positive denominator show that $G_i''=\partial_vm_i$ is uniformly
equicontinuous under condition (W), with a common modulus of order
$\omega_f(Ca)+a$, which is dominated by the function $\omega$ fixed in
Section~\ref{sec:assumptions}. Hence
\[
 G_i(\tau+s;d)
 =G_i(\tau;d)+m_i(\tau;d)s
     +\tfrac12\partial_vm_i(\tau;d)s^2
     +O\{s^2\omega(|s|)\}.
\]
The linear term has mean zero because $\E H_T=0$, and the remainder has
expectation $o(T^{-1})$ uniformly. To see this, split at $|H_T|\leq\delta$, where the remainder is at most
$\tfrac12\omega(\delta)H_T^2$, and note that on the complement, whose
probability is $O(e^{-cT\delta^2})$, both $G_i$ and its quadratic Taylor
polynomial are bounded because $m_i$ and $\partial_vm_i(\tau;d)$ are
bounded and $|H_T|\leq1$; then
let $\delta\downarrow0$ after $T\to\infty$. The $O(T^{-2})$ difference between $\Var(Q_T)$ and
$v_\tau/T$ is negligible. Since $\tau\E x_{i1}-G_i(\tau;d)=S_i(d)$ by the
definitions of $G_i$ and $S_i$, this proves \eqref{eq:meanexp} in regime
W.

Under condition (H$_\eta$), $\partial_vm_i$ is uniformly H\"older of
order $\eta$, so the Taylor remainder is $O(|s|^{2+\eta})$ for
$|s|\leq\delta_0$ and, with the exponentially small tails treated as in
regime W, its expectation is of order $T^{-1-\eta/2}$. Condition (I), or merely
(H$_1$), makes $\partial_vm_i$ uniformly Lipschitz and gives the rate of
regime I; no second density derivative is needed.

For regime II, the rising-factorial formula for beta moments gives
\begin{align}
 \E H_T^3&=
 \frac{2Tv_\tau(1-2\tau)+3(1-2\tau)r(1-r)+r(1-r)(1-2r)}
        {T(T+1)(T+2)}=O(T^{-2}),\label{eq:higher-beta-moments}\\
 \E H_T^4&=\frac{3v_\tau^2}{T^2}+O(T^{-3}).\label{eq:fourth-beta-moment}
\end{align}
Under condition (II), $G_i^{(4)}=\partial_v^3m_i$ is uniformly bounded. The
Taylor expansion through the cubic term therefore has remainder $O(s^4)$
for $|s|\leq\delta_0$, and the tails are again exponentially small.
The cubic term has mean $O(T^{-2})$ by the \emph{signed} third moment in
\eqref{eq:higher-beta-moments}, and the remainder has mean of the same
order by \eqref{eq:fourth-beta-moment}. This
proves \eqref{eq:meanexp} with $\epsilon_T=O(T^{-2})$. It suffices instead
that $G_i'''$ be uniformly Lipschitz, which follows from condition (I) with
a uniformly Lipschitz $f_i''$. An absolute third-moment bound alone would
only give $O(T^{-3/2})$.

Finally, differentiating $g_{i,0}/f_{i,0}$ with respect to its quantile
argument gives
\[
 \partial_vm_i(\tau;0)
 =\frac{\E[x_{i1}f_i'(0\mid x_{i1})]-\gamma_i f_i'(0)}{a_i^2}
 =\frac{c_i}{a_i^2}.
\]
The other identities in \eqref{eq:scorebias} follow from the definitions
and the conditional quantile restriction. \qed

\subsection{Movement of the Bias Coefficient}

\begin{lemma}[Population curvature and movement of the bias]\label{lem:bias-modulus}
Suppose that Assumptions~\ref{ass:sampling}, \ref{ass:density} and
condition (W) of Assumption~\ref{ass:smooth} hold. Then, uniformly for
small deterministic $b$ and $d$,
\[
 q_i(0)=0,\quad \nabla q_i(d)=-m_i(\tau;d),\quad
 |q_i(b+d)-q_i(b)|+\norm{m_i(\tau;b+d)-m_i(\tau;b)}\leq C\norm d,
\]
so that in particular $|q_i(d)|+\norm{m_i(\tau;d)-\gamma_i}\leq C\norm d$,
and, with $\Gamma_i=\E[f_i(0\mid x_{i1})z_{i1}z_{i1}\trans]$, so that
$\Gamma_n=n^{-1}\sum_i\Gamma_i$,
\[
 \nabla S_i(d)=-\Gamma_i+O(\norm d),\qquad
 S_i(d)=-\Gamma_i d+O(\norm d^2).
\]
Moreover, with the modulus $\omega$ of Section~\ref{sec:assumptions},
\begin{equation}\label{eq:bias-slope-modulus}
 \norm{\partial_vm_i(\tau;b+d)-\partial_vm_i(\tau;b)}
       \leq C\omega(\norm d).
\end{equation}
One can take $\omega(a)=Ca^\eta$ in regime H$_\eta$ and $\omega(a)=Ca$ in
regimes I and II.
\end{lemma}

\begin{proof}
Apply the implicit function theorem to
$\E F_i(q_i(d)+x_{i1}\trans d\mid x_{i1})=\tau$. The density-weighted mean
$m_i(\tau;d)$ is Lipschitz in $d$ because $f_i'$ is bounded and the
denominator is bounded below. Differentiating the population score gives
\[
 \nabla S_i(d)=
 -\E\left[
   f_i(q_i(d)+x_{i1}\trans d\mid x_{i1})
   (x_{i1}-m_i(\tau;d))(x_{i1}-m_i(\tau;d))\trans
 \right].
\]
This matrix is Lipschitz at zero under the same first-derivative bound.
For \eqref{eq:bias-slope-modulus}, differentiate the density ratio once in
its scalar argument; the resulting expression involves $f_i$, $f_i'$,
their weighted moments, and denominators bounded away from zero. When $d$
changes by $a$, the density arguments change by at most $Ca$. Uniform
continuity of $f_i'$ and Lipschitz continuity of $f_i$ give a modulus of
order $\omega_f(Ca)+a$. The stated quantitative moduli follow under the
stronger conditions.
\end{proof}

\subsection{Proof of Lemma~\ref{lem:remainder}}

Let $w_{it}=w_{it}(0)$ be the rank weight of Remark~\ref{rem:outline},
equal to one for ranks at most $\ell$, to
$r$ for rank $\ell+1$, and to zero otherwise. Because $\sum_t w_{it}=T\tau$,
\begin{equation}\label{eq:R-exact}
 \mathcal R_{iT}(0)=-\frac1T\sum_{t=1}^T(x_{it}-\gamma_i)
               \{w_{it}-\ind(u_{it}\leq0)\}.
\end{equation}
Let $K_i=\sum_t|w_{it}-\ind(u_{it}\leq0)|$ and
$N=\sum_t\ind(u_{it}\leq0)$, which is $\operatorname{Bin}(T,\tau)$ because
$\Prb(u_{it}\leq0)=\tau$. Then $K_i=|T\tau-N|$: ordered by the errors, the
indicators equal one for the $N$ smallest errors and zero otherwise, and the
rank weights equal one for the $\ell$ smallest, $r$ for the next and zero
otherwise, so one of the two sequences dominates the other. The binomial
central moments $\E(N-T\tau)^2=Tv_\tau$ and
$\E(N-T\tau)^4=Tv_\tau\{1+3(T-2)v_\tau\}$ and the Cauchy--Schwarz
inequality then give
\[
 \E K_i\leq(Tv_\tau)^{1/2}=O(T^{1/2}),\qquad \E K_i^4=O(T^2).
\]
Conditionally on the errors of individual $i$, its regressors remain
independent over $t$.
Given the errors, the sum in \eqref{eq:R-exact} therefore has conditional
variance at most $CK_i/T^2$, since its weights are bounded by one in
absolute value and sum to $K_i$; the expected conditional variance is
$O(T^{-3/2})$.

For the conditional mean, the density ratio
$\E[x_{i1}\mid u_{i1}=u]=g_{i,0}(u)/f_{i,0}(u)$ is uniformly Lipschitz in
$u$ on any subinterval of $(-\epsilon_0,\epsilon_0)$ on which the marginal density is bounded below, and
equals $\gamma_i$ at $u=0$. Every
observation with a non-zero weight difference in \eqref{eq:R-exact} lies
between zero and $q_{iT}=R_{i(\ell+1)}(0)$, in the closed interval. Let
$\epsilon_1\in(0,\epsilon_0)$ be such that $f_i(u)\geq a_i/2$ for
$|u|\leq\epsilon_1$ and all $i$; such an $\epsilon_1$ exists because
$f_i(u)\geq a_i-L|u|$ with $L$ the bound on $f_i'$. On
$\{|q_{iT}|\leq\epsilon_1\}$ the squared conditional mean is bounded by
$Cq_{iT}^2K_i^2/T^2$. Binomial concentration and the uniformly positive
marginal density give
\[
 \Prb(|q_{iT}|>\epsilon_1)\leq Ce^{-cT},\qquad
 \E[q_{iT}^4\ind(|q_{iT}|\leq\epsilon_1)]\leq CT^{-2}.
\]
The Cauchy--Schwarz inequality bounds the expected squared conditional
mean by $O(T^{-2})$. On the exceptional event, the remainder is bounded.
This proves the second-moment bound without moments of the error tails.

The same argument applies with $(u_{it},\gamma_i)$ replaced by
$(e_{it}(d),m_i(\tau;d))$: the $\tau$-quantile of $e_{it}(d)$ is zero, its
density is uniformly positive near zero, and $\E(x_{it}\mid e_{it}(d)=e)$
is uniformly Lipschitz near $e=0$, where it equals $m_i(\tau;d)$. This proves the displaced
second-moment bound. Lemma~\ref{lem:beta} gives the mean at zero.
Independence across individuals then yields
\[
 \E\left\|\frac1n\sum_{i=1}^n\{\mathcal R_{iT}(0)-\E\mathcal R_{iT}(0)\}\right\|^2
       \leq\frac{C}{nT^{3/2}}.
\]
The normalised error is $O_p(T^{-1/4})$, and the normalised mean error
vanishes by \eqref{eq:growth}. \qed

\subsection{Fluctuations at the Displaced Centre}

\begin{lemma}[Fluctuations at a displaced centre]\label{lem:shifted-score}
Suppose that Assumptions~\ref{ass:sampling}, \ref{ass:density} and
condition (W) of Assumption~\ref{ass:smooth} hold. If the deterministic
sequence $b_{nT}$ satisfies $\norm{b_{nT}}\leq C/T$, then
\begin{equation}\label{eq:shifted-fluctuation}
 \frac{\sqrt{nT}}n\sum_{i=1}^n
   \{S_{iT}(b_{nT})-\E S_{iT}(b_{nT})\}
 =\frac1{\sqrt{nT}}\sum_{i=1}^n\sum_{t=1}^Tz_{it}\ps(u_{it})+\op(1).
\end{equation}
This assertion imposes no upper growth rate on $n$.
\end{lemma}

\begin{proof}
Lemma~\ref{lem:remainder} bounds the contribution of
$\mathcal R_{iT}(b_{nT})-\E\mathcal R_{iT}(b_{nT})$ by $O_p(T^{-1/4})$ at
the displayed scale. Lemma~\ref{lem:bias-modulus} and bounded densities
give, for small $b$ and $d$,
\[
 \E\norm{\xi_{it}(b+d)-\xi_{it}(b)}^2\leq C\norm d.
\]
Indeed, the regressor mean changes by at most $C\norm d$, and the two
indicators can differ only in an interval of conditional probability at
most $C\norm d$. Centring these differences and using independence of all
observations makes their normalised sum
$O_p(\norm{b_{nT}}^{1/2})=O_p(T^{-1/2})$. Since
$\xi_{it}(0)=z_{it}\ps(u_{it})$ has mean zero, the result follows.
\end{proof}

\subsection{The Score at the Displaced Centre}

Combining
Lemmas \ref{lem:beta} and \ref{lem:bias-modulus} gives
\begin{equation}\label{eq:center-mean}
 \frac1n\sum_{i=1}^n\E S_{iT}(B_n/T)
 =-\Gamma_nB_n/T+\frac{b_n}{T}
   +O\{\epsilon_T+T^{-2}+T^{-1}\omega(C/T)\}.
\end{equation}
The first two terms cancel because $\Gamma_nB_n=b_n$. The term $T^{-1}\omega(C/T)$ is essential
under weak smoothness: mere uniform continuity of $f_i'$ does not give an
$O(T^{-1})$ change in the bias coefficient, that is, an $O(T^{-2})$ change
in the mean score. All three residual
terms in \eqref{eq:center-mean}, multiplied by $\sqrt{nT}$, vanish in each
regime. In regime W, this follows from bounded $n/T$ and
$\omega(C/T)\to0$. In regime H$_\eta$, the modulus term is
$O(\sqrt{n/T}\,T^{-\eta})=o(1)$; in regimes I and II, it is
$O(\sqrt{n/T^3})=o(1)$. Together with Lemma~\ref{lem:shifted-score}
applied with $b_{nT}=B_n/T$, this proves \eqref{eq:center-score-main}:
\[
 \Delta^c_{nT}=\frac{\sqrt{nT}}n\sum_{i=1}^n S_{iT}(B_n/T)
 =\frac1{\sqrt{nT}}\sum_{i=1}^n\sum_{t=1}^Tz_{it}\ps(u_{it})+\op(1)=O_p(1).
\]

\subsection{Proof of Lemma~\ref{lem:local}}\label{app:proof-local}

Let $d\in\R^p$ be a deterministic displacement of small norm; the choice
$d=\theta/\sqrt{nT}$ is made in Step 3. Write $b=B_n/T$ and
$a=\norm d$.

\emph{Step 1: mean.} The mean-score expansion \eqref{eq:meanexp} and
Lemma~\ref{lem:bias-modulus} imply
\begin{equation}\label{eq:meanscorelocal}
 \E\{S_{iT}(b+d)-S_{iT}(b)\}
 =-\Gamma_i d+
 O\{a^2+a/T+\omega(a)/T+\epsilon_T\}.
\end{equation}
For
\[
 D^c_{iT}(d)=\mathbb M_{iT}(b+d)-\mathbb M_{iT}(b)+d\trans S_{iT}(b),
\]
integration along the segment from $b$ to $b+d$ gives
\begin{equation}\label{eq:Dmean}
 \E D^c_{iT}(d)=\tfrac12d\trans\Gamma_i d+
 O\{a^3+a^2/T+a\omega(a)/T+a\epsilon_T\}.
\end{equation}
The integration is exact for every realisation. By the subgradient
inequality of Appendix~\ref{app:lemmas}, $g(s)=-d\trans S_{iT}(b+sd)$ is a
subgradient at $s$ of the convex function $\varphi(s)=\mathbb M_{iT}(b+sd)$,
so $g$ is non-decreasing on $[0,1]$. Over a partition of $[0,1]$ with mesh
$1/J$, the increments of $\varphi$ therefore lie between the lower and upper
Riemann sums of $g$, which differ by at most $\{g(1)-g(0)\}/J$. Hence
$\varphi(1)-\varphi(0)=\int_0^1g(s)\,ds$, that is,
\[
 D^c_{iT}(d)=-\int_0^1d\trans\{S_{iT}(b+sd)-S_{iT}(b)\}\,ds,
\]
with no almost-everywhere argument and no convention for ties. Since
$\norm{S_{iT}}\leq M$, Fubini's theorem allows expectations to be taken
inside the integral, and \eqref{eq:meanscorelocal}, with $sd$ in place of
$d$, gives \eqref{eq:Dmean}.

\emph{Step 2: variance.} A centred variance bound follows from the
displaced remainder representation. Independence over time and the
squared-increment bound in the proof of Lemma~\ref{lem:shifted-score}
yield, uniformly over $0\leq s\leq1$,
\[
 \left(\E\norm{\{S_{iT}(b+sd)-S_{iT}(b)\}
       -\E\{S_{iT}(b+sd)-S_{iT}(b)\}}^2\right)^{1/2}
 \leq C\{\sqrt{a/T}+T^{-3/4}\}.
\]
The two individual remainders each have $L^2$ norm $O(T^{-3/4})$ by
Lemma~\ref{lem:remainder}. Integrating the last bound using Minkowski's
inequality gives
\begin{equation}\label{eq:Dvariance}
 \Var\{D^c_{iT}(d)\}\leq C\{a^3/T+a^2T^{-3/2}\}.
\end{equation}

\emph{Step 3: pointwise approximation.} For fixed $\theta$, put
$d=\theta/\sqrt{nT}$. The total mean error in $T\sum_i D^c_{iT}(d)$ is bounded
by $C_\theta$, a constant depending only on $\norm\theta$ and on the constants in
Assumptions~\ref{ass:sampling}--\ref{ass:nondeg}, times
\[
 (nT)^{-1/2}+T^{-1}
       +\sqrt{n/T}\,\omega(C_\theta/\sqrt{nT})
       +\sqrt{nT}\,\epsilon_T.
\]
This tends to zero in regime W because $n/T$ is bounded and
$\omega(a)\to0$. In regime H$_\eta$, the modulus term is at most
$C_\theta\sqrt{n/T^{1+\eta}}$, using $n\geq1$, and tends to zero. In regimes I
and II, it is $O(T^{-1})$. Independence across individuals and
\eqref{eq:Dvariance} give
\[
 \Var\left\{T\sum_{i=1}^nD^c_{iT}(\theta/\sqrt{nT})\right\}
       \leq C_\theta\{(nT)^{-1/2}+T^{-1/2}\}\longrightarrow0.
\]
Since $T\sum_iD^c_{iT}(\theta/\sqrt{nT})=H^c_{nT}(\theta)+\theta\trans\Delta^c_{nT}$, this
convex function of $\theta$ differs pointwise in probability from
$\tfrac12\theta\trans\Gamma_n\theta$ by $o_p(1)$.

\emph{Step 4: uniformity.} To obtain compact-uniform convergence with a
moving $\Gamma_n$, take any subsequence. Assumption~\ref{ass:nondeg}
supplies a further subsequence with $\Gamma_n\to\Gamma_*$, where
$\Gamma_*$ is positive definite. Lemma~1 of \citet{HjortPollard93}
(see also \citealp[Section~6]{Pollard91}), by which pointwise convergence
in probability of convex random functions is uniform on compact sets,
gives compact-uniform convergence to $\tfrac12\theta\trans\Gamma_*\theta$
along that subsequence. The quadratic forms with $\Gamma_n$ also converge uniformly on
compacts to this form. By the subsequence principle, the difference between
$T\sum_iD^c_{iT}(\theta/\sqrt{nT})$ and $\tfrac12\theta\trans\Gamma_n\theta$
therefore tends to zero in probability uniformly on compact sets along
the full sequence.
This proves \eqref{eq:LAQ}. \qed

\subsection{Proof of Theorem~\ref{thm:main}}

\emph{Step 1: stochastic expansion.} By Assumption~\ref{ass:nondeg} and
\eqref{eq:center-score-main},
$\theta^*_{nT}:=\Gamma_n^{-1}\Delta^c_{nT}=O_p(1)$ is the unique minimiser
of $G_{nT}(\theta)=-\theta\trans\Delta^c_{nT}+\tfrac12\theta\trans\Gamma_n\theta$,
and $\inf_{\norm{\theta-\theta^*_{nT}}=\delta}G_{nT}(\theta)-G_{nT}(\theta^*_{nT})
\geq c_\Gamma\delta^2/2$. Since $H^c_{nT}$ is convex, Lemma~2 of
\citet{HjortPollard93}, their (1.1) with (1.2), gives for every $\delta>0$
\[
 \Prb\{\norm{\widehat\theta_{nT}-\theta^*_{nT}}\geq\delta\}
 \leq\Prb\Bigl\{\sup_{\norm{\theta-\theta^*_{nT}}\leq\delta}
     |H^c_{nT}(\theta)-G_{nT}(\theta)|\geq c_\Gamma\delta^2/4\Bigr\},
\]
which tends to zero by Lemma~\ref{lem:local} because $\theta^*_{nT}$ is
stochastically bounded, as in the proof of their Basic Corollary, (1.3),
here with a matrix $\Gamma_n$ that need not converge. Hence
\[
 \sqrt{nT}(\widehat\beta-\beta_0-B_n/T)
       =\Gamma_n^{-1}\Delta^c_{nT}+\op(1).
\]
Substitution of \eqref{eq:center-score-main} proves \eqref{eq:expansion}.

\emph{Step 2: Gaussian approximation for the linear term.} Define
\begin{equation}\label{eq:Lbeta}
 L^\beta_{nT}=\Gamma_n^{-1}\frac1{\sqrt{nT}}
       \sum_{i=1}^n\sum_{t=1}^Tz_{it}\ps(u_{it}),
\end{equation}
so that \eqref{eq:expansion} reads
$\sqrt{nT}(\widehat\beta-\beta_0-B_n/T)=L^\beta_{nT}+\op(1)$. The
conditional quantile restriction gives $\E[z_{it}\ps(u_{it})]=0$ and
$\E[z_{it}z_{it}\trans\ps(u_{it})^2]=v_\tau\E[z_{it}z_{it}\trans]$. Thus
$\Cov(L^\beta_{nT})=\Omega_n$ exactly. Whiten by $\Omega_n^{-1/2}$. The
independent centred summands then have total covariance $I_p$ and total
third absolute moment at most $C_p(nT)^{-1/2}$, by Assumptions~\ref{ass:density} and \ref{ass:nondeg}. The Berry--Esseen bound over
measurable convex sets of \citet[Theorem~1.1, eq.~(1.2)]{Raic19} bounds
the error by $(42p^{1/4}+16)$ times the total third absolute moment for
independent, non-identically distributed summands, and an invertible
linear image of a lower orthant is convex. Consequently,
\begin{equation}\label{eq:linear-BE}
 \dK\{\mathcal L(L^\beta_{nT}),N(0,\Omega_n)\}
       \leq C_p(nT)^{-1/2}.
\end{equation}

\emph{Step 3: transfer to the estimator.} Write
$W_{nT}=\sqrt{nT}(\widehat\beta-\beta_0-B_n/T)$ for the normalised
estimator, so that $W_{nT}-L^\beta_{nT}=o_p(1)$ by \eqref{eq:expansion}. For every
$\varepsilon>0$, coordinatewise bracketing and the uniformly bounded
marginal Gaussian densities give
\begin{equation}\label{eq:K-transfer}
 \dK\{\mathcal L(W_{nT}),N(0,\Omega_n)\}
 \leq C_p\{(nT)^{-1/2}+\varepsilon\}
       +\Prb(\norm{W_{nT}-L^\beta_{nT}}_\infty>\varepsilon).
\end{equation}
For example, the Gaussian probability of the difference between the
orthants at $x+\varepsilon(1,\ldots,1)\trans$ and $x-\varepsilon(1,\ldots,1)\trans$ is at
most $C_p\varepsilon$ by a union bound over coordinates. First take a limit
superior, then let $\varepsilon\downarrow0$. This proves
\eqref{eq:centered}. Translation by $\sqrt{n/T}\,B_n$ proves
\eqref{eq:shifted}, even if that shift diverges. \qed

\subsection{Proofs of Corollaries \ref{cor:limit} and \ref{cor:centred}}

For Corollary~\ref{cor:limit}, \eqref{eq:shifted} and the convergence of
$\sqrt{n/T}\,B_n$ to $\sqrt\upsilon\,B$ and of $\Omega_n$ to $\Omega$ give the
claim, since the Gaussian distribution functions converge uniformly when
their means and covariance matrices converge to a non-singular limit; $\Omega$ is
non-singular because the eigenvalues of $\Omega_n$ are bounded away from
zero.

For Corollary~\ref{cor:centred}, if \eqref{eq:vanishing-bias} holds, then
\eqref{eq:centered} and the bracketing argument in \eqref{eq:K-transfer},
applied with the deterministic shift $\sqrt{n/T}\,B_n\to0$, give the first
claim. Under \eqref{eq:zerobias}, $c_i=0$ for every $i$, so $b_n=0$ and
$B_n=0$. In case (i), $\gamma_i=\E x_{i1}$ and
$f_i'(0\mid x_{i1})=f_i'(0)$, so \eqref{eq:zerobias} holds; moreover,
$m_i(v;0)=\E x_{i1}$ for every $v$; since $x_{i1}$ is independent of
$u_{i1}$, each regressor $x_{i[j]}(0)$ has mean $\E x_{i1}$ and the rank
weights sum to $T\tau$, so $\E S_{iT}(0)=0$ for every $T$.
In case (ii), symmetry and differentiability at zero give
$f_i'(0\mid x)=0$ for every $x$, so that $c_i=0$. \qed

\subsection{The Growth Conditions}\label{app:growth}

As noted in Remark~\ref{rem:outline}, since the stochastic averaging step costs
only $O_p(T^{-1/4})$ at the $\sqrt{nT}$ scale, whatever the relative
size of $n$ and $T$, the restrictions in Assumption~\ref{ass:growth}
come from the unremoved deterministic error: a remainder of order $T^{-s}$ contributes
$\sqrt{nT}\,T^{-s}=T^{(q+1)/2-s}$ when $n=T^q$, which explains $q<1+\eta$
for $s=1+\eta/2$, $q<2$ for $s=3/2$ and $q<3$ for $s=2$; regime W gives
only $o(T^{-1})$, hence bounded $n/T$. These are sufficient ranges, not
claims of optimality. At $q=3$ even the regime II residual can be of
order one, so a further correction, or a condition eliminating its
leading coefficient, would be needed there; we do not pursue $q\geq3$.
The range $n=o(T^3)$ parallels the requirement in smooth models that $T$
grow faster than $n^{1/3}$ for the analytical correction
\citep{HahnNewey04}, in
both cases because a remaining $O(T^{-2})$ error is negligible at the
$\sqrt{nT}$ scale; here the range is attained by the bias-centred
approximation and by the jackknife of Section~\ref{sec:inference}. At $n=T$,
Theorem~\ref{thm:main} needs neither a second derivative of the
densities nor convergence of $\Gamma_n$ and $V_n$; since $B_n$ involves
$f_i'(0\mid x)$, however, dispensing with density differentiability
altogether would require a different bias argument.

\section{Proofs for Section~\ref{sec:inference}}\label{app:inference}

\subsection{Proof of Theorem~\ref{thm:spj}}

Since $T_j/T\to\lambda_j\in(0,1)$, each regime is preserved when $T$ is
replaced by $T_j$. Theorem~\ref{thm:main} therefore applies to all three fits.
They have the same $\Gamma_n$ and $B_n$, because observations are
identically distributed over time within each individual. Write
\[
 L_j=\Gamma_n^{-1}\frac1{nT_j}\sum_{i=1}^n\sum_{t\in\mathcal T_j}
       z_{it}\ps(u_{it}),\qquad
 L=\Gamma_n^{-1}\frac1{nT}\sum_{i=1}^n\sum_{t=1}^Tz_{it}\ps(u_{it})
       =\frac{L^\beta_{nT}}{\sqrt{nT}}
\]
for the unnormalised linear terms of the two subsample fits and of the
full-sample fit. Then
\[
 L=\frac{T_1}{T}L_1+\frac{T_2}{T}L_2,\qquad
 \frac{T_1}{T}\frac{B_n}{T_1}
       +\frac{T_2}{T}\frac{B_n}{T_2}=\frac{2B_n}{T}.
\]
Thus the bias cancels, and the retained linear term is
$2L-L=L$. The three remainders are all
$o_p((nT)^{-1/2})$, since $nT_j\geq cnT$. This proves the expansion in
\eqref{eq:spj-expansion}. The Gaussian transfer argument in Steps 2 and 3
of the proof of Theorem~\ref{thm:main} then proves the Gaussian approximation in \eqref{eq:spj-expansion}.
\qed

\subsection{Proof of Proposition~\ref{prop:cov}}

Theorem~\ref{thm:main} and Theorem~\ref{thm:alpha-main}, whose proof uses
no result of Section~\ref{sec:inference}, give
\[
 \max_{i\leq n}|\widehat\alpha_i-\alpha_{i0}|
       +\norm{\widehat\beta-\beta_0}
       =O_p\{\sqrt{L_{nT}/T}\}.
\]
For the true errors, and $T$ so large that $h$ times the support of $K$
lies in $(-\epsilon_0,\epsilon_0)$, the kernel expectations differ from their
zero-density targets by $O(h)$ uniformly; this only requires a bounded
first density derivative. Bernstein's inequality \citep[e.g.][Lemma
2.2.9]{VanDerVaarWellner1996EmpiricalProcess} and a union bound over
individuals give uniform fluctuations of order
$O_p\{\sqrt{L_{nT}/(Th)}+L_{nT}/(Th)\}$ for the kernel-weighted zeroth,
first and second regressor moments. Replacing errors by residuals adds at
most $Ch^{-2}\max_{i,t}|\widehat u_{it}-u_{it}|=O_p\{h^{-2}\sqrt{L_{nT}/T}\}$,
by the bounded derivative of $K$. All error terms are $o_p(1)$ whenever
$h\to0$, $L_{nT}/(Th)\to0$ and $h^{-2}\sqrt{L_{nT}/T}\to0$, in particular
under Assumption~\ref{ass:kernel}. Hence, since $a_i\geq c$, all
denominators are positive with probability tending to one. Since
$\sum_tK(\widehat u_{it}/h)\widehat z_{it}=0$, $\widehat\Gamma_n$ is the
average over $i$ of the kernel-weighted second moments minus
$\widehat a_i\widehat\gamma_i\widehat\gamma_i\trans$. Uniform consistency of
$\widehat a_i$ and $\widehat\gamma_i$, bounded regressors, and a law of large numbers for the
unweighted second moments then give \eqref{eq:cov-consistency}, since the
eigenvalues of $\Gamma_n$ are bounded away from zero by
Assumption~\ref{ass:nondeg}, and hence so are those of $\widehat\Gamma_n$
with probability tending to one. \qed

\subsection{Proof of Corollary~\ref{cor:slope-CI}}

For any fixed non-zero $v\in\R^p$, apply the scalar Berry--Esseen bound to
$v\trans L^\beta_{nT}$ standardised by $\sqrt{v\trans\Omega_nv}$, which is
bounded away from zero by Assumption~\ref{ass:nondeg}. By
\eqref{eq:spj-expansion} and the bracketing argument in
\eqref{eq:K-transfer},
$\sqrt{nT}\,v\trans(\widehat\beta_{\rm SPJ}-\beta_0)/\sqrt{v\trans\Omega_nv}
\Rightarrow N(0,1)$. By \eqref{eq:cov-consistency},
$(v\trans\widehat\Omega_nv)/(v\trans\Omega_nv)\longrightarrow_p1$, and Slutsky's theorem
completes the proof. \qed

\section{Theory for the Individual Effects}\label{app:alpha}

This appendix develops the theory for the individual effects, reports
simulations and gives the proofs. Each intercept is estimated from its
own $T$ observations, so its rate is $\sqrt T$ whatever the size of $n$.
Appendix~\ref{sec:alpha} gives a Bahadur representation that holds
uniformly over the $n$ intercepts (Theorem~\ref{thm:alpha-main}), a
Gaussian approximation jointly with the slope
(Corollary~\ref{cor:alpha-joint}) and a Gumbel limit for the maximal
standardised deviation (Theorem~\ref{thm:alpha-maximum}). The estimated
slope enters the intercepts only at order $T^{-1}+(nT)^{-1/2}$, by
\eqref{eq:alpha-slope-negligible}, so that, in every regime and for every
fixed $i$, $\sqrt T(\widehat\alpha_i-\alpha_{i0})$ is asymptotically
normal with variance $v_\tau/a_i^2$; the slope bias does not shift this
distribution. Appendix~\ref{sec:alpha-inference} makes these results
feasible with the kernel estimators $\widehat a_i$ of
Section~\ref{sec:covariance}, and Appendix~\ref{sec:alpha-sim} reports
simulations.

\subsection{Results}\label{sec:alpha}

Write $\widehat d=\widehat\beta-\beta_0$ and
$\Delta_{iT}=\widehat\alpha_i-\alpha_{i0}$. The first-order intercept
problem is a sample-quantile problem, as in \citet{Bahadur66}; the
additional tasks are to control the estimated common slope, to allow all
admissible intercept minimisers, and to obtain bounds uniform over a
growing number of individuals. For any $b\in\R^p$, let
\[
 \widehat\alpha_i(b)\in\arg\min_{\alpha\in\R}
       \frac1T\sum_{t=1}^T\rho_\tau(y_{it}-\alpha-x_{it}\trans b),
\]
so that $\widehat\alpha_i=\widehat\alpha_i(\widehat\beta)$. For
$d=b-\beta_0$, let $\mathcal I_{iT}(d)$ be the set of sample
$\tau$-quantiles of $\{u_{it}-x_{it}\trans d:t\leq T\}$; then
$\widehat\alpha_i(b)-\alpha_{i0}\in\mathcal I_{iT}(d)$. This set is a
singleton, the order statistic of rank $\lceil T\tau\rceil$, if
$T\tau$ is not an integer, and otherwise the closed interval between the
order statistics of ranks $T\tau$ and $T\tau+1$, which also
accommodates tied residuals.

\begin{lemma}[Oracle transfer for an individual intercept]\label{lem:alpha-oracle}
Fix $i$. Suppose that the $u_{it}$ are i.i.d.\ over $t$ with a density
that is continuous at zero, where it equals $a_i>0$, $\Prb(u_{it}\leq0)=\tau$, and
$\norm{x_{it}}\leq M$. For any fitted slope $\widetilde\beta$ with
$\sqrt T\norm{\widetilde\beta-\beta_0}=o_p(1)$, every measurable selection
of $\widehat\alpha_i(\widetilde\beta)$ satisfies
\begin{equation}\label{eq:alpha-oracle-CLT}
 \sqrt T\{\widehat\alpha_i(\widetilde\beta)-\alpha_{i0}\}
 \Rightarrow N\left(0,\frac{v_\tau}{a_i^2}\right).
\end{equation}
No independence between $\widetilde\beta$ and the observations of
individual $i$ is required.
\end{lemma}

Theorem~\ref{thm:main} supplies the required slope rate:
\begin{equation}\label{eq:alpha-slope-negligible}
 \norm{\widehat d}=O_p\{T^{-1}+(nT)^{-1/2}\},\qquad
 \sqrt T\norm{\widehat d}=O_p(T^{-1/2}+n^{-1/2})=o_p(1),
\end{equation}
so the slope bias $B_n/T$ does not shift the first-order intercept
distribution, even when $n/T\to\infty$; the same holds after
re-estimating each intercept at $\widehat\beta_{\rm SPJ}$ of
Section~\ref{sec:inference}.

\begin{theorem}[Uniform Bahadur representation]\label{thm:alpha-main}
Under Assumptions~\ref{ass:sampling}--\ref{ass:growth}, every measurable
joint minimiser of \eqref{eq:estimator} satisfies
\begin{equation}\label{eq:alpha-uniform-rate}
 \max_{i\leq n}|\Delta_{iT}|
       =O_p\left(\sqrt{L_{nT}/T}\right)
\end{equation}
and, more precisely,
\begin{equation}\label{eq:alpha-Bahadur}
 \Delta_{iT}
 =\frac1{a_iT}\sum_{t=1}^T\ps(u_{it})
       -\gamma_i\trans\widehat d+R^\alpha_{iT},\qquad
 \max_{i\leq n}|R^\alpha_{iT}|
       =O_p\left\{\left(\frac{L_{nT}}{T}\right)^{3/4}\right\}.
\end{equation}
Consequently, for every fixed $i$,
$\sqrt T(\widehat\alpha_i-\alpha_{i0})\Rightarrow N(0,v_\tau/a_i^2)$, and
uniformly in the label,
\begin{equation}\label{eq:alpha-uniform-Gaussian}
 \sup_{i\leq n}\sup_{x\in\R}
 \left|\Prb\left\{
 \frac{a_i\sqrt T(\widehat\alpha_i-\alpha_{i0})}{\sqrt{v_\tau}}
       \leq x\right\}-\Phi(x)\right|\longrightarrow0.
\end{equation}
These statements hold in every regime, including regime W at $n=T$.
\end{theorem}

The sign in \eqref{eq:alpha-Bahadur} is negative because the slope error enters
as $-x_{it}\trans\widehat d$ in the residual whose quantile
defines the intercept, and the coefficient is the density-weighted mean
$\gamma_i$ rather than the unweighted regressor mean.

\begin{corollary}[Slope and finitely many individual effects]\label{cor:alpha-joint}
Let $m$ be fixed and let $i_1,\ldots,i_m\leq n$ be distinct
deterministic labels, which may vary with $n$ and $T$. Under Assumptions~\ref{ass:sampling}--\ref{ass:growth},
\begin{equation}\label{eq:alpha-joint-limit}
 \dK\left\{\mathcal L\begin{pmatrix}
  \sqrt{nT}(\widehat\beta-\beta_0-B_n/T)\\[2pt]
  a_{i_1}\sqrt T\,\Delta_{i_1T}/\sqrt{v_\tau}\\
  \vdots\\
  a_{i_m}\sqrt T\,\Delta_{i_mT}/\sqrt{v_\tau}
 \end{pmatrix},
 N\left(0,\begin{pmatrix}\Omega_n&0\\0&I_m\end{pmatrix}\right)\right\}
           \longrightarrow0.
\end{equation}
The conclusion is unchanged if the slope component is replaced by
$\sqrt{nT}(\widehat\beta_{\rm SPJ}-\beta_0)$ of Section~\ref{sec:inference}.
\end{corollary}

The approximating covariance is block diagonal, so that the slope and each
intercept are asymptotically independent; at $n=T$ the approximation
reads
\begin{equation}\label{eq:alpha-equal-panel}
 \dK\left\{
 \mathcal L\begin{pmatrix}
 n(\widehat\beta-\beta_0)\\
 \sqrt n(\widehat\alpha_i-\alpha_{i0})
 \end{pmatrix},
 N\left[
 \begin{pmatrix}B_n\\0\end{pmatrix},
 \begin{pmatrix}\Omega_n&0\\0&v_\tau/a_i^2\end{pmatrix}
 \right]\right\}\longrightarrow0,
\end{equation}
since a positive diagonal rescaling preserves lower-orthant
approximation. For $n=\lfloor T^q\rfloor$, the bias-centred slope has
normalisation $T^{(q+1)/2}$ while each intercept retains $\sqrt T$; the
intercept approximation stays centred even when the normalised
uncorrected slope bias diverges.

The maximal deviation over all $n$ intercepts has a Gumbel limit. Let
\[
 M_{nT}=\max_{i\leq n}
       \frac{a_i\sqrt T|\widehat\alpha_i-\alpha_{i0}|}{\sqrt{v_\tau}}.
\]

\begin{theorem}[Maximal deviation]\label{thm:alpha-maximum}
Under Assumptions~\ref{ass:sampling}--\ref{ass:growth}, for every fixed
$x\in\R$,
\begin{equation}\label{eq:alpha-Gumbel}
 \Prb\{M_{nT}^2-2\log n+\log\log n\leq x\}
 \longrightarrow
 \exp\{-\pi^{-1/2}e^{-x/2}\}.
\end{equation}
\end{theorem}

The limit is the Gumbel law $G_0(x)=\exp(-e^{-x})$ evaluated at $(x+\log\pi)/2$, and
the norming constants are those of the maximum of $n$ independent absolute
standard normal variables; the normal law lies in the domain of attraction
of $G_0$ \citep{resnick2007heavy}. The proof rests on
a moderate-deviation approximation to binomial tails at the threshold
$\sqrt{2\log n}$, available because $(\log n)^3/T\to0$ under
Assumption~\ref{ass:growth}.

\begin{remark}[Scope of the intercept results]\label{rem:alpha-scope}
An intercept is estimated from its own $T$ observations; a larger $n$
improves the common slope but not the $\sqrt T$ rate or the variance
$v_\tau/a_i^2$ of an intercept, and estimating all intercepts at once
costs the $\sqrt{L_{nT}}$ factor in \eqref{eq:alpha-uniform-rate} and the
$\sqrt{2\log n}$ critical value of the simultaneous bands of
Appendix~\ref{sec:alpha-inference}. No $T^{-1}$ formula for
$\E(\widehat\alpha_i-\alpha_{i0})$ is asserted: it would have to account
for the sample-quantile convention, the intercept's own curvature bias and
slope estimation, and our assumptions do not even guarantee that the
intercept estimators have finite moments; the probability
statements above do not require them. In particular, \eqref{eq:alpha-Bahadur} does
not justify replacing a second-order intercept bias by
$-\gamma_i\trans B_n/T$. \hfill $\lozenge$
\end{remark}

\subsection{Inference}\label{sec:alpha-inference}

Replacing $a_i$ by its kernel estimator $\widehat a_i$ of
Section~\ref{sec:covariance} gives feasible pointwise intervals and
simultaneous bands.

\begin{corollary}[Studentised individual effects]\label{cor:alpha-studentized}
Under Assumptions~\ref{ass:sampling}--\ref{ass:growth} and the kernel
conditions of Assumption~\ref{ass:kernel}, for any deterministic bandwidth
$h\to0$,
\begin{equation}\label{eq:alpha-density-rate}
 \max_{i\leq n}|\widehat a_i-a_i|
 =O_p\left\{
 h+\sqrt{\frac{L_{nT}}{Th}}+\frac{L_{nT}}{Th}
       +\frac1{h^2}\sqrt{\frac{L_{nT}}T}\right\},
\end{equation}
so that the bandwidth of Assumption~\ref{ass:kernel} gives
\begin{equation}\label{eq:alpha-density-log}
 \log(n+1)\max_{i\leq n}|\widehat a_i-a_i|=o_p(1)
\end{equation}
and, uniformly in the label,
\begin{equation}\label{eq:alpha-studentized-uniform}
 \sup_{i\leq n}\sup_{x\in\R}
 \left|\Prb\left\{
 \frac{\widehat a_i\sqrt T(\widehat\alpha_i-\alpha_{i0})}{\sqrt{v_\tau}}
       \leq x\right\}-\Phi(x)\right|\longrightarrow0.
\end{equation}
Hence, for $\delta\in(0,1)$,
\begin{equation}\label{eq:alpha-pointwise-CI}
 \widehat\alpha_i\ \pm\
 \Phi^{-1}(1-\delta/2)\frac{\sqrt{v_\tau}}{\widehat a_i\sqrt T}
\end{equation}
is a pointwise asymptotic $(1-\delta)$ confidence interval for
$\alpha_{i0}$, with a coverage error that tends to zero uniformly in
$i\leq n$.
\end{corollary}

A common fixed critical value does not give simultaneous coverage over
the $n$ intercepts. Let
\begin{equation}\label{eq:alpha-critical}
 c_{n,\delta}=
 \Phi^{-1}\left(\frac{1+(1-\delta)^{1/n}}2\right),\qquad 0<\delta<1,
\end{equation}
be the exact $(1-\delta)$-quantile of the maximum of $n$ independent
absolute standard normal variables.

\begin{theorem}[Simultaneous confidence bands]\label{thm:alpha-band}
Under Assumptions~\ref{ass:sampling}--\ref{ass:kernel}, the Gumbel limit
\eqref{eq:alpha-Gumbel} of Theorem~\ref{thm:alpha-maximum} holds with
$M_{nT}$ replaced by its studentised counterpart
$\widehat M_{nT}=\max_{i\leq n}\widehat a_i\sqrt T|\widehat\alpha_i
-\alpha_{i0}|/\sqrt{v_\tau}$, and
\begin{equation}\label{eq:alpha-simultaneous}
 \Prb\left\{|\widehat\alpha_i-\alpha_{i0}|\leq
  c_{n,\delta}\frac{\sqrt{v_\tau}}{\widehat a_i\sqrt T}
  \ \text{ for every } i\leq n\right\}\longrightarrow1-\delta.
\end{equation}
\end{theorem}

The simultaneous half-width is of order $\sqrt{\log n/T}$, that is,
$\sqrt{q\log T/T}$ at $n=\lfloor T^q\rfloor$. The band relies on
\eqref{eq:alpha-density-log}, that is, on $\max_i|\widehat a_i-a_i|$
being $o_p(1/\log n)$. The simulations in Appendix~\ref{sec:alpha-sim}
show that this is demanding when $T$ is small or $\tau$ is far from
$1/2$; at $\tau=0.9$ the Gumbel approximation itself is also slow.

\subsection{Simulation Results}\label{sec:alpha-sim}

Table~\ref{tab:sim-fe} reports the coverage of the intervals and bands
of Appendix~\ref{sec:alpha-inference} for the design of
Section~\ref{sec:sim-design} with $n=T\in\{50,100,200,400,800\}$ and
2000 replications per design.

\begin{table}[!htb]
\centering
\begin{threeparttable}
\caption{Coverage of 95 per cent pointwise intervals and simultaneous bands for the individual effects}
\label{tab:sim-fe}
\footnotesize
\setlength{\tabcolsep}{5pt}
\begin{tabular}{rrrrrrrrrrr}
\toprule
 & & \multicolumn{3}{c}{$\tau=0.5$} & \multicolumn{3}{c}{$\tau=0.75$} & \multicolumn{3}{c}{$\tau=0.9$}\\
\cmidrule(lr){3-5}\cmidrule(lr){6-8}\cmidrule(lr){9-11}
$n=T$ & $c_{n,0.05}$ & Pointwise & Band & Oracle & Pointwise & Band & Oracle & Pointwise & Band & Oracle\\
\midrule
50 & 3.28 & 0.943 & 0.873 & 0.918 & 0.935 & 0.802 & 0.884 & 0.926 & 0.640 & 0.793\\
100 & 3.47 & 0.950 & 0.920 & 0.933 & 0.948 & 0.866 & 0.918 & 0.935 & 0.658 & 0.810\\
200 & 3.66 & 0.952 & 0.937 & 0.946 & 0.950 & 0.884 & 0.920 & 0.938 & 0.677 & 0.849\\
400 & 3.83 & 0.954 & 0.952 & 0.950 & 0.951 & 0.907 & 0.929 & 0.942 & 0.726 & 0.862\\
800 & 4.00 & 0.954 & 0.944 & 0.938 & 0.952 & 0.923 & 0.937 & 0.943 & 0.772 & 0.908\\
\bottomrule
\end{tabular}
\begin{tablenotes}[flushleft]\footnotesize
\item \emph{Notes:} Full-sample FEQR fit with $n=T$; $\widehat a_i$ is
the kernel estimate of the marginal error density at zero, $a_i$.
Pointwise: coverage of the nominal 95 per cent interval
\eqref{eq:alpha-pointwise-CI}, averaged over the $n$ individuals and the
replications. Band: proportion of replications in which the band
\eqref{eq:alpha-simultaneous} with $\delta=0.05$ covers all $n$
intercepts; its critical value $c_{n,0.05}$ of \eqref{eq:alpha-critical}
is the $0.95$-quantile of the maximum of $n$ independent absolute
standard normal variables. Oracle: the same band with the true $a_i$ in
place of $\widehat a_i$. With 2000 replications per design, the Monte
Carlo standard error of a band coverage rate is $0.0049$ near $0.95$ and
at most $0.011$; that of a pointwise rate is smaller.
\end{tablenotes}
\end{threeparttable}
\end{table}

In Table~\ref{tab:sim-fe}, the pointwise interval
\eqref{eq:alpha-pointwise-CI} of Corollary~\ref{cor:alpha-studentized}
covers $0.943$--$0.954$ at $\tau=0.5$, $0.935$--$0.952$ at $\tau=0.75$
and $0.926$--$0.943$ at $\tau=0.9$, rising with $T$; with the true $a_i$
it covers $0.942$--$0.950$ (not tabulated).

The band \eqref{eq:alpha-simultaneous} of Theorem~\ref{thm:alpha-band}
uses a critical value far in the tails ($3.28$--$4.00$), where the Gumbel
limit of Theorem~\ref{thm:alpha-maximum} requires $(\log n)^3/T\to0$;
at $n=T$ this ratio falls only from $1.20$ to $0.37$ between $T=50$ and
$T=800$, so coverage should approach $0.95$ slowly, most slowly at
$\tau=0.9$, where the skewness of the $\operatorname{Bin}(T,\tau)$ count of
non-positive errors and the slope of the error density at zero relative to
its squared value, which both enter the tail probabilities at first order
and here partly offset each other, are largest. At $\tau=0.5$ the feasible band covers $0.873$ at
$T=50$ and $0.937$--$0.952$ for $T\geq200$, within $0.013$ of the oracle
band with the true $a_i$ for $T\geq100$; at $\tau=0.75$ the gap to the
oracle shrinks from $0.082$ to $0.014$. At $\tau=0.9$ coverage rises only
from $0.640$ to $0.772$, and that of the oracle band from $0.793$ to
$0.908$; the remaining gap of $0.14$--$0.17$ between the two shows that
the uniform accuracy \eqref{eq:alpha-density-log} of the density
estimates on which the band relies is demanding there.

\subsection{Proof of Lemma~\ref{lem:alpha-oracle}}

Every order statistic changes by at most the largest change of an input.
The endpoints of $\mathcal I_{iT}(d)$ therefore differ from the
corresponding endpoints of $\mathcal I_{iT}(0)$ by at most $M\norm{d}$. For
any fixed oracle selection $q_{iT}\in\mathcal I_{iT}(0)$,
\begin{equation}\label{eq:alpha-oracle-distance}
 |\widehat\alpha_i(b)-\alpha_{i0}-q_{iT}|
 \leq M\norm{b-\beta_0}+\operatorname{diam}\mathcal I_{iT}(0).
\end{equation}
The diameter is zero at non-integer ranks and is $O_p(T^{-1})$ at integer
ranks. To see the latter, apply the probability integral transform: an
adjacent uniform order-statistic spacing has the
$\operatorname{Beta}(1,T)$ distribution, and the inverse distribution
function is locally Lipschitz at $\tau$. Both order statistics lie in that
neighbourhood with probability tending to one.

For fixed $s\in\R$, the probability that the oracle sample quantile is at
most $s/\sqrt T$ is bounded by binomial tails for the two endpoint ranks,
with success probability $F_i(s/\sqrt T)=\tau+a_is/\sqrt T+o(T^{-1/2})$,
where $F_i$ is the marginal error distribution function of
Assumption~\ref{ass:density}. The binomial central limit theorem gives
$\sqrt Tq_{iT}\Rightarrow N(0,v_\tau/a_i^2)$. Equation
\eqref{eq:alpha-oracle-distance} and the assumed slope rate give
\eqref{eq:alpha-oracle-CLT}. \qed

\subsection{Proof of Theorem~\ref{thm:alpha-main}}

The proof proceeds in five steps; Step 3 treats residual ties. The density
expansion uses only a bounded $f_i'$ and a
positive marginal density, not $f_i''$ or $f_i'''$. The
regime enters only through the slope rate \eqref{eq:alpha-slope-negligible}
and through $L_{nT}=O(\log T)$ in \eqref{eq:growth}.

\emph{Step 1: localisation.} The density assumptions imply that the
marginal error density satisfies $f_i(u)\geq c/2$ on a common
neighbourhood of zero. Binomial concentration at the two thresholds $\pm v$
gives
\[
 \Prb\left\{\sup_{q\in\mathcal I_{iT}(0)}|q|>v\right\}
 \leq C\exp(-cTv^2)
\]
for $v\leq\epsilon_1$, where $\epsilon_1$ is the half-width on which the
marginal densities are at least $c/2$, and $Tv^2$ sufficiently large. A
union bound over $i$ at $v\asymp\sqrt{\log n/T}$, which lies in this
range for large $T$, followed by the order-statistic perturbation bound used in
\eqref{eq:alpha-oracle-distance}, gives
\[
 \max_{i\leq n}|\Delta_{iT}|
 =O_p\left\{\sqrt{\frac{\log n}T}+\norm{\widehat d}\right\}.
\]
Equation \eqref{eq:alpha-slope-negligible} proves
\eqref{eq:alpha-uniform-rate}. In particular, with
$\varrho_{nT}=(L_{nT}/T)^{1/2}$, we have
$\max_i|\Delta_{iT}|+\norm{\widehat d}=O_p(\varrho_{nT})$.

\emph{Step 2: a localised empirical bound.} For a scalar $\alpha$ and a vector
$d$, define
\[
 F_{iT}(\alpha,d)=\frac1T\sum_{t=1}^T\ind(u_{it}\leq \alpha+x_{it}\trans d),
 \qquad F_i(\alpha,d)=\Prb(u_{i1}\leq \alpha+x_{i1}\trans d).
\]
The two-argument $F_i$ extends the marginal distribution function of
Assumption~\ref{ass:density}, in the sense that $F_i(\alpha,0)=F_i(\alpha)$. For
every fixed $C_0$, bounded regressors and densities imply
\begin{equation}\label{eq:alpha-empirical}
 \begin{split}
 &\max_{i\leq n}\sup_{|\alpha|+\norm{d}\leq C_0\varrho_{nT}}
 \left|(F_{iT}-F_i)(\alpha,d)-(F_{iT}-F_i)(0,0)\right|\\
 &\hspace{25mm}
 =O_p\left\{\sqrt{\frac{\varrho_{nT}L_{nT}}T}+\frac{L_{nT}}T\right\}.
 \end{split}
\end{equation}
To verify \eqref{eq:alpha-empirical}, note that a grid of mesh $T^{-2}$
on the set $\{|\alpha|+\norm d\leq C_0\varrho_{nT}\}$ has
$O(T^{2(p+1)})$ points.
For parameters in the indicated neighbourhood, the indicator difference is
bounded by one in absolute value and has variance at most $C\varrho_{nT}$,
because it can be non-zero only if $|u_{it}|\leq |\alpha|+M\norm{d}$.
Bernstein's inequality gives, for a deviation of the order of the
right-hand side of \eqref{eq:alpha-empirical} with a large constant, a
tail bound $\exp(-C'L_{nT})$ with $C'$ as large as desired, which beats
the union over the $O(nT^{2(p+1)})$ pairs of grid points and individuals
because $L_{nT}=\log(nT)$. For an
arbitrary parameter value, round $d$ and $\alpha$ to the grid and shift the
rounded intercept up or down by $CT^{-2}$. The two resulting indicators
bracket the original one, since the regressors are bounded, and their
expectation gap is at most $CT^{-2}$. Including those shifted grid points
in the finite union proves \eqref{eq:alpha-empirical} on the entire
neighbourhood.

\emph{Step 3: quantile optimality and ties.} Almost surely, for no individual do there exist $(\alpha,d)$ and
$p+2$ periods $t$ with $u_{it}=\alpha+x_{it}\trans d$ for all of them. Indeed, conditionally on
all regressors, any selected $p+2$ errors have a joint density by
Assumptions~\ref{ass:sampling} and \ref{ass:density}, whereas their being
fitted by $p+1$ linear coefficients puts them in a subspace of dimension
at most $p+1$. A finite union over the relevant subsets has probability
zero. The quantile subgradient inequalities at the joint fit therefore
imply
\begin{equation}\label{eq:alpha-score-gap}
 |F_{iT}(\Delta_{iT},\widehat d)-\tau|\leq(p+1)/T
 \quad\text{for every }i.
\end{equation}
Indeed, since $\widehat\alpha_i$ minimises the criterion of individual $i$
at $\widehat\beta$, the one-sided derivatives of that criterion give
$N_i^-\leq T\tau\leq N_i^-+N_i^0$, where $N_i^-$ and $N_i^0$ are the numbers
of negative and of zero residuals of individual $i$, while
$TF_{iT}(\Delta_{iT},\widehat d)=N_i^-+N_i^0$ and $N_i^0\leq p+1$. Since
only optimality in $\alpha_i$ is used,
this remains true for any choice within a non-unique intercept interval;
compare the computational properties of quantile regression discussed
in \citet{koenker2005quantile}.

\emph{Step 4: expansion.} A Taylor expansion of the conditional
distribution functions gives, uniformly in a neighbourhood,
\begin{equation}\label{eq:alpha-population}
 F_i(\alpha,d)-\tau
 =a_i(\alpha+\gamma_i\trans d)+O\{(|\alpha|+\norm{d})^2\}.
\end{equation}
Also, $F_{iT}(0,0)-\tau=-T^{-1}\sum_t\ps(u_{it})$. Combine
\eqref{eq:alpha-empirical}--\eqref{eq:alpha-population} on the localised
parameter set. After dividing by $a_i\geq c$, the uniform remainder is
\[
 O_p\left\{\varrho_{nT}^2+
       \sqrt{\frac{\varrho_{nT}L_{nT}}T}+\frac{L_{nT}}T+\frac1T\right\}
 =O_p\{(L_{nT}/T)^{3/4}\},
\]
which proves \eqref{eq:alpha-Bahadur}.

\emph{Step 5: Gaussian approximations.} Because \eqref{eq:growth} gives
$L_{nT}=O(\log T)$, the remainder in \eqref{eq:alpha-Bahadur} multiplied by
$\sqrt T$ tends to zero. Equation \eqref{eq:alpha-slope-negligible} then
reduces each intercept to its i.i.d.\ Bernoulli-score sum, and the ordinary
central limit theorem proves the pointwise convergence. For the uniform
statement, put $Z_{iT}=(Tv_\tau)^{-1/2}\sum_t\ps(u_{it})$. The preceding
bounds give
\[
 \max_{i\leq n}\left|
  \frac{a_i\sqrt T\,\Delta_{iT}}{\sqrt{v_\tau}}-Z_{iT}\right|
 =O_p\{T^{-1/2}+n^{-1/2}+L_{nT}^{3/4}T^{-1/4}\}=o_p(1).
\]
Every $Z_{iT}$ has the same standardised Bernoulli-score law, whose scalar
Berry--Esseen error is $O(T^{-1/2})$. Bracketing by a fixed $\varepsilon$
and using the bounded normal density, uniformly in $i$, proves
\eqref{eq:alpha-uniform-Gaussian}. This argument also covers deterministic
labels changing with $n$ and $T$. \qed

\subsection{Proof of Corollary~\ref{cor:alpha-joint}}

Theorem~\ref{thm:main} and the uniform representation in
Theorem~\ref{thm:alpha-main} replace the vector in
\eqref{eq:alpha-joint-limit}, with an $o_p(1)$ error, by the joint
influence vector
\[
 \left((L^\beta_{nT})\trans,Z_{i_1T},\ldots,Z_{i_mT}\right)\trans,
 \qquad Z_{iT}=\frac1{\sqrt{Tv_\tau}}\sum_{t=1}^T\ps(u_{it}).
\]
Here $L^\beta_{nT}$ is the slope influence term \eqref{eq:Lbeta}. The intercept scores are independent across the selected individuals,
each with mean zero and variance one. The covariance of the slope term and
the $j$th intercept score is
\[
 \Cov(L^\beta_{nT},Z_{i_jT})
   =\frac{\sqrt{v_\tau}}{\sqrt n}
           \Gamma_n^{-1}\E z_{i_j1}=O(n^{-1/2})
\]
uniformly in the labels. This covariance need not be zero for finite $n$: the density-weighted centring that defines $\gamma_i$ does not
imply $\E z_{i1}=0$. The complete covariance matrix therefore differs from
$\operatorname{diag}(\Omega_n,I_m)$ by $O(n^{-1/2})$ and has eigenvalues uniformly bounded
above and away from zero for sufficiently large $n$.

The joint vector is a sum of independent centred vectors indexed by
$(i,t)$. After whitening by its exact covariance, the total third absolute
moment of the summands is $O\{(nT)^{-1/2}+T^{-1/2}\}$: only the $m$
selected rows contribute intercept components, each of order $T^{-1/2}$.
The Berry--Esseen bound over convex sets of \citet[Theorem~1.1]{Raic19} gives a
fixed-dimensional Gaussian approximation with the exact joint covariance.
Gaussian laws are uniformly continuous in total variation as their
covariance varies over a compact set of positive definite matrices, so
that covariance can be replaced by $\operatorname{diag}(\Omega_n,I_m)$ with an $o(1)$ error.
The bracketing argument in \eqref{eq:K-transfer}, in dimension $p+m$,
transfers the approximation to the original vector. The jackknife estimator has the
same slope influence term by Theorem~\ref{thm:spj}, which proves the last
claim. \qed

\subsection{A Moderate-Deviation Approximation to Binomial Tails}

\begin{lemma}[Binomial tails at a moderate-deviation threshold]\label{lem:alpha-binomial-tail}
Let $N_T\sim\operatorname{Bin}(T,\tau)$ and
$Z_T=(N_T-T\tau)/\sqrt{Tv_\tau}$. For any $x_T\to\infty$ with
$x_T^3/\sqrt T\to0$,
\[
 \Prb(Z_T>x_T)\sim1-\Phi(x_T),\qquad
 \Prb(Z_T<-x_T)\sim1-\Phi(x_T).
\]
\end{lemma}

\begin{proof}
Choose an integer $k_T$ at the upper-tail threshold and write
$q_T=k_T/T=\tau+x_T\sqrt{v_\tau/T}+O(T^{-1})$. Let
\[
 D(q\|\tau)=q\log(q/\tau)+(1-q)\log\{(1-q)/(1-\tau)\},
 \qquad
 \lambda_T=\log\frac{q_T(1-\tau)}{\tau(1-q_T)}.
\]
Stirling's formula, uniformly for $q_T$ in a compact subset of $(0,1)$,
gives
\[
 \Prb(N_T=k_T)=
 \frac{\exp\{-TD(q_T\|\tau)\}}{\sqrt{2\pi Tq_T(1-q_T)}}\{1+o(1)\}.
\]
Moreover, $TD(q_T\|\tau)=x_T^2/2+O(x_T^3/\sqrt T+x_T/\sqrt T)$ and
$\lambda_T\sim x_T/\sqrt{Tv_\tau}$. The ratio $\Prb(N_T=k+1)/\Prb(N_T=k)=(T-k)\tau/\{(k+1)(1-\tau)\}$ is
decreasing in $k$ and at $k=k_T$ equals $e^{-\lambda_T}k_T/(k_T+1)$, so
successive binomial masses above $k_T$ have ratios at most
$e^{-\lambda_T}$. For every fixed $A>0$ and
$0\leq j\leq A/\lambda_T$, their exact product formula also gives
\[
 \log\frac{\Prb(N_T=k_T+j)}{\Prb(N_T=k_T)}
 =-\lambda_T j+O\{(j+j^2)/T\}
 =-\lambda_T j+o(1)
\]
uniformly in that range, because $T\lambda_T^2$ is of order $x_T^2$. An
upper geometric sum and a lower sum truncated at $A/\lambda_T$, followed by
$A\to\infty$, imply $\Prb(N_T\geq k_T)\sim\Prb(N_T=k_T)/\lambda_T$. The
displayed expansions give
$\Prb(N_T\geq k_T)\sim e^{-x_T^2/2}/(\sqrt{2\pi}x_T)\sim1-\Phi(x_T)$.
Integer rounding changes the standardised threshold by $O(T^{-1/2})$ and
is negligible in this calculation. Apply the same argument to $T-N_T$ for
the lower tail.
\end{proof}

\subsection{Proof of Theorem~\ref{thm:alpha-maximum}}

Let $Z_{iT}=(Tv_\tau)^{-1/2}\sum_t\ps(u_{it})$ as before, and set
$Z_{nT}^{\max}=\max_i|Z_{iT}|$. Theorem~\ref{thm:alpha-main} and
\eqref{eq:alpha-slope-negligible} imply
\begin{equation}\label{eq:alpha-max-approx}
 |M_{nT}-Z_{nT}^{\max}|
 =O_p\left\{
       T^{-1/2}+n^{-1/2}+\frac{L_{nT}^{3/4}}{T^{1/4}}\right\}.
\end{equation}
The right-hand side multiplied by $\sqrt{\log n}$ tends to zero, since
$\log n/n\to0$ and \eqref{eq:growth} gives $L_{nT}=O(\log T)$. Binomial
concentration gives $Z_{nT}^{\max}=O_p(\sqrt{\log n})$. It follows that
\begin{equation}\label{eq:alpha-max-square}
 M_{nT}^2-(Z_{nT}^{\max})^2=o_p(1).
\end{equation}

The variables $Z_{iT}$ are independent and have the same standardised
binomial law, regardless of heterogeneity in the marginal error densities.
For fixed $x$, put $u_n(x)=\sqrt{2\log n-\log\log n+x}$. Since
$(\log n)^3/T\to0$, Lemma~\ref{lem:alpha-binomial-tail} and the normal tail
asymptotics yield
\[
 n\Prb\{|Z_{iT}|>u_n(x)\}
 \sim 2n\{1-\Phi(u_n(x))\}
 \longrightarrow \pi^{-1/2}e^{-x/2}.
\]
Independence, and $(1-p_n)^n\to e^{-c}$ when $np_n\to c$, the step behind
the equivalence of (4.28) and (4.29) in \citet[p.~92]{resnick2007heavy},
prove the limit for $(Z_{nT}^{\max})^2$. Equation \eqref{eq:alpha-max-square} then proves
\eqref{eq:alpha-Gumbel}. \qed

\subsection{Proof of Corollary~\ref{cor:alpha-studentized}}

For the true errors, the kernel expectation is $\int K(s)f_i(hs)\,ds=a_i+O(h)$
uniformly. The individual kernel summands are bounded by $C/h$ and have
variance at most $C/h$. Bernstein's inequality and a union bound yield the
second and third terms in \eqref{eq:alpha-density-rate}. The change to
fitted residuals is bounded, using the bounded derivative of $K$, by
\[
 \frac C{h^2}\max_{i,t}|\widehat u_{it}-u_{it}|
 \leq\frac C{h^2}
       \{\max_i|\Delta_{iT}|+M\norm{\widehat d}\}.
\]
Theorem~\ref{thm:alpha-main} gives the last term in
\eqref{eq:alpha-density-rate}. Substituting $h=C_hT^{-1/8}$ and
$L_{nT}=O(\log T)$ proves \eqref{eq:alpha-density-log}. In particular, all
$\widehat a_i$ are positive with probability tending to one; the intervals
can be defined arbitrarily on the complementary event. For the studentised
statistic, let $\varepsilon\in(0,1/2)$. By \eqref{eq:alpha-density-log} and
$a_i\geq c$, the event $\max_{i\leq n}|\widehat a_i/a_i-1|\leq\varepsilon$ has
probability tending to one, and on it the studentised statistic lies
between $1-\varepsilon$ and $1+\varepsilon$ times
$a_i\sqrt T(\widehat\alpha_i-\alpha_{i0})/\sqrt{v_\tau}$. Since
$\sup_x|\Phi(x/(1\pm\varepsilon))-\Phi(x)|\leq C\varepsilon$,
\eqref{eq:alpha-uniform-Gaussian} gives
\eqref{eq:alpha-studentized-uniform} on letting $\varepsilon\downarrow0$. \qed

\subsection{Proof of Theorem~\ref{thm:alpha-band}}

Because $a_i\geq c$,
$|\widehat M_{nT}-M_{nT}|\leq C\max_{i\leq n}|\widehat a_i-a_i|\,M_{nT}$.
Theorem~\ref{thm:alpha-maximum} gives $M_{nT}^2=O_p(\log n)$, so
\eqref{eq:alpha-density-log} yields $\widehat M_{nT}^2-M_{nT}^2=o_p(1)$,
and \eqref{eq:alpha-Gumbel} carries over to $\widehat M_{nT}$. When all
$\widehat a_i>0$, which holds with probability tending to one, the event
in \eqref{eq:alpha-simultaneous} is $\{\widehat M_{nT}\leq c_{n,\delta}\}$. Finally, the critical value
\eqref{eq:alpha-critical} is the exact $(1-\delta)$-quantile of the maximum
of $n$ independent absolute standard normal variables, and
\[
 c_{n,\delta}^2-2\log n+\log\log n
 \longrightarrow-\log\pi-2\log\{-\log(1-\delta)\}.
\]
Substitution into the continuous limiting distribution function proves
the simultaneous coverage \eqref{eq:alpha-simultaneous}. \qed

\section{Proofs for Section~\ref{sec:dependence}}\label{app:dependence}

Throughout this appendix, $\psi=\ps$, the index $i$ is dropped where no
confusion arises, and all bounds are uniform in $i$. Except in the proofs
of Proposition~\ref{prop:kappa} and Corollary~\ref{cor:dep-special}, the
conditions of Theorem~\ref{thm:dep} hold.

\emph{Rates.} Under (a)(G) of Assumption~\ref{ass:dep}, let
$\Theta(k)=\varrho^k$, $\varrho_1=\varrho^{1/(1+\varsigma)}$,
$\theta_1(k)=\varrho_1^k$, $\lambda(s)=1+\log^+(1/s)$ for $s\in(0,1]$ and
$m=m_T=\lceil C_m\log T\rceil$, where $C_m$ is so large that
$\varrho_1^{m/4}\leq T^{-(10+p)}$. Under (a)(P), let $\Theta(k)=k^{-\nu}$,
$\lambda_\nu=\nu/(1+\varsigma)$, which exceeds $9$, $\theta_1(k)=k^{-\lambda_\nu}$
and $\lambda(s)=s^{-1/\lambda_\nu}$, fix
$\mu\in(2/(\lambda_\nu-1),1/4)$, which is possible because
$\lambda_\nu>9$, and let $m=m_T=\lceil T^\mu\rceil$. In both cases,
$\E\norm{W_t-W_t^*}^\varsigma\leq C\Theta(t)$ for $t\geq1$, by
Lemma~\ref{lem:dep-poly} under (P); $m\leq C\bar m_T$; and, as
$T\to\infty$,
\begin{equation}\label{eq:dep-rates}
 \sum_{k\geq1}\theta_1(k)^{1/4}<\infty,\qquad
 T^2m\,\theta_1(\lfloor m/2\rfloor)+T^{3/2}m^2\theta_1(m)\to0,\qquad
 m\,\theta_1(m)\leq T^{-2}.
\end{equation}
Under (P), the second and third claims hold because
$\mu(\lambda_\nu-1)>2$ and hence $\mu(\lambda_\nu-2)>3/2$. Moreover,
for $s\in(0,1]$ and $C_\xi\geq1$,
\begin{equation}\label{eq:dep-lambda}
 \sum_{h\geq1}\min\{s,C_\xi\theta_1(h)\}\leq Cs\lambda(s/C_\xi),
\end{equation}
by splitting the sum where $C_\xi\theta_1(h)=s$, and $s\mapsto s\lambda(s/C_\xi)$
is non-decreasing. For integers $k<t$, let
$W_{it}^{*k}=H_i(\ldots,\eta'_{i,k-1},\eta'_{ik},\eta_{i,k+1},\ldots,\eta_{it})$
be $W_{it}$ with the innovations up to time $k$ replaced; it has the law
of $W_{it}$, it is independent of $\mathcal F_{ik}$, $W_{it}^{*0}=W_{it}^*$,
and stationarity gives
$\E\norm{W_{it}-W_{it}^{*k}}^\varsigma\leq C\Theta(t-k)$. We write
$W_{it}^{**k}$ for the same construction with a second copy
$(\eta''_{it})_{t\in\mathbb Z}$, independent of the innovations and of
$(\eta'_{it})_t$, so that $W_{it}^{*k}$ and $W_{is}^{**k'}$ are independent
when the periods $(k,t]$ and $(k',s]$ of their unreplaced innovations are
disjoint. Functions of coupled observations carry the same
superscript.

\subsection{Coupling, Moments and Decay}

\begin{lemma}[Polynomial coupling]\label{lem:dep-poly}
Under (a)(P) of Assumption~\ref{ass:dep},
$\E\norm{W_{it}-W_{it}^*}^\varsigma\leq Ct^{-\nu}$ for $t\geq1$.
\end{lemma}

\begin{proof}
For $j\geq0$, let $U_j$ be $W_t$ with $\eta_0,\eta_{-1},\ldots,\eta_{-j+1}$
replaced by $\eta'_0,\eta'_{-1},\ldots,\eta'_{-j+1}$, so that $U_0=W_t$.
Since $U_j$ and $U_{j+1}$ differ only in the innovation at time $-j$,
stationarity gives
$\E\norm{U_j-U_{j+1}}^\varsigma=\E\norm{W_{t+j}-W_{t+j}^{\langle0\rangle}}^\varsigma
\leq C(t+j)^{-\nu-1}$. Next, $U_j-W_t^*\to0$ in probability. The pair
$(U_j,W_t^*)$ has the law of $(W_t,W_t^{*(-j)})$, and, given
$\mathcal G_j=\sigma(\eta_{-j+1},\ldots,\eta_t)$, $W_t$ and
$W_t^{*(-j)}$ are conditionally i.i.d. With $\varphi(w)=w/(1+\norm w)$, a
homeomorphism onto the open unit ball, the martingale convergence theorem
gives $\E\{\varphi(W_t)\mid\mathcal G_j\}\to\varphi(W_t)$ almost surely; since
$\norm{\E\{\varphi(W_t)\mid\mathcal G_j\}}\leq\E\{\norm{\varphi(W_t)}\mid\mathcal G_j\}<1$
almost surely, $V_j=\varphi^{-1}[\E\{\varphi(W_t)\mid\mathcal G_j\}]$ is well defined,
$\mathcal G_j$-measurable and $V_j\to W_t$ almost surely; since
$(W_t^{*(-j)},V_j)$ has the law of $(W_t,V_j)$,
$\Prb(\norm{W_t-W_t^{*(-j)}}>2\epsilon)\leq2\Prb(\norm{W_t-V_j}>\epsilon)\to0$
for every $\epsilon>0$. Along a subsequence $j_k$, $U_{j_k}\to W_t^*$
almost surely, and, since $\norm{a+b}^\varsigma\leq\norm a^\varsigma+\norm b^\varsigma$
for $\varsigma\leq1$,
$\norm{W_t-W_t^*}^\varsigma\leq\sum_{j\geq0}\norm{U_j-U_{j+1}}^\varsigma$.
Taking expectations gives
$\E\norm{W_t-W_t^*}^\varsigma\leq C\sum_{k\geq t}k^{-\nu-1}\leq Ct^{-\nu}$.
\end{proof}

\begin{lemma}[Coupling and moment bounds]\label{lem:dep-moments}
\begin{enumerate}[label=(\roman*),leftmargin=2.4em]
\item For $k<t$ and $v\in\R$,
$\Prb\{\ind(u_t\leq v)\neq\ind(u_t^{*k}\leq v)\}
+\E\norm{x_t-x_t^{*k}}\leq C\theta_1(t-k)$.
\item Let $\xi_t$ be a fixed measurable function of
$(W_t,\ldots,W_{t+h_0})$, for a fixed $h_0\geq0$, with $\norm{\xi_t}\leq1$,
$\E\xi_t=0$ and $\E\norm{\xi_t-\xi_t^{*k}}\leq C_\xi\theta_1(t-k)$
for $k<t$, where $C_\xi\geq1$. Let $p_\xi$ be the dimension of $\xi_t$,
$\sigma^2=\E\norm{\xi_0}^2$ and $\mathcal A\subseteq\{1,\ldots,T\}$. Then
$\E\norm{\sum_{t\in\mathcal A}\xi_t}^2\leq C|\mathcal A|\sigma^2\lambda(\sigma^2/C_\xi)$,
$\E\norm{\sum_{t\in\mathcal A}\xi_t}^4\leq C|\mathcal A|^2$ and, for $x\geq1$,
\[
 \Prb\Bigl(\Bigl\|\sum_{t\in\mathcal A}\xi_t\Bigr\|\geq x\Bigr)
 \leq4p_\xi\exp\left[-\frac{x^2}
   {C|\mathcal A|\{\sigma^2\lambda(\sigma^2/C_\xi)+C_\xi T^{-2}\}+Cmx}\right]
   +\frac{4C_\xi T\theta_1(m)}x,
\]
where $C$ depends only on $h_0$, $p_\xi$ and the constants in the
assumptions and, in the fourth-moment bound, also on $C_\xi$.
\item Fix $t_0\in\{1,\ldots,T\}$ and let each $W_s$,
$s\in\{1,\ldots,T\}$, be replaced by some $\widetilde W_s=(\tilde x_s\trans,\tilde u_s)\trans$ with
$\E\norm{W_s-\widetilde W_s}^\varsigma\leq C\Theta(k_s)$ for some $k_s\geq m/2$,
such as $W_s^{*(s-k_s)}$, $W_s^{**(s-k_s)}$ or $W_s$ itself. Then, for all
sufficiently large $T$, outside
an event of probability at most $\pi_{3T}=CT\theta_1(\lceil m/2\rceil)$,
$\max_s\norm{W_s-\widetilde W_s}\leq\bar\theta_T=\theta_1(\lceil m/2\rceil)$ and, if
$|u_{t_0}|\leq\epsilon_0/2$, $\ind(u_s<u_{t_0})=\ind(\tilde u_s<\tilde u_{t_0})$ for
all $s\neq t_0$.
\end{enumerate}
\end{lemma}

\begin{proof}
(i) The indicators differ only if $|u_t-v|\leq\delta$ or
$|u_t-u_t^{*k}|>\delta$. The bounded density and Markov's inequality
bound the probability by $C\delta+C\Theta(t-k)\delta^{-\varsigma}$, and
$\delta=\Theta(t-k)^{1/(1+\varsigma)}=\theta_1(t-k)$ gives the claim; for the
regressors, $\norm{x_t-x_t^{*k}}\leq(2M)^{1-\varsigma}\norm{x_t-x_t^{*k}}^\varsigma$
and $\Theta\leq C\theta_1$.

(ii) For $h>h_0$, $\xi_h^{*h_0}$ is independent of $\xi_0$ and centred,
so $\norm{\E\xi_0\xi_h\trans}=\norm{\E\xi_0(\xi_h-\xi_h^{*h_0})\trans}\leq
C_\xi\theta_1(h-h_0)$, while $\norm{\E\xi_0\xi_h\trans}\leq\sigma^2$; summing the
minimum of the two bounds over $h$ and using \eqref{eq:dep-lambda} gives
the second-moment bound. For the fourth moment, let
$\mathcal P_s=\E(\cdot\mid\mathcal F_s)-\E(\cdot\mid\mathcal F_{s-1})$
be the projections of \citet[Section~2]{Wu07}. Since
$\E(\xi_k^{*0}\mid\mathcal F_0)=\E\xi_k=0$, Jensen's inequality gives
$\norm{\E(\xi_k\mid\mathcal F_0)}_4\leq\norm{\xi_k-\xi_k^{*0}}_4\leq
2\{C_\xi\theta_1(k)\}^{1/4}$, and likewise with $\mathcal F_{-1}$, so that
$\norm{\mathcal P_0\xi_k}_4\leq4\{C_\xi\theta_1(k)\}^{1/4}$ and
$\E(\xi_t\mid\mathcal F_{t-k})\to0$ in $L^4$ as $k\to\infty$. Hence
$\sum_{t\in\mathcal A}\xi_t=\sum_{k\geq-h_0}\sum_{t\in\mathcal A}\mathcal P_{t-k}\xi_t$,
where each inner sum is a sum of martingale differences. Burkholder's
inequality \citep{Burkholder73}, Minkowski's inequality in $L^2$ and
stationarity give
$\norm{\sum_{t\in\mathcal A}\mathcal P_{t-k}\xi_t}_4\leq C|\mathcal A|^{1/2}\norm{\mathcal P_0\xi_k}_4$,
and summation over $k$, by \eqref{eq:dep-rates}, proves the bound. For the
exponential bound, let
$\xi_t^{(m)}=\E(\xi_t\mid\eta_{t-m+1},\ldots,\eta_{t+h_0})$. Jensen's
inequality and the coupling give $\E\norm{\xi_t-\xi_t^{(m)}}\leq
\E\norm{\xi_t-\xi_t^{*(t-m)}}\leq C_\xi\theta_1(m)$, so that, by Markov's
inequality, $\sum_{t\in\mathcal A}\norm{\xi_t-\xi_t^{(m)}}\geq x/2$ with probability
at most $2C_\xi T\theta_1(m)/x$. The $\xi_t^{(m)}$ are centred, bounded by
one and $(m+h_0)$-dependent. Divide the integers into consecutive
blocks of length $m+h_0$ and group the summands with $t\in\mathcal A$
accordingly; the sums over the odd-numbered blocks are independent, as
are those over the even-numbered blocks, and each block sum is bounded
by $m+h_0$ and, by the second-moment bound applied to $\mathcal A\cap B$,
where $B$ is the block, and
$\E\norm{\xi_t-\xi_t^{(m)}}^2\leq2C_\xi\theta_1(m)$, has variance at most
$C|\mathcal A\cap B|\{\sigma^2\lambda(\sigma^2/C_\xi)+C_\xi m\theta_1(m)\}$,
which is at most
$C|\mathcal A\cap B|\{\sigma^2\lambda(\sigma^2/C_\xi)+C_\xi T^{-2}\}$ by
\eqref{eq:dep-rates}; these variances sum to at most
$C|\mathcal A|\{\sigma^2\lambda(\sigma^2/C_\xi)+C_\xi T^{-2}\}$ in each
group. Bernstein's inequality
\citep[Lemma 2.2.9]{VanDerVaarWellner1996EmpiricalProcess}, applied to
each group and coordinate, completes the proof.

(iii) By Markov's inequality,
$\Prb(\norm{W_s-\widetilde W_s}>\bar\theta_T)\leq C\Theta(\lceil m/2\rceil)\bar\theta_T^{-\varsigma}
=C\theta_1(\lceil m/2\rceil)$ for each $s$. On the complement of the union of
these events, the indicators can differ only if $|u_s-u_{t_0}|\leq2\bar\theta_T$, and,
for $|u_{t_0}|\leq\epsilon_0/2$ and large $T$, this has probability at most
$C\bar\theta_T$ by Assumption~\ref{ass:dep}(b), applied to the conditional
density of $u_s$ given $x_s$ and $W_{t_0}$; a union bound over $s$ completes
the proof.
\end{proof}

\begin{lemma}[Decay]\label{lem:dep-decay}
Uniformly in $i$ and $h\geq1$,
\begin{gather*}
 |\E\psi(u_{i0})\psi(u_{ih})|
 +\norm{\E z_{i0}z_{ih}\trans\psi(u_{i0})\psi(u_{ih})}\leq C\theta_1(h),\\
 \norm{\E\{z_{i0}\psi(u_{ih})\mid u_{i0}=0\}}
 +\norm{\E\{z_{i0}\psi(u_{i,-h})\mid u_{i0}=0\}}\leq C\theta_1(h)^{1/2}.
\end{gather*}
Hence the series in \eqref{eq:dep-bias} converge absolutely and uniformly
in $i$, and
\begin{gather*}
 V_{iT}:=\Var\Bigl(T^{-1/2}\sum_{t=1}^Tz_{it}\psi(u_{it})\Bigr)
 =\Lambda_i+O(T^{-1}),\\
 \Lambda_i=\sum_{h\in\mathbb Z}\E\{z_{i0}z_{ih}\trans\psi(u_{i0})\psi(u_{ih})\},
\end{gather*}
so that $\Lambda_n=n^{-1}\sum_i\Lambda_i$.
The same bounds hold, with the same constants, when $(u_{it},z_{it})$ is
replaced by $(e_{it}(d),x_{it}-m_i(\tau;d))$ of Appendix~\ref{app:lemmas}
for small deterministic $d$, provided that the products in the first
display are centred, since
$\E[\{x_{i0}-m_i(\tau;d)\}\psi(e_{i0}(d))]=S_i(d)$ need not vanish.
\end{lemma}

\begin{proof}
Since $u_h^{*0}$ is independent of $u_0$ and $\E\psi(u_h^{*0})=0$,
$|\E\psi(u_0)\psi(u_h)|=|\E\psi(u_0)\{\psi(u_h)-\psi(u_h^{*0})\}|\leq
C\theta_1(h)$ by Lemma~\ref{lem:dep-moments}(i), and the matrix bound
follows in the same way. For the conditional means, let
$\Xi_h(v)=\E\{z_0\ind(u_0\leq v)\psi(u_h)\}$. By the same coupling,
$\sup_v|\Xi_h(v)|=\sup_v|\E[z_0\ind(u_0\leq v)\{\psi(u_h)-\psi(u_h^{*0})\}]|\leq
C\theta_1(h)$. By Assumption~\ref{ass:dep}(b) with $J=\{h\}$, $\Xi_h$ is
twice differentiable on $(-\epsilon_0,\epsilon_0)$ with
$\Xi_h'(v)=\E\{z_0\psi(u_h)f_{\{h\}}(v\mid x_0,W_h)\}$ and
$|\Xi_h''|\leq C$, so that, for $0<\delta<\epsilon_0$,
$|\Xi_h'(0)|\leq\{|\Xi_h(\delta)|+|\Xi_h(-\delta)|\}/(2\delta)+C\delta$.
The choice $\delta=\min\{(\sup_v|\Xi_h(v)|)^{1/2},\epsilon_0/2\}$ gives
$|\Xi_h'(0)|\leq C\theta_1(h)^{1/2}$, which also holds if $\Xi_h$ vanishes
identically, and
$\Xi_h'(0)=a\,\E\{z_0\psi(u_h)\mid u_0=0\}$ with $a\geq c$. For the negative
lag, stationarity gives
$\E\{z_0\ind(u_0\leq v)\psi(u_{-h})\}=\E\{z_h\ind(u_h\leq v)\psi(u_0)\}$,
and the coupling of $W_h$ with $W_h^{*0}$ gives the same
bound. The series converge because
$\sum_h\theta_1(h)^{1/2}<\infty$ by \eqref{eq:dep-rates}. The covariance
$V_{iT}$ is the sum over $|h|<T$ of the lag-$h$ terms of $\Lambda_i$
multiplied by $1-|h|/T$, so it differs from $\Lambda_i$ by
$O\{\sum_h\min(1,|h|/T)\theta_1(|h|)\}=O(T^{-1})$, since
$\sum_hh\theta_1(h)<\infty$. The displaced process
satisfies Assumption~\ref{ass:dep}(a) and, with
$f_{i,J}(\cdot+x_{i0}\trans d+q_i(d)\mid\cdot)$ in place of $f_{i,J}$,
condition (b) on $(-\epsilon_0/2,\epsilon_0/2)$ for small $d$; the
products $\{x_{i0}-m_i(\tau;d)\}\psi(e_{i0}(d))$ are centred before the
covariance bound is applied.
\end{proof}

\subsection{Order Statistics and Counts}

Let $\delta_T=C_\delta(\log T/T)^{1/2}$ with $C_\delta$ large, let
$\mathcal A\subseteq\{1,\ldots,T\}$ omit at most $3m$ elements, let
$Q_{(1)}\leq\cdots\leq Q_{(|\mathcal A|)}$ be the order statistics of
$\{u_s:s\in\mathcal A\}$, and let $N_{\mathcal A}(v)=\sum_{s\in\mathcal A}\ind(u_s\leq v)$, so that
$N_{\mathcal A}(v)=j$ if and only if $Q_{(j)}\leq v<Q_{(j+1)}$. Put
$p_j(v)=\Prb\{N_{\mathcal A}(v)=j\}$ and $\sigma_j=\int_{-\epsilon_0}^{\epsilon_0}p_j(v)\,dv$,
the expected length of the part of the $j$th spacing that lies in
$(-\epsilon_0,\epsilon_0)$. Let
$\chi_T=C'(\log T)^{C'}T^{-1}$ be the bound in Assumption~\ref{ass:dep}(d),
$\omega_T=(\log T)^{C}\{T\delta_T\lambda(\delta_T)\}^{1/2}$ for a fixed $C>1/2$,
$\pi_{1T}=2T^{-2}+CT^{1/2}\theta_1(m)$ and
$\pi_{2T}=2\pi_{1T}+CT(\log T)\theta_1(m)$. Then $\delta_T\omega_T\to0$ and,
by \eqref{eq:dep-rates}, $T\pi_{2T}\to0$ and $Tm^2\pi_{1T}\to0$.

\begin{lemma}[Order statistics of large subsamples]\label{lem:dep-quantile}
For all sufficiently large $T$, uniformly in $i$, in such sets $\mathcal A$
and in $|j-T\tau|\leq T^{1/2}$:
\begin{enumerate}[label=(\roman*),leftmargin=2.4em]
\item $\Prb(|Q_{(j)}|>\delta_T)\leq\pi_{1T}$;
\item with probability at least $1-\pi_{2T}$,
$Q_{(j)}=a^{-1}\{(j/|\mathcal A|-\tau)+|\mathcal A|^{-1}\sum_{s\in\mathcal A}\psi(u_s)\}+R_j$,
where $|R_j|\leq C\omega_T/T$; consequently, for
$|j-T\tau|\leq3m+1$,
$\E\{Q_{(j)}^2\ind(|Q_{(j)}|\leq\epsilon_0)\}=\Omega_\psi/(a^2T)+o(T^{-1})$;
\item $\sigma_j=(aT)^{-1}+O(\epsilon_{\sigma T})$ for $|j-T\tau|\leq3m+1$, where
$\epsilon_{\sigma T}=mT^{-2}+\omega_TT^{-11/8}+T^{3/8}\delta_T\chi_T
+\pi_{1T}+\pi_{2T}T^{-3/8}=o(T^{-1})$.
\end{enumerate}
\end{lemma}

\begin{proof}
(i) Since $\E N_{\mathcal A}(\delta_T)=|\mathcal A|F(\delta_T)\geq j+cT\delta_T/2$ for
large $T$, $\Prb(Q_{(j)}>\delta_T)=\Prb\{N_{\mathcal A}(\delta_T)<j\}$ is bounded by
Lemma~\ref{lem:dep-moments}(ii), with bounded $C_\xi$ and
$x=cT\delta_T/2$: as $m\delta_T\to0$, the exponential term is at most
$T^{-2}$ when $C_\delta$ is large, and the last term is at most
$CT^{1/2}\theta_1(m)$; the lower tail is analogous.

(ii) On a grid of mesh $T^{-3/4}$ in $[-\delta_T,\delta_T]$, apply
Lemma~\ref{lem:dep-moments}(ii) to
$\ind(0<u_s\leq v)-\Prb(0<u_s\leq v)$ and its analogue for $v<0$, with
$\sigma^2\leq C\delta_T$ and $x=\omega_T$. The exponential term is at most
$CT^{-3}$, because $\omega_T^2/\{T\delta_T\lambda(\delta_T)\}=(\log T)^{2C}$
and $\omega_T/m\geq T^{1/4-\mu}$ under (P) and
$\omega_T/m\geq T^{1/4}/\log T$ under (G), and the last term is at most
$CT^{3/4}\theta_1(m)$. Monotonicity between grid points, which costs at
most $C|\mathcal A|T^{-3/4}\leq\omega_T$, and a union bound over the
$O\{T^{1/4}(\log T)^{1/2}\}$ grid points show that, outside an event of
probability at most $T^{-2}+CT(\log T)\theta_1(m)$, the oscillation
$\sup_{|v|\leq\delta_T}|N_{\mathcal A}(v)-N_{\mathcal A}(0)-|\mathcal A|\{F(v)-F(0)\}|$ is at most
$C\omega_T$. On this event and that of (i), $N_{\mathcal A}(Q_{(j)})=j$,
$N_{\mathcal A}(0)-|\mathcal A|\tau=-\sum_{s\in\mathcal A}\psi(u_s)$ and $F(v)-F(0)=av+O(v^2)$ give
the representation, with $|R_j|\leq C(\omega_T+T\delta_T^2)/T\leq C\omega_T/T$.
Since $Q_{(j)}^2\ind(|Q_{(j)}|\leq\epsilon_0)\leq\epsilon_0^2$, the
complement of these events contributes $O(\pi_{2T})=o(T^{-1})$ to the
truncated second moment. The linear term is bounded, centred and has
second moment $\Omega_\psi/(a^2T)+O(mT^{-2})$ by
Lemma~\ref{lem:dep-decay}, since $\mathcal A$ omits at most $3m$ periods; the
deterministic term is $O(m/T)$; and the remainder contributes
$O(\delta_T\omega_T/T)=o(T^{-1})$ on the event. This gives the second
claim.

(iii) By Assumption~\ref{ass:dep}(d) with $J=\emptyset$, for which the
probabilities are not random, so that the exceptional event can be taken to
be empty once $C'T^{-2}<1$, the
probabilities $p_j(v)$ change by at most $\chi_T$
between neighbouring $j$ for $|v|<\epsilon_0$, and (i) gives
$\sup_{\delta_T<|v|<\epsilon_0}p_j(v)\leq\pi_{1T}$, so that
$|\sigma_j-\sigma_{j'}|\leq2\delta_T|j-j'|\chi_T+2\epsilon_0\pi_{1T}$. Let
$J_T=\lfloor T^{3/8}\rfloor$. The sum $\sum_{j'=j-J_T}^{j+J_T-1}\sigma_{j'}$
is the expected length of
$[Q_{(j-J_T)},Q_{(j+J_T)})\cap(-\epsilon_0,\epsilon_0)$, which lies between
$0$ and $2\epsilon_0$ and, on the events of (i) and (ii) for $j\pm J_T$,
equals $Q_{(j+J_T)}-Q_{(j-J_T)}=2J_T/(a|\mathcal A|)+R_{j+J_T}-R_{j-J_T}$, because
the linear terms cancel. Hence the sum is
$2J_T/(a|\mathcal A|)+O(\omega_T/T+\pi_{2T})$, and averaging $\sigma_{j'}$ over
$j-J_T\leq j'<j+J_T$, with $|\mathcal A|\geq T-3m$, gives the claim. Finally,
$\omega_TT^{-11/8}=o(T^{-1})$ because $\omega_T=O(T^{1/4+1/(4\lambda_\nu)}(\log T)^C)$ under (P),
and $T^{3/8}\delta_T\chi_T=o(T^{-1})$.
\end{proof}

\subsection{The Mean of the Profiled Score}

\begin{lemma}[Mean of the profiled score under dependence]\label{lem:dep-mean}
Let $b_i^{\rm dep}=-(\Omega_{\psi,i}/2)c_i/a_i^2-\zeta_i$, so that
$b_n^{\rm dep}=n^{-1}\sum_ib_i^{\rm dep}$. For every sequence
$\varepsilon_T\to0$, uniformly in $i$ and deterministic $d$ with
$\norm d\leq\varepsilon_T$,
\begin{equation}\label{eq:dep-mean}
 \E S_{iT}(d)=S_i(d)+\frac{b_i^{\rm dep}}T+o(T^{-1}).
\end{equation}
\end{lemma}

\begin{proof}
\emph{Step 1: reduction.} The rank weights of $R_{it}(d)$ are those of
$e_{it}(d)=R_{it}(d)-q_i(d)$, which satisfies
$\Prb\{e_{it}(d)\leq0\}=\tau$, and, because the weights sum to $T\tau$,
$S_{iT}(d)=T^{-1}\sum_t\{x_{it}-m_i(\tau;d)\}\{\tau-w_{it}(d)\}$. For small
$d$, the process $(x_{it},e_{it}(d))_t$ satisfies
Assumption~\ref{ass:density}, condition (W) with $A_i(d)\geq c/2$ in place
of $a_i\geq c$, and conditions (a), (b) and
(d) of Assumption~\ref{ass:dep}, with $\epsilon_0/2$ in place of
$\epsilon_0$ and constants that do not depend on $d$, although not the
conditional restriction in \eqref{eq:model}: see the proof of
Lemma~\ref{lem:dep-decay} for (a) and (b),
and its counts of errors below $v$ are $N_{i,\mathcal A}(v+q_i(d),d)$, to which (d)
applies. Its coefficient
$b_i^{\rm dep}(d)$, defined as in \eqref{eq:dep-bias}, tends to $b_i^{\rm dep}$
uniformly in $i$ as $d\to0$: the series converge uniformly by
Lemma~\ref{lem:dep-decay}, their terms are continuous in $d$ because the
conditional densities have bounded derivatives, and
$c_i(d)/A_i(d)^2=\partial_vm_i(\tau;d)$ is continuous by
\eqref{eq:bias-slope-modulus}. It therefore suffices to prove
$\E S_{iT}(0)=b_i^{\rm dep}/T+o(T^{-1})$ with constants that depend only on
those in the assumptions; at $d\neq0$ the same argument gives
\eqref{eq:dep-mean}, the constant term
$\E[\{x_{i0}-m_i(\tau;d)\}\{\tau-\ind(e_{i0}(d)\leq0)\}]=S_i(d)$ arising in
Step 3 from $\Psi(0)$ below. At $d=0$, the weights sum to $T\tau$, so
$\E S_{iT}(0)=-T^{-1}\sum_{t=1}^T\E\{z_t(w_t-\tau)\}$.

\emph{Step 2: near and far periods.} Fix $t$ and let
$\mathcal N_t=\{s\in\{1,\ldots,T\}:0<|s-t|\leq m\}$ and
$\mathcal A_t=\{1,\ldots,T\}\setminus(\mathcal N_t\cup\{t\})$. Let $N_t^{\rm F}(v)$
be the number of $s\in \mathcal A_t$ with $u_s<v$ and $D_t(v)$ the number of
$s\in\mathcal N_t$ with $u_s<v$. Ties with $u_t$ have probability zero
when $|u_t|<\epsilon_0$, by Assumption~\ref{ass:dep}(b); when
$|u_t|\geq\epsilon_0$, both $w_t$ and $w_t^{\rm F}$ below equal
$\ind(u_t<0)$ outside an event of probability $O(\pi_{1T})$, by
Lemma~\ref{lem:dep-quantile}(i), so ties can be ignored. The rank of
$u_t$ is then $1+N_t^{\rm F}(u_t)+D_t(u_t)$, and
\[
 w_t=\ind\{N_t^{\rm F}(u_t)+D_t(u_t)\leq\ell-1\}
     +r\,\ind\{N_t^{\rm F}(u_t)+D_t(u_t)=\ell\}.
\]
Let $w_t^{\rm F}$ be the same expression with $D_t\equiv0$, the weight of
$u_t$ among the far observations alone. Then
$\E\{z_t(w_t-\tau)\}=\E\{z_t(w_t^{\rm F}-\tau)\}-\E\{z_t(w_t^{\rm F}-w_t)\}$,
and we treat the two terms in turn. Below, $\bar\pi_T=\pi_{1T}+\pi_{3T}+\bar\theta_T$,
with $\pi_{3T}$ and $\bar\theta_T$ from Lemma~\ref{lem:dep-moments}(iii); by
\eqref{eq:dep-rates}, $Tm\bar\pi_T\to0$.

\emph{Step 3: the far term.} Replace $W_t$ by $W_t^{*(t-m)}$
and each $W_s$ with $s>t+m$ by $W_s^{**t}$. After the replacement,
$(x_t,u_t)$ is independent of the far observations. Let $Q_{(j)}$ be the
order statistics of $\{u_s:s\in\mathcal A_t\}$ and $\widetilde Q_{(j)}$
those of the same errors after the replacement. By
Lemma~\ref{lem:dep-moments}(iii) with $t_0=t$, outside an event of
probability $\pi_{3T}$, all observations move by at most $\bar\theta_T$,
so that $|\widetilde Q_{(j)}-Q_{(j)}|\leq\bar\theta_T$, because order
statistics are 1-Lipschitz in the maximum norm, and, if
$|u_t|\leq\epsilon_0/2$, all indicators $\ind(u_s<u_t)$ are unchanged. If
$|u_t|>\epsilon_0/2$, then, outside that event and an event of
probability $2\pi_{1T}$ on which $|Q_{(\ell)}|$ or $|Q_{(\ell+1)}|$
exceeds $\delta_T$ (Lemma~\ref{lem:dep-quantile}(i)), these two order
statistics and their replacements are at most $\delta_T+\bar\theta_T$ in
absolute value, whereas $u_t$ and its replacement exceed
$\epsilon_0/2-\bar\theta_T>\delta_T+\bar\theta_T$ in absolute value for
large $T$, so that $w_t^{\rm F}=\ind(u_t<0)$ before and after the
replacement. Hence the replacement changes $\E\{z_tw_t^{\rm F}\}$ by
$O(\bar\pi_T)$. After the replacement,
$w_t^{\rm F}=(1-r)\ind(u_t<\widetilde Q_{(\ell)})+r\ind(u_t<\widetilde Q_{(\ell+1)})$
almost surely. Since $(x_t,u_t)$ is then independent of
the $\widetilde Q_{(j)}$ and $\Psi(q)=\E\{z_0\ind(u_0<q)\}$ is Lipschitz,
it follows that
$\E\{z_tw_t^{\rm F}\}=\E\{(1-r)\Psi(Q_{(\ell)})+r\Psi(Q_{(\ell+1)})\}+O(\bar\pi_T)$.
Here $\Psi(0)=\tau\E z_0$, $\Psi'(0)=\E\{z_0f(0\mid x_0)\}=0$ by the
definition of $\gamma_i$, and $\Psi''(q)=\E\{z_0f'(q\mid x_0)\}$ is
uniformly equicontinuous under condition (W), with $\Psi''(0)=c_i$. Since
$\Psi$ is bounded, Lemma~\ref{lem:dep-quantile}(i) and (ii), applied with
$\mathcal A=\mathcal A_t$ and $j\in\{\ell,\ell+1\}$, give
\[
 \E\{z_t(w_t^{\rm F}-\tau)\}=\frac{c_i}2\frac{\Omega_{\psi,i}}{a_i^2T}+o(T^{-1}).
\]

\emph{Step 4: the near term.} By Step 2,
\[
 w_t^{\rm F}-w_t=\ind\{\ell-D_t(u_t)\leq N_t^{\rm F}(u_t)\leq\ell-1\}
  +r\bigl[\ind\{N_t^{\rm F}(u_t)=\ell\}-\ind\{N_t^{\rm F}(u_t)+D_t(u_t)=\ell\}\bigr].
\]
Let $\mathcal W_t$ be the $\sigma$-field generated by the $W_s$ with
$|s-t|\leq m$, which determines $z_t$, $u_t$ and $D_t$, and let
$\pi_j(v)=\Prb\{N_t^{\rm F}(v)=j\mid\mathcal W_t\}$. By
Assumption~\ref{ass:dep}(d), with $\mathcal A=\mathcal A_t$ and $J=\{s:|s-t|\leq m\}$,
there is an event $E_t\in\mathcal W_t$ of probability at most $C'T^{-2}$
outside which
$|\pi_j(v)-\pi_{j'}(v)|\leq|j-j'|\chi_T$ for $|v|<\epsilon_0$. Since $D_t\leq2m$,
\[
 \E\{z_t(w_t^{\rm F}-w_t)\mid\mathcal W_t\}
 =z_t\sum_{k=1}^{D_t(u_t)}\pi_{\ell-k}(u_t)
   +rz_t\{\pi_\ell(u_t)-\pi_{\ell-D_t(u_t)}(u_t)\}
 =z_tD_t(u_t)\pi_\ell(u_t)+\rho_t,
\]
where
$|\rho_t|\leq Cm^2\chi_T\ind(|u_t|\leq\delta_T)
+Cm\sum_{k=0}^{2m}\pi_{\ell-k}(u_t)\ind(|u_t|>\delta_T)$ outside $E_t$, and
$|\rho_t|\leq Cm$ on $E_t$, because the $\pi_j$ are probabilities. If
$N_t^{\rm F}(u_t)=j$ and $u_t>\delta_T$, then $Q_{(j+1)}>\delta_T$, and
similarly for $u_t<-\delta_T$; hence Lemma~\ref{lem:dep-quantile}(i) and
the bounded density give
$\E|\rho_t|\leq Cm^2\chi_T\delta_T+Cm^2\pi_{1T}+CmT^{-2}=o(T^{-1})$, since
$m^2=o\{T^{1/2}(\log T)^{-C'-1/2}\}$ and $Tm^2\pi_{1T}\to0$. By the tower
property, $\E\{z_tD_t(u_t)\pi_\ell(u_t)\}$ is the sum over
$s\in\mathcal N_t$ of
$\E[z_t\ind(u_s<u_t)\ind\{N_t^{\rm F}(u_t)=\ell\}]$. Let
$g(v)=\E\{z_0f(v\mid x_0)\}$, so that $g(0)=0$ and $|g(v)|\leq C|v|$ for
$|v|<\epsilon_0$, and, for $h\neq0$,
$g_h(v)=\E\{z_0\ind(u_h<v)f_{\{h\}}(v\mid x_0,W_h)\}$, which is uniformly
Lipschitz near zero: the derivative of $f_{\{h\}}$ is bounded by
Assumption~\ref{ass:dep}(b), and the conditional density of $u_h$ given
$x_0$ and $x_h$ is bounded near zero by the same condition with
$J=\{-h\}$ and stationarity.

If $m/2<|s-t|\leq m$, replace $W_t$ by $W_t^{*(t-\lceil m/2\rceil)}$
and the $W_{s'}$ with $s'>t+m/2$ by $W_{s'}^{**t}$, which makes $(x_t,u_t)$
independent of all observations more than $m/2$ periods away; as in
Step 3, the term changes by $O(\bar\pi_T)$. After the replacement, it
equals $\E\int g(v)\ind(\tilde u_s<v)\ind\{\widetilde N_t^{\rm F}(v)=\ell\}\,dv$,
with tildes marking replaced quantities; outside the event of
Lemma~\ref{lem:dep-moments}(iii), the two indicators agree with
$\ind(u_s<v)$ and $\ind\{N_t^{\rm F}(v)=\ell\}$ except for $v$ in a set of
Lebesgue measure at most $3\bar\theta_T$, because
$|\tilde u_s-u_s|\leq\bar\theta_T$ and
$|\widetilde Q_{(j)}-Q_{(j)}|\leq\bar\theta_T$. Hence, since
$\sup_v|g(v)|\leq C$ and $\int|g(v)|\,dv\leq2M$ by
Assumption~\ref{ass:density}, the term equals
$\E\int g(v)\ind(u_s<v)\ind\{N_t^{\rm F}(v)=\ell\}\,dv+O(\bar\pi_T)$,
which, because $\Prb\{N_t^{\rm F}(v)=\ell\}\leq\pi_{1T}$ for
$|v|>\delta_T$, is at most
$C\delta_T\sigma_\ell+C\bar\pi_T=O(\delta_T/T+\bar\pi_T)$ by
Lemma~\ref{lem:dep-quantile}(i) and (iii), so these at most $2m$ terms contribute
$O(m\delta_T/T+m\bar\pi_T)=o(T^{-1})$. If $0<|s-t|\leq m/2$, replace $W_t$ and $W_s$ by $W_t^{*k}$ and $W_s^{*k}$
with $k=\min(s,t)-\lceil m/2\rceil$,
and the $W_{s'}$ with $s'>t+m$ by $W_{s'}^{**\max(s,t)}$, which makes $(W_t,W_s)$
independent of the observations in $\mathcal A_t$, more than $m/2$ periods from
both; as in Step 3 and the previous case, with $h=s-t$ and with the part
on which $|u_t|\geq\epsilon_0$ set aside, which contributes $O(\pi_{1T})$
by Lemma~\ref{lem:dep-quantile}(i), the term equals
\begin{align*}
 \int_{-\epsilon_0}^{\epsilon_0}g_h(v)\,\Prb\{N_t^{\rm F}(v)=\ell\}\,dv+O(\bar\pi_T)
 &=g_h(0)\sigma_\ell+O(\delta_T\sigma_\ell+\bar\pi_T)\\
 &=\frac{g_h(0)}{aT}+O\Bigl\{\frac{\delta_T}T
   +|g_h(0)|\epsilon_{\sigma T}+\bar\pi_T\Bigr\},
\end{align*}
by Lemma~\ref{lem:dep-quantile}(iii), and
$g_h(0)=a\E\{z_0\ind(u_h<0)\mid u_0=0\}=-a\E\{z_0\psi(u_h)\mid u_0=0\}$,
because $\E\{z_0\mid u_0=0\}=g(0)/a=0$. For $m<t\leq T-m$, summing over
$0<|h|\leq m/2$, with total error
$O(m\delta_T/T+\epsilon_{\sigma T}+m\bar\pi_T)=o(T^{-1})$ because
$\sum_h|g_h(0)|<\infty$ by Lemma~\ref{lem:dep-decay}, and using that
lemma for $|h|>m/2$ gives
$\E\{z_t(w_t^{\rm F}-w_t)\}=-\zeta_i/T+o(T^{-1})$. For the other $2m$
periods, the same bounds give $|\E\{z_t(w_t^{\rm F}-w_t)\}|\leq Cm/T$.

\emph{Step 5: conclusion.} By Steps 3 and 4,
$\E\{z_t(w_t-\tau)\}=(c_i/2)\Omega_{\psi,i}/(a_i^2T)+\zeta_i/T+o(T^{-1})$ for
$m<t\leq T-m$ and $|\E\{z_t(w_t-\tau)\}|\leq Cm/T$ otherwise, uniformly in
$i$ and $t$. Step 1 gives
$\E S_{iT}(0)=b_i^{\rm dep}/T+o(T^{-1})+O(m^2T^{-2})$, which proves the
lemma.
\end{proof}

\begin{lemma}[Remainder under dependence]\label{lem:dep-remainder}
Let $\epsilon_{RT}=(\omega_T/T)^2+\pi_{2T}+(\log T)^2T^{-2}$, so that
$T\epsilon_{RT}\to0$. Uniformly in $i$ and small deterministic $d$,
$\E\norm{\mathcal R_{iT}(d)-\E\mathcal R_{iT}(d)}^2\leq C\epsilon_{RT}$,
with $\mathcal R_{iT}(d)$ as in Appendix~\ref{app:lemmas}, and
$\E\mathcal R_{iT}(0)=b_i^{\rm dep}/T+o(T^{-1})$. Consequently,
\[
 \frac1n\sum_{i=1}^nS_{iT}(0)
 =\frac1{nT}\sum_{i=1}^n\sum_{t=1}^Tz_{it}\psi(u_{it})+\frac{b_n^{\rm dep}}T
  +\Op\{(\epsilon_{RT}/n)^{1/2}\}+o(T^{-1}).
\]
\end{lemma}

\begin{proof}
At $d=0$, let $q_T$ be the $(\ell+1)$th order statistic of
$u_1,\ldots,u_T$, $t^*$ the period with $u_{t^*}=q_T$, and
$\widehat\Psi(v)=T^{-1}\sum_tz_t\ind(u_t<v)$. Then
$\mathcal R_{iT}(0)=-\{\widehat\Psi(q_T)-\widehat\Psi(0)\}-rz_{t^*}/T$
almost surely on the event $|q_T|<\epsilon_0$, on which ties with $q_T$
have probability zero by Assumption~\ref{ass:dep}(b); the complement of
this event has probability at most $\pi_{1T}$ by
Lemma~\ref{lem:dep-quantile}(i), and $\mathcal R_{iT}(0)$ is bounded. The
term $rz_{t^*}/T$ has second moment $O(T^{-2})$. With $\Psi$
as in the proof of Lemma~\ref{lem:dep-mean},
$\widehat\Psi(q_T)-\widehat\Psi(0)=\Psi(q_T)-\Psi(0)+\Delta_T$, where, on
the event $|q_T|\leq\delta_T$, whose probability is at least $1-\pi_{1T}$
by Lemma~\ref{lem:dep-quantile}(i),
$\norm{\Delta_T}\leq\sup_{|v|\leq\delta_T}\norm{(\widehat\Psi-\Psi)(v)-(\widehat\Psi-\Psi)(0)}$.
Lemma~\ref{lem:dep-moments}(ii), applied as in the proof of
Lemma~\ref{lem:dep-quantile}(ii) to the positive and negative parts of the
coordinates of $z_t\{\ind(u_t<v)-\ind(u_t<0)\}$, centred, on a grid of
mesh $T^{-3/4}$, shows that this supremum exceeds $C\omega_T/T$ with
probability at most $\pi_{2T}$; since it is bounded, its second moment is
at most $C\{(\omega_T/T)^2+\pi_{2T}\}$. Finally,
$\norm{\Psi(q_T)-\Psi(0)}\leq Cq_T^2$, so that its variance is
$O\{(\log T)^2T^{-2}+\pi_{1T}\}$. This proves the variance bound at $d=0$, and the
displaced case is identical with $(x_t-m_i(\tau;d),e_t(d))$ in place of
$(z_t,u_t)$. The mean is given by Lemma~\ref{lem:dep-mean}, because
$\E\{z_t\psi(u_t)\}=0$, and independence across individuals gives the
averaged expansion.
\end{proof}

\subsection{Proof of Proposition~\ref{prop:kappa}}

The proof uses Lemma~\ref{lem:bias-modulus}, which involves only the law
of $(x_{i0},u_{i0})$, and Lemmas~\ref{lem:dep-poly},
\ref{lem:dep-moments}(i) and~\ref{lem:dep-decay}, whose proofs use only
Assumptions~\ref{ass:density} and~\ref{ass:causal} and conditions (a)
and (b) of Assumption~\ref{ass:dep}, with condition (W); the conditions of the proposition
therefore suffice. Fix $i$ and drop it. For small $d$, the implicit
function theorem applied to $\E F\{q(d)+x_0\trans d\mid x_0\}=\tau$ gives
$\nabla q(d)=-m(\tau;d)$ (Lemma~\ref{lem:bias-modulus}), so that, with
$\tilde q_t(d)=q(d)+x_t\trans d$ and $z_t(d)=x_t-m(\tau;d)=\nabla\tilde q_t(d)$,
$\nabla A(d)=\E[f'\{\tilde q_0(d)\mid x_0\}z_0(d)]$ and $\nabla A(0)=c$ under
condition (W). For $h\neq0$, the lag-$h$ term of $\Omega_\psi(d)$ is
$\Upsilon_h(d)=\Prb\{u_0\leq \tilde q_0(d),u_h\leq \tilde q_h(d)\}-\tau^2$, because
$e_t(d)=u_t-\tilde q_t(d)$ has $\tau$-quantile zero. Both thresholds move with
$d$, and we move them one at a time. By Assumption~\ref{ass:dep}(b), the
conditional law of $u_0$ given $(x_0,W_h)$ has the density $f_{\{h\}}$ on
$(-\epsilon_0,\epsilon_0)$ and, by stationarity, that of $u_h$ given
$(x_h,W_0)$ has the density $f_{\{-h\}}$, with bounded derivatives.
Hence, for $d$ and $d+\delta$ in a neighbourhood of zero that does not
depend on $h$, moving the first threshold from $\tilde q_0(d)$ to $\tilde q_0(d+\delta)$
changes $\Upsilon_h$ by
$\E[f_{\{h\}}\{\tilde q_0(d)\mid x_0,W_h\}z_0(d)\trans\delta\,\ind\{e_h(d)\leq0\}]
+O(\norm\delta^2)$; moving the second changes it by the same expression
with the roles of $0$ and $h$ interchanged and $f_{\{-h\}}$ in place of
$f_{\{h\}}$, up to $O(\norm\delta^2)$, since replacing
$\ind\{u_0\leq \tilde q_0(d+\delta)\}$ by $\ind\{u_0\leq \tilde q_0(d)\}$ in this
expression costs $O(\norm\delta^2)$, because $u_0$ has a bounded density
and no atom near zero. The remainders are uniform in $h$, so $\Upsilon_h$ is
differentiable on this neighbourhood with gradient
\begin{align*}
 &\E\bigl[f_{\{h\}}\{\tilde q_0(d)\mid x_0,W_h\}z_0(d)\ind\{e_h(d)\leq0\}\bigr]\\
 &\qquad+\E\bigl[f_{\{-h\}}\{\tilde q_h(d)\mid x_h,W_0\}z_h(d)\ind\{e_0(d)\leq0\}\bigr].
\end{align*}
Write $\ind\{e_h(d)\leq0\}=\tau-\psi\{e_h(d)\}$ and, extending the
notation of \eqref{eq:dep-bias} to the displaced errors,
$\E[z_0(d)\psi\{e_h(d)\}\mid e_0(d)=0]
=A(d)^{-1}\E[z_0(d)\psi\{e_h(d)\}f_{\{h\}}\{\tilde q_0(d)\mid x_0,W_h\}]$,
where $A(d)$ is the density of $e_0(d)$ at zero. Since
$\E[f_{\{h\}}\{\tilde q_0(d)\mid x_0,W_h\}\mid x_0]=f\{\tilde q_0(d)\mid x_0\}$ and
$\E[f\{\tilde q_0(d)\mid x_0\}z_0(d)]=0$ by the definition of $m(\tau;d)$, the
first term equals $-A(d)\E[z_0(d)\psi\{e_h(d)\}\mid e_0(d)=0]$, and the
second equals $-A(d)\E[z_h(d)\psi\{e_0(d)\}\mid e_h(d)=0]$. By
Lemma~\ref{lem:dep-decay} for the displaced process, both are bounded by
$C\theta_1(|h|)^{1/2}$ uniformly for small $d$, so the series of
gradients converges uniformly near zero, while $|\Upsilon_h(d)|\leq
C\theta_1(|h|)$. Hence $\Omega_\psi$, whose lag-zero term is the constant
$v_\tau$, is differentiable at zero with
\[
 \nabla\Omega_\psi(0)=-a\sum_{h\neq0}
 \bigl[\E\{z_0\psi(u_h)\mid u_0=0\}+\E\{z_h\psi(u_0)\mid u_h=0\}\bigr]
 =-2a\zeta,
\]
where the two sums are equal by stationarity. Hence
$\nabla\kappa(0)=\nabla\Omega_\psi(0)/a-\Omega_\psi\nabla A(0)/a^2
=-2\zeta-\Omega_\psi c/a^2$. \qed

\subsection{Proof of Theorem~\ref{thm:dep}}

All displacements below, $B_n^{\rm dep}/T$ and
$B_n^{\rm dep}/T+\theta/\sqrt{nT}$ for fixed $\theta$, have norm
$O\{T^{-1}+(nT)^{-1/2}\}=o(1)$, so Lemma~\ref{lem:dep-mean} applies to
them. Lemma~\ref{lem:bias-modulus} involves only the law of
$(x_{i0},u_{i0})$ and holds unchanged.

\emph{Step 1: the score at the displaced centre.} By
Lemmas~\ref{lem:dep-mean} and~\ref{lem:bias-modulus},
$n^{-1}\sum_i\E S_{iT}(B_n^{\rm dep}/T)$ equals $-\Gamma_nB_n^{\rm dep}/T+b_n^{\rm dep}/T
+o(T^{-1})=o(T^{-1})$, which is $o\{(nT)^{-1/2}\}$ because $n/T=O(1)$. For
the fluctuations,
the remainders contribute $\Op\{(T\epsilon_{RT})^{1/2}\}=\op(1)$ at the
$\sqrt{nT}$ scale by Lemma~\ref{lem:dep-remainder}, and for
$\norm b\leq C/T$ the increments $\xi_{it}(b)-\xi_{it}(0)$ have second
moments at most $C\norm b$ (proof of Lemma~\ref{lem:shifted-score}) and
satisfy the coupling condition of Lemma~\ref{lem:dep-moments}(ii), by
part (i) applied to the displaced errors. By that lemma and independence
across individuals, their normalised sum has variance
$O\{\norm b\lambda(\norm b)\}=o(1)$. Hence
$\Delta^c_{nT}=\sqrt{nT}\,n^{-1}\sum_iS_{iT}(B_n^{\rm dep}/T)
=(nT)^{-1/2}\sum_i\sum_tz_{it}\psi(u_{it})+\op(1)=\Op(1)$, the
counterpart of \eqref{eq:center-score-main}.

\emph{Step 2: the local quadratic approximation.} Lemma~\ref{lem:local}
holds with $B_n^{\rm dep}$ in place of $B_n$. In Step 1 of its proof,
Lemmas~\ref{lem:dep-mean} and~\ref{lem:bias-modulus} give
$\E\{S_{iT}(b+d)-S_{iT}(b)\}=-\Gamma_id+O(\norm d^2+\norm b\norm d)
+o(T^{-1})$, uniformly for $\norm b\leq C/T$ and $\norm d\leq\varepsilon_T$
with $\varepsilon_T\to0$, and hence
$\E D^c_{iT}(d)=\tfrac12d\trans\Gamma_id+O(\norm d^3+\norm b\norm d^2)
+o(\norm d/T)$. In Step 2,
Lemmas~\ref{lem:dep-moments}(ii) and~\ref{lem:dep-remainder} replace
independence over time and give
$\Var\{D^c_{iT}(d)\}\leq C\{\norm d^3\lambda(\norm d)/T+\norm d^2\epsilon_{RT}\}$. At
$d=\theta/\sqrt{nT}$, the mean error of $T\sum_iD^c_{iT}(d)$ is
$O\{(nT)^{-1/2}+T^{-1}\}+o(\sqrt{n/T})=o(1)$, because $n/T$ is bounded,
and its variance is
$O\{(nT)^{-1/2}\lambda((nT)^{-1/2})+T\epsilon_{RT}\}=o(1)$. Steps 3 and 4 are
unchanged.

\emph{Step 3: expansion and Gaussian approximation.} Step 1 of the proof
of Theorem~\ref{thm:main} now gives \eqref{eq:dep-expansion}. In the
linear term
$L^{\rm dep}_{nT}=\Gamma_n^{-1}(nT)^{-1/2}\sum_is_i$, the individual sums
$s_i=\sum_tz_{it}\psi(u_{it})$ are independent and centred, the
covariances of the $(nT)^{-1/2}s_i$ add up to
$n^{-1}\sum_iV_{iT}=\Lambda_n+O(T^{-1})$ by Lemma~\ref{lem:dep-decay},
whose eigenvalues are bounded away from zero by
Assumption~\ref{ass:dep}(c), and
$\sum_i\E\norm{(nT)^{-1/2}s_i}^3\leq n(nT)^{-3/2}(CT^2)^{3/4}=Cn^{-1/2}$ by
Lemma~\ref{lem:dep-moments}(ii). After whitening, the bound of
\citet[Theorem~1.1]{Raic19} gives a Gaussian approximation over convex
sets with error $O(n^{-1/2})$, and replacing $n^{-1}\sum_iV_{iT}$ by
$\Lambda_n$ changes the normal law by $O(T^{-1})$ in total variation.
Step 3 of the proof of Theorem~\ref{thm:main} then proves
\eqref{eq:dep-shifted}.

\emph{Step 4: the jackknife.} The contiguous sets are stationary panels
of lengths $T_1$ and $T_2=T-T_1$ that satisfy the conditions of the
theorem with $n/T_j=O(1)$ and, by stationarity, have the same
$\Gamma_n$ and $B_n^{\rm dep}$. Hence $\widehat\beta^{(j)}-\beta_0
=L_j+B_n^{\rm dep}/T_j+\op((nT)^{-1/2})$, with $L_j$ the linear
terms of the proof of Theorem~\ref{thm:spj}, and the cancellation there
gives $\widehat\beta_{\rm SPJ}-\beta_0=L+\op((nT)^{-1/2})$; Step 3
completes the proof. For the interleaved sets of
Remark~\ref{rem:dep-split}, the same argument applies to each subsampled
process under the condition stated there, with constant
$B_n^{{\rm dep},(2)}$ and length $T/2$, so the jackknife has bias
$2B_n^{\rm dep}/T-2B_n^{{\rm dep},(2)}/T$. \qed

\subsection{Proof of Corollary~\ref{cor:dep-special}}

(i) For $h\neq0$, $\E\{\psi(u_0)\psi(u_h)\}=0$ and
$\E\{z_0\psi(u_h)\mid u_0=0\}=\E\{z_0\mid u_0=0\}\E\psi(u_h)=0$, and
$\Lambda_n=v_\tau V_n$. (ii) Independence gives $f_i(u\mid x)=f_i(u)$,
so $\gamma_i=\E x_{i0}$, $\E z_{i0}=0$ and $c_i=f_i'(0)\E z_{i0}=0$; given
$u_{i0}=0$, $z_{i0}$ is independent of $u_{ih}$ with mean zero, so every
term of $\zeta_i$ vanishes; and since the rank weights are functions of
the errors, $\E(x_{it}\mid(u_{is})_s)=\E x_{i0}$ and the weights sum to
$T\tau$, the profiled score has mean zero for every $T$. (iii) For
$h\geq1$, $\E\{\psi(u_h)\mid x_h,(W_s)_{s<h}\}=0$ gives
$\E\{\psi(u_0)\psi(u_h)\}=0$ and $\E\{z_0z_h\trans\psi(u_0)\psi(u_h)\}=0$,
and the function $\Xi_h$ in the proof of Lemma~\ref{lem:dep-decay}
vanishes identically, so that $\E\{z_0\psi(u_h)\mid u_0=0\}=0$; stationarity turns
the terms with negative lags into those displayed. \qed

\subsection{Proof of Proposition~\ref{prop:cluster}}

\emph{The matrix $\widehat\Gamma_n$.} With $\widehat d=\widehat\beta-\beta_0$
and $\Delta_{iT}=\widehat\alpha_i-\alpha_{i0}$,
$\max_i|\Delta_{iT}|+\norm{\widehat d}=\Op\{(\log T/T)^{1/2}\}$, by
Theorem~\ref{thm:dep}, \eqref{eq:alpha-oracle-distance},
Lemma~\ref{lem:dep-quantile}(i) and a union bound over $n=O(T)$
individuals, since $T\pi_{1T}\to0$. Let
$\widehat M_{ki}=(Th)^{-1}\sum_tK(\widehat u_{it}/h)x_{it}^{\otimes k}$ for
$k=0,1,2$, with $x^{\otimes0}=1$, $x^{\otimes1}=x$ and
$x^{\otimes2}=xx\trans$, so that
$\widehat\Gamma_n=n^{-1}\sum_ig(\widehat M_i)$ with
$g(M)=M_2-M_1M_1\trans/M_0$, and $\Gamma_n=n^{-1}\sum_ig(M_i)$ with
$M_{ki}=\E\{f_i(0\mid x_{i0})x_{i0}^{\otimes k}\}$ and $M_{0i}=a_i\geq c$.
Replacing residuals by errors changes $\widehat M_{ki}$ by at most
$Ch^{-2}(|\Delta_{iT}|+M\norm{\widehat d})=\op(1)$, uniformly in $i$, by
the bounded derivative of $K$. For the errors, the expectations of the
moments $\widetilde M_{ki}=(Th)^{-1}\sum_tK(u_{it}/h)x_{it}^{\otimes k}$ are
within $O(h)$ of $M_{ki}$, as in the proof of Proposition~\ref{prop:cov},
and Lemma~\ref{lem:dep-moments}(ii), applied to the centred
$K(u_t/h)x_t^{\otimes k}$ divided by $2\norm K_\infty M^k$, with
$\sigma^2\leq Ch$ and $C_\xi\leq Ch^{-1}$, bounds their variances by
$CT^{-1}h^{-1}\lambda(h^2)\to0$, uniformly in $i$; here
$\E|K(u_t/h)-K(u_t^{*j}/h)|\leq2\norm K_\infty\Prb\{|u_t-u_t^{*j}|>\theta_1(t-j)\}
+\norm{K'}_\infty\theta_1(t-j)/h\leq Ch^{-1}\theta_1(t-j)$, by the proof of
part (i), and $\sigma^2\lambda(\sigma^2/C_\xi)\leq Ch\lambda(ch^2)$
because $s\lambda(s/C_\xi)$ is non-decreasing in $s$ and in
$C_\xi$. Since
$\norm{g(M)-g(M')}\leq C\norm{M-M'}$ when $M_0,M_0'\geq c/2$, and
$\norm{g(\widehat M_i)}\leq C(\widehat M_{0i}+\norm{\widehat M_{2i}})$ because
$\norm{\widehat M_{1i}}\leq M\widehat M_{0i}$, the Cauchy--Schwarz
inequality and $\Prb(\widetilde M_{0i}<c/2)\to0$ uniformly in $i$ give
$n^{-1}\sum_i\norm{g(\widehat M_i)-g(M_i)}\to_p0$. Hence
$\norm{\widehat\Gamma_n-\Gamma_n}\to_p0$. Moreover, the exponential bound
of Lemma~\ref{lem:dep-moments}(ii), applied with $x$ a small multiple of
$Th$, and a union bound over the $n=O(T)$ individuals give
$\min_{i\leq n}\widehat M_{0i}\geq c/2$ with probability tending to one: the
exponent is at least a multiple of $Th/\{\lambda(ch^2)+m\}$, which is at
least $cT^{5/8}$ under both (G) and (P), and $n$ times the last term is at most
$CTh^{-2}\theta_1(m)\to0$ by \eqref{eq:dep-rates}. On this event, every
$\widehat\gamma_i=\widehat M_{1i}/\widehat M_{0i}$ is a weighted average of
the $x_{it}$, so that $\norm{\widehat\gamma_i}\leq M$.

\emph{The matrix $\widehat\Lambda_n^{\rm CR}$.} With
$s_i=\sum_tz_{it}\psi(u_{it})$ as in the proof of Theorem~\ref{thm:dep},
Lemma~\ref{lem:dep-moments}(ii) gives
$\E\norm{s_i}^4\leq CT^2$, so that
$(nT)^{-1}\sum_is_is_i\trans-n^{-1}\sum_iV_{iT}=\Op(n^{-1/2})$, and
$n^{-1}\sum_iV_{iT}=\Lambda_n+O(T^{-1})$. By the Cauchy--Schwarz
inequality it remains to show that
$(nT)^{-1}\sum_i\norm{\widehat s_i-s_i}^2\to_p0$. With probability
tending to one, at most $p+1$ residuals of each individual vanish. Indeed,
by the first step, with probability tending to one every vanishing
residual has $|u_{it}|\leq|\Delta_{iT}|+M\norm{\widehat d}<\epsilon_0$. If
$p+2$ residuals of an individual vanish, the vector of the corresponding
errors lies in the span of the columns of the matrix with rows
$(1,x_{it}\trans)$ at these periods, and is therefore orthogonal to a
non-zero vector that is orthogonal to these columns and can be chosen as a
measurable function of the regressors. An error with a non-zero coefficient
in this vector is then a function of its own regressor and of the
observations of the other periods, which has probability zero when that
error lies in $(-\epsilon_0,\epsilon_0)$, by Assumption~\ref{ass:dep}(b)
with $J$ the set of the other periods and stationarity; a union over the
finitely many sets of periods, the choices of this error and the
individuals completes the argument. Hence
the subgradient conditions behind \eqref{eq:alpha-score-gap} give
$|\sum_t\psi(\widehat u_{it})|\leq p+1$, and $\psi(\widehat u_{it})$
differs from $\tau-w_{it}(\widehat d)$ for at most $p+1$ periods; hence
$\widehat s_i=TS_{iT}(\widehat d)+O(1)$ uniformly in $i$, while
$TS_{iT}(0)=s_i+T\mathcal R_{iT}(0)$. Lemma~\ref{lem:dep-remainder} gives
$\E\norm{T^{1/2}\mathcal R_{iT}(0)}^2\leq C(T\epsilon_{RT}+T^{-1})\to0$. Let
$\bar d_T=C_1\{T^{-1}+(nT)^{-1/2}\}$; by Theorem~\ref{thm:dep}, the event
$\norm{\widehat d}\leq \bar d_T$ has probability arbitrarily close to one for
large $C_1$. By \eqref{eq:profile-score},
$S_{iT}(d)=\tau\overline x_i-\widehat{\mathcal H}_i(\widehat q_i(d),d)
-rx_{i[\ell+1]}(d)/T$, up to an error of at most $pM/T$ when up to $p+1$
residuals are tied at $\widehat q_i(d)$, as may happen at $d=\widehat d$,
where $\widehat q_i(d)$ is the $(\ell+1)$th order
statistic of the $R_{it}(d)$ and
$\widehat{\mathcal H}_i(v,d)=T^{-1}\sum_tx_{it}\ind(u_{it}-x_{it}\trans d<v)$, whose
mean $\mathcal H_i(v,d)$ is Lipschitz in $(v,d)$. Order statistics move by at most
$M\norm d$ when $d$ changes, so
$\norm{\mathcal H_i(\widehat q_i(d),d)-\mathcal H_i(\widehat q_i(0),0)}\leq C\norm d$, and
$|\widehat q_i(d)|\leq2\delta_T$ for $\norm d\leq \bar d_T$ outside an event
of probability $\pi_{1T}$, by Lemma~\ref{lem:dep-quantile}(i). Split each
coordinate of the regressor into its positive and negative parts; since
$\ind(u_{it}<v-M\norm d)\leq\ind(u_{it}-x_{it}\trans d<v)\leq\ind(u_{it}<v+M\norm d)$,
the oscillation
$\sup_{|v|\leq2\delta_T,\,\norm d\leq \bar d_T}
\norm{(\widehat{\mathcal H}_i-\mathcal H_i)(v,d)-(\widehat{\mathcal H}_i-\mathcal H_i)(0,0)}$ is at most the
oscillation at $d=0$ over $|v|\leq3\delta_T$ plus $C\bar d_T$, and, as in the
proof of Lemma~\ref{lem:dep-remainder}, the former has second moment at
most $C\epsilon_{RT}$. Hence, on the event $\norm{\widehat d}\leq \bar d_T$,
$T\norm{S_{iT}(\widehat d)-S_{iT}(0)}^2$ is at most $C(T\bar d_T^2+T^{-1})$
plus a term whose mean is $O(T\epsilon_{RT})$, uniformly in $i$;
this does not use Lemma~\ref{lem:dep-mean}. Averaging over $i$ proves the
claim, and the coverage
statement follows from Theorem~\ref{thm:dep}, as in the proof of
Corollary~\ref{cor:slope-CI}. \qed

\subsection{Proof of the Claim in Remark~\ref{rem:hac}}\label{app:hac}

Let $s_{it}=z_{it}\psi(u_{it})$ and $\Lambda_i(j)=\E s_{ij}s_{i0}\trans$, so
that $\Lambda_i=\Lambda_i(0)+\sum_{j\geq1}\{\Lambda_i(j)+\Lambda_i(j)\trans\}$, and
let $\widetilde\Lambda_n$ be $\widehat\Lambda_n^{\rm HAC}$ computed with
the $s_{it}$ in place of the $\widehat s_{it}$. By
Lemma~\ref{lem:dep-decay}, $\norm{\Lambda_i(j)}\leq C\theta_1(j)$ for
$j\geq1$, and $\E\widetilde\Lambda_n$ has the coefficient
$(1-j/T)\{1-j/(b_T+1)\}$ on $n^{-1}\sum_i\Lambda_i(j)$ for
$j\leq b_T$, so that
$\norm{\E\widetilde\Lambda_n-\Lambda_n}\leq
C\sum_{j\geq1}\min\{1,j/(b_T+1)+j/T\}\theta_1(j)\to0$ by
dominated convergence, since $b_T\to\infty$ and
$\sum_j\theta_1(j)<\infty$. For each $j\leq b_T$, the products
$s_{it}s_{i,t-j}\trans$ are bounded, and for $|t-t'|>j$ the later
one can be coupled with the observations up to the earlier time, so that
Lemma~\ref{lem:dep-moments}(i) bounds their covariance by
$C\theta_1(|t-t'|-j)$; hence
$\Var\{T^{-1}\sum_ts_{it}s_{i,t-j}\trans\}\leq C(j+1)/T$, with $C$
not depending on $j$. By
Minkowski's inequality and independence across individuals,
$\E\norm{\widetilde\Lambda_n-\E\widetilde\Lambda_n}^2\leq
Cn^{-1}\{\sum_{j\leq b_T}(j+1)^{1/2}T^{-1/2}\}^2\leq Cb_T^3/(nT)\to0$.
Finally,
$\norm{\widehat s_{it}-s_{it}}\leq\norm{\widehat\gamma_i-\gamma_i}
+2M\ind(|u_{it}|\leq|\Delta_{iT}|+M\norm{\widehat d})$, and all scores are
bounded on the event $\max_i\norm{\widehat\gamma_i}\leq M$, whose
probability tends to one by the proof of Proposition~\ref{prop:cluster},
so that
\[
 \norm{\widehat\Lambda_n^{\rm HAC}-\widetilde\Lambda_n}\leq
 Cb_T\Bigl\{\frac1n\sum_{i=1}^n\norm{\widehat\gamma_i-\gamma_i}
 +\frac1{nT}\sum_{i,t}\ind(|u_{it}|\leq C_1\delta_T)\Bigr\}
\]
on this event and the event
$\max_i|\Delta_{iT}|+M\norm{\widehat d}\leq C_1\delta_T$,
whose probability is arbitrarily close to one for large $C_1$ by the first
step of the proof of Proposition~\ref{prop:cluster}. The last average has mean at most
$CC_1\delta_T$ by the bounded density, and, since
$\widehat\gamma_i=\widehat M_{1i}/\widehat M_{0i}$ and $\gamma_i=M_{1i}/M_{0i}$,
the same proof gives
$n^{-1}\sum_i\norm{\widehat\gamma_i-\gamma_i}=\Op\{h+(T^{-1}h^{-1}\lambda(h^2))^{1/2}+h^{-2}\delta_T\}
=\Op(T^{-1/8})$, because $h=C_hT^{-1/8}$. Hence
$\norm{\widehat\Lambda_n^{\rm HAC}-\widetilde\Lambda_n}=\Op(b_TT^{-1/8})=\op(1)$,
because $b_T=o(T^{1/8})$. \qed

\section{Special Cases in Serially Correlated Panels}\label{app:special}

This appendix treats two special classes of serially dependent panels in
which, unlike in Section~\ref{sec:dependence}, $n/T$ need not be
bounded. In the first, the conditional mean of the regressors given the
whole residual path is affine in the current residual, as in Gaussian
panels whose regressors and errors share a common linear filter. The
estimator then needs no bias correction, whatever the growth of $n$
relative to $T$ (Appendix~\ref{sec:dep-nobias}). The second, a 1-dependent
model whose one-period error density is constant near zero, shows that
condition (b) of Assumption~\ref{ass:dep} cannot be dropped: when $n/T$
tends to a positive constant, the estimator
has a non-Gaussian limit with an atom at the true slope
(Appendix~\ref{sec:dep-cusp}). The proofs are in
Appendices~\ref{app:nobias}--\ref{app:cusp}, in the notation of
Appendix~\ref{app:dependence}.

\subsection{Panels without Bias}\label{sec:dep-nobias}

By Corollary~\ref{cor:dep-special}(ii), the profiled score has mean zero at
the true slope for every $T$ when the regressor and error processes are
independent. This alone does not remove the restriction $n/T=O(1)$ in
Theorem~\ref{thm:dep}, because the local quadratic approximation also
involves the mean of the profiled score at slopes within $(nT)^{-1/2}$ of
the true slope, where Lemma~\ref{lem:dep-mean} controls it only up to
$o(T^{-1})$. The cusp model of Proposition~\ref{prop:cusp} below, whose
regressor and error processes are independent but which violates condition
(b), shows that, without (b), this independence does not even ensure a
Gaussian limit. In this subsection, we give
a class of panels in which the mean of the profiled score is known exactly
near the true slope. In this class, the estimator needs no bias correction,
and $n$ may grow at any rate relative to $T$.

For a deterministic $d$, let $R_{it}(d)=u_{it}-x_{it}\trans d$, $t\in\mathbb
Z$, be the residuals of Remark~\ref{rem:outline}. The condition is that the
conditional mean of the regressors given the whole residual path is affine
in the current residual: for some $r_{\mathcal D}>0$ and every $d$ in the
ball $\mathcal D=\{d\in\R^p:\norm d<r_{\mathcal D}\}$, there are
deterministic $\mu_i(d),\varphi_i(d)\in\R^p$ such that
\begin{equation}\label{eq:affine}
 \E\{x_{it}\mid(R_{is}(d))_{s\in\mathbb Z}\}=\mu_i(d)+\varphi_i(d)R_{it}(d)
 \qquad\text{for all }t.
\end{equation}
At $d=0$, the residuals are the errors, and \eqref{eq:affine} and
\eqref{eq:model} force $\varphi_i(0)=0$ (Appendix~\ref{app:nobias-proof}), so that
the regressors are mean independent of the whole error path. This excludes
feedback from past errors to current regressors, the source of the bias in
Corollary~\ref{cor:dep-special}(iii). At $d\neq0$, \eqref{eq:affine}
restricts the joint law of the regressor and error processes. With bounded
regressors, \eqref{eq:affine} with $\varphi_i(d)\neq0$ would force the
residuals, and hence the errors, to be bounded. Our leading example,
Corollary~\ref{cor:gauss} below, has Gaussian regressors, so we replace the
bound on the regressors by a moment condition.

\begin{assumption}[Affine conditional means]\label{ass:affine}
Assumption~\ref{ass:causal} holds without the bound on $\norm{x_{it}}$,
condition (a)(G) of Assumption~\ref{ass:dep} holds, and \eqref{eq:affine}
holds with $r_{\mathcal D}$ not depending on $i$. Moreover, there are
constants $\varkappa>4$, $C$ and $c>0$, uniform over $i$, such that:
\begin{enumerate}[label=(\alph*),leftmargin=2.4em]
\item $\E(\norm{x_{i0}}+|u_{i0}|)^\varkappa\leq C$;
\item the conditional density $f_i(\cdot\mid x)$ of $u_{i0}$ given
$x_{i0}=x$ is continuously differentiable on $\R$, with
$|f_i|+|f_i'|\leq C$, and $a_i=\E f_i(0\mid x_{i0})\geq c$;
\item for every $d\in\mathcal D$, the residuals $R_{it}(d)$,
$t\in\mathbb Z$, are almost surely distinct;
\item $\lambda_{\min}(\Gamma_n)\geq c$ and $\lambda_{\min}(\Lambda_n)\geq c$
for all sufficiently large $n$.
\end{enumerate}
\end{assumption}

Under the other conditions, the matrices $\Gamma_n$ of
\eqref{eq:matrices} and $\Lambda_n$ of \eqref{eq:dep-bias}, with $x_{i0}$
in place of $x_{i1}$, are well defined and bounded
(Lemmas~\ref{lem:nb-pop} and~\ref{lem:nb-moments} in Appendix~\ref{app:nobias}).
Condition (c) rules
out errors that coincide with positive probability. Let
$\mathbb M_i(d)=\min_{\alpha\in\R}\E\rho_\tau\{R_{i0}(d)-\alpha\}$ be the population
counterpart of the profiled criterion $\mathbb M_{iT}(d)$, and recall that
$S_i(d)$ is that of the profiled score $S_{iT}(d)$
(Appendix~\ref{app:lemmas}).

\begin{proposition}[Panels without bias]\label{prop:nobias}
Suppose that Assumption~\ref{ass:affine} holds.
\begin{enumerate}[label=(\roman*),leftmargin=2.4em]
\item For every $T\geq1$, $i$ and $d\in\mathcal D$,
\[
 \E\mathbb M_{iT}(d)=k_{iT}\mathbb M_i(d)\qquad\text{and}\qquad
 \E S_{iT}(d)=k_{iT}S_i(d),
\]
where $k_{iT}=\E\mathbb M_{iT}(0)/\mathbb M_i(0)$ does not depend on $d$ and
$1-C/T\leq k_{iT}\leq1$.
\item If $n,T\to\infty$, with no restriction on their relative growth, then
the slope component of every measurable minimiser in \eqref{eq:estimator}
satisfies
\[
 \widehat\beta-\beta_0
 =\Gamma_n^{-1}\frac1{nT}\sum_{i=1}^n\sum_{t=1}^Tz_{it}\ps(u_{it})
   +\op\bigl((nT)^{-1/2}\bigr)
\]
and
$\dK\{\mathcal L(\sqrt{nT}(\widehat\beta-\beta_0)),N(0,\Omega_n^{\rm dep})\}
\to0$.
\end{enumerate}
\end{proposition}

Part (i) holds exactly, for every $T$. The expected profiled criterion is a
multiple of the population criterion, and so is minimised at the true slope,
and the mean of the profiled score is the population score multiplied by a
factor that depends on $T$ but not on $d$. This is the expansion of the
mean in Remark~\ref{rem:dep-growth} with $\epsilon_T=0$, except that the
function $b_i(d)=T(k_{iT}-1)S_i(d)$, which is uniformly bounded and
Lipschitz and vanishes at $d=0$ because $S_i(0)=0$, depends on $T$. By part
(ii), the estimator is
approximately normal and centred, with the long-run covariance, whatever the
growth of $n$ relative to $T$, and no jackknife is needed. Unlike
Theorem~\ref{thm:dep}, part (ii) requires neither bounded regressors nor
conditions (b) and (d) of Assumption~\ref{ass:dep}, because the exact
identity in part (i) replaces the expansion of the mean in
Lemma~\ref{lem:dep-mean}. The class is narrow, however. For $T=2$,
$\mathbb M_{i2}(d)=\{\min(\tau,1-\tau)/2\}|R_{i1}(d)-R_{i2}(d)|$, so part (i)
requires $\E|R_{i1}(d)-R_{i2}(d)|/\mathbb M_i(d)$ not to depend on $d$, a property
of scale families of residual laws such as those of the next corollary.

\begin{corollary}[Gaussian panels with a common filter]\label{cor:gauss}
Let $G_{it}^{(k)}$, $k=0,\ldots,p$, be independent standard normal variables
over $i$, $t$ and $k$, and let
\[
 Z_{it}^{(k)}=\sum_{j\geq0}c_{ij}G_{i,t-j}^{(k)},\qquad
 x_{it}=Q_i^{1/2}\bigl(Z_{it}^{(1)},\ldots,Z_{it}^{(p)}\bigr)\trans,\qquad
 u_{it}=\sigma_{u,i}\bigl(Z_{it}^{(0)}-q_\tau\bigr),
\]
where $q_\tau=\Phi^{-1}(\tau)$, the $Q_i$ are symmetric $p\times p$
matrices with $cI_p\preceq Q_i\preceq CI_p$, $c\leq\sigma_{u,i}\leq C$,
$\sum_{j\geq0}c_{ij}^2=1$, $|c_{i0}|\geq c$ and $|c_{ij}|\leq C\varrho^j$
for some $\varrho\in(0,1)$. Then Assumption~\ref{ass:affine} holds for every
$r_{\mathcal D}>0$, with
\[
 \varphi_i(d)=-\frac{Q_id}{\sigma_i(d)^2},\qquad
 \mu_i(d)=\sigma_{u,i}q_\tau\varphi_i(d),\qquad
 \sigma_i(d)=(\sigma_{u,i}^2+d\trans Q_id)^{1/2}.
\]
Moreover, $\mathbb M_i(d)=\phi(q_\tau)\sigma_i(d)$,
$\E\mathbb M_{iT}(d)=\sigma_i(d)\E\min_{\alpha\in\R}T^{-1}
\sum_{t=1}^T\rho_\tau(Z_{it}^{(0)}-\alpha)$,
$\Gamma_n=n^{-1}\sum_i\phi(q_\tau)Q_i/\sigma_{u,i}$ and
\[
 \Lambda_n=\frac1n\sum_{i=1}^nQ_i\sum_{h\in\mathbb Z}r_i(h)c_\tau\{r_i(h)\},
 \qquad r_i(h)=\sum_{j\geq0}c_{ij}c_{i,j+|h|},
\]
where $c_\tau(r)=\Prb(Z\leq q_\tau,Z'\leq q_\tau)-\tau^2$ for standard
normal $Z$ and $Z'$ with correlation $r$. Every term of the inner series is
non-negative, and $\Lambda_n\succeq v_\tau n^{-1}\sum_iQ_i$.
\end{corollary}

An instance is $c_{ij}=(1-\rho_i^2)^{1/2}\rho_i^j$ with
$\sup_i|\rho_i|<1$, in which each coordinate of the regressors and the
errors follows a Gaussian autoregression of order one with coefficient
$\rho_i$, and $r_i(h)=\rho_i^{|h|}$. Here the regressors are unbounded,
so Theorem~\ref{thm:dep} does not apply. Moreover,
$r_i(h)c_\tau\{r_i(h)\}>0$ for $h\neq0$ when $\rho_i\neq0$, so that the
smallest eigenvalue of $\Lambda_n-v_\tau V_n$, and hence of
$\Omega_n^{\rm dep}-\Omega_n$, is bounded away from zero when a
non-vanishing fraction of the $\rho_i$ stay away from zero; standard errors
that estimate $\Omega_n$ are then too small. Clustering targets $\Lambda_n$, but
Proposition~\ref{prop:cluster} assumes bounded regressors and bounded $n/T$,
and we do not study it under Assumption~\ref{ass:affine}. The common filter
is what makes \eqref{eq:affine} hold: with different filters for the
regressors and the errors, the conditional mean of $x_{it}$ given the
residual path in general depends on the residuals of other periods.

\subsection{When Condition (b) Fails}\label{sec:dep-cusp}

Fast decay of the dependence and a smooth one-period error density do not
by themselves give a Gaussian approximation. In the following model,
adjacent errors share the same innovation with positive probability, so
the joint law of $(u_{i0},u_{i1})$ has a component on the diagonal and
condition (b) of Assumption~\ref{ass:dep} fails.

\begin{proposition}[A cusp model: a kink and an atom at the true slope]\label{prop:cusp}
Let $p=1$, $\tau=1/2$ and $T$ be even, and let
\[
 u_{it}=B_{it}\varepsilon_{it}+(1-B_{it})\varepsilon_{i,t-1},
\]
where the $B_{it}$ are i.i.d.\ Bernoulli$(1/2)$; the $\varepsilon_{it}$
are i.i.d., bounded and symmetric about zero, with a bounded density that equals
$f_0>0$ on an interval $[-a_0,a_0]$; the $x_{it}$ are i.i.d., bounded,
symmetric about zero and continuously distributed; and the three arrays
are independent. Put
$\sigma_x^2=\E x_{i0}^2$, $D_x=\E|x_{i0}-x_{i1}|$, $\Gamma=f_0\sigma_x^2$ and
$\Lambda=\sigma_x^2/4$. If $n,T\to\infty$ with
$n/T\to\upsilon\in[0,\infty)$, then every measurable slope minimiser in
\eqref{eq:estimator} satisfies
\[
 \sqrt{nT}(\widehat\beta-\beta_0)\Rightarrow
 \Gamma^{-1}\operatorname{sgn}(Z)\bigl(|Z|-c_\upsilon\bigr)_+,
 \qquad Z\sim N(0,\Lambda),\qquad c_\upsilon=\frac{D_x}8\sqrt\upsilon,
\]
and $\Prb(\widehat\beta=\beta_0)\to2\Phi\{D_x\sqrt\upsilon/(4\sigma_x)\}-1$.
If instead $n/T\to\infty$, then $\Prb(\widehat\beta=\beta_0)\to1$, however
fast $n$ grows relative to $T$.
\end{proposition}

\begin{figure}[!htb]
\centering
\includegraphics[width=\textwidth]{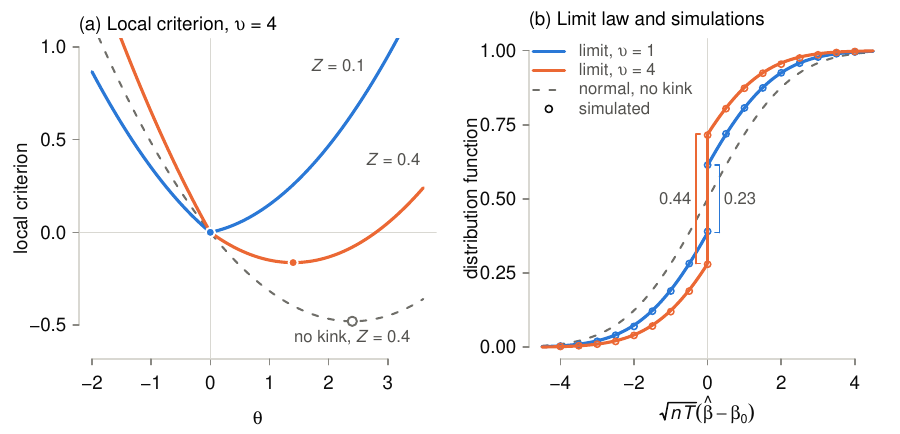}
\caption{The cusp model of Proposition~\ref{prop:cusp} with $x_{it}$ and
$\varepsilon_{it}$ uniform on $[-1,1]$, so that $\Gamma=1/6$ and
$\Lambda=1/12$. (a) The limiting local criterion
$-\theta Z+\Gamma\theta^2/2+c_\upsilon|\theta|$ at $\upsilon=4$, so that
$c_\upsilon=1/6$, for two values of $Z$, and the criterion without the
kink at $Z=0.4$ (dashed); dots mark the minimisers. When $|Z|\leq c_\upsilon$,
the minimiser stays at zero; otherwise the kink moves it towards zero by
$c_\upsilon/\Gamma=1$. (b) Distribution functions of
$\sqrt{nT}(\widehat\beta-\beta_0)$: the soft-thresholded limits, with atoms
of mass 0.23 and 0.44 at zero; the empirical distribution functions from
10,000 replications with $(n,T)=(100,100)$ and $(400,100)$, in which
$\widehat\beta=\beta_0$ exactly in 22.4 and 43.6 per cent of the
replications (minimisers computed by bisection on the averaged profiled
score, which is monotone); and the normal law $N(0,\Lambda/\Gamma^2)$
that Theorem~\ref{thm:dep} would give.}
\label{fig:cusp}
\end{figure}

Each $W_{it}$ is a function of $(\eta_{it},\eta_{i,t-1})$, with
$\eta_{it}=(B_{it},\varepsilon_{it},x_{it})$, so that $W_{it}^*=W_{it}$ for
$t\geq2$. The model is therefore 1-dependent and satisfies (a)(G), as well as
Assumptions~\ref{ass:density}, \ref{ass:nondeg} and~\ref{ass:causal},
condition (W) and conditions (c) and (d). For (d), given the Bernoulli
variables and $(W_{is})_{s\in J}$, the count $N_{i,\mathcal A}(v,d)$ is, apart
from the at most $|J|$ periods whose error coincides with that of a period
in $J$, whose indicators then depend only on their own regressors and are
independent of the rest, a sum of independent
counts over clusters of one or two errors, which keep their unconditional
laws. Since the Bernoulli variables of the periods outside $J$ are
independent of $(W_{is})_{s\in J}$, at least $cT$ of these clusters are
single errors outside an event of conditional probability $Ce^{-cT}$. Their
counts are Bernoulli with success probabilities bounded away from
zero and one, so that the conditional characteristic function of $N_{i,\mathcal A}(v,d)$ is
at most $\exp\{-cT\sin^2(t/2)\}+Ce^{-cT}$ in modulus and Fourier
inversion gives the bound $C/T$. Since the regressors are independent of
the errors, have mean zero and face a flat error density near zero,
$c_i=0$ and $\Lambda_n=\Lambda$, and $\zeta_i=0$ once the expectations
given $u_{i0}=0$ in \eqref{eq:dep-bias} are read as conditional
expectations, because $x_{i0}$ is independent of $(u_{i0},u_{ih})$. Thus
$b_n^{\rm dep}=0$, and the conclusion of Theorem~\ref{thm:dep} would be a
centred normal limit with variance $\Lambda/\Gamma^2$.
Proposition~\ref{prop:nobias} does not apply either: condition (c)
of Assumption~\ref{ass:affine} fails at $d=0$, and \eqref{eq:affine} fails
at every small $d\neq0$, because the flat error density makes
$\E\{x_{it}\mid R_{it}(d)\}$ vanish for $R_{it}(d)$ near zero, which would
force $\mu_i(d)=\varphi_i(d)=0$, whereas
$\E\{x_{it}R_{it}(d)\}=-d\sigma_x^2\neq0$. Adjacent errors coincide with probability $1/4$,
however, and the shared innovation gives every individual's expected
profiled criterion a cusp, a kink at the true slope, and about a quarter
of the realised criteria a kink exactly there: the proof
(Appendix~\ref{app:cusp}) shows that
$\E S_{iT}(d)=-\Gamma d-\operatorname{sgn}(d)D_x/(8T)
+O\{T^{-3/2}(\log T)^{1/2}\}$ for small $d\neq0$, and, correspondingly,
$\kappa_i(d)=3/(8f_0)-D_x|d|/4$ for small $d$, which is not
differentiable at zero. When
$n/T\to\upsilon>0$, the kink is of the order of the standard deviation,
and the limit is a soft-thresholded normal law with an atom at zero, so
no Gaussian approximation holds. Unlike a bias, the kink pulls the
estimator towards the true slope, and when $n/T\to\infty$ it dominates the
sampling error, so that the estimator equals the true slope with
probability tending to one. If $x_{it}$ and $\varepsilon_{it}$ are
uniform on $[-1,1]$, the mass of the limiting atom is $2\Phi\{(\upsilon/12)^{1/2}\}-1$,
or $0.23$ at $\upsilon=1$ and $0.44$ at $\upsilon=4$.
Figure~\ref{fig:cusp} shows how the kink produces the atom and compares
the limit with simulations.

\FloatBarrier
\subsection{Proofs for Appendix~\ref{sec:dep-nobias}}\label{app:nobias}

Throughout this subsection and Appendix~\ref{app:nobias-proof},
Assumption~\ref{ass:affine} holds; $c$ and $C$ depend only on $p$, $\tau$ and the constants in
Assumption~\ref{ass:affine}.

\emph{Rates.} Let $\iota\in(0,1)$ solve
$1/3=(1-\iota)/\varsigma+\iota/\varkappa$, which is possible because
$\varsigma\leq1<3<\varkappa$. Hölder's inequality, with exponents
$\varsigma/\{3(1-\iota)\}$ and $\varkappa/(3\iota)$, gives
$\norm X_3\leq(\E\norm X^\varsigma)^{(1-\iota)/\varsigma}\norm X_\varkappa^\iota$
for every random vector $X$.
Put $\varrho_3=\varrho^{(1-\iota)/\varsigma}$ and $\varrho_2=\varrho_3^{3/4}$.
In this subsection, $\theta_1(k)=\varrho_2^k$, and $\lambda$ and $m$ are
defined as under (a)(G) at the start of Appendix~\ref{app:dependence}, with $\varrho_2$ in
place of $\varrho_1$, so that $\theta_1(m)\leq T^{-40}$. Then
\eqref{eq:dep-rates} and \eqref{eq:dep-lambda} hold. Since its proof uses
only its own hypotheses, the coupled variables and these two displays,
Lemma~\ref{lem:dep-moments}(ii) holds with this $\theta_1$ and $m$.

\emph{Population quantities.} Let $F(\cdot\mid x)$ be the conditional
distribution function of $u_0$ given $x_0=x$. For $d\in\R^p$, let $q(d)$ be
a $\tau$-quantile of $R_0(d)$, $e_t(d)=R_t(d)-q(d)$,
$f_d(x)=f\{q(d)+x\trans d\mid x\}$, $A(d)=\E f_d(x_0)$,
$\gamma(d)=\E\{x_0f_d(x_0)\}/A(d)$, $z_t(d)=x_t-\gamma(d)$,
$\Gamma(d)=\E\{f_d(x_0)z_0(d)z_0(d)\trans\}$,
$S(d)=\E[x_0\psi\{e_0(d)\}]$ and $\mathbb M(d)=\E\rho_\tau\{e_0(d)\}$. These agree
with the definitions in Section~\ref{sec:dep-bias} and
Appendix~\ref{app:lemmas}; in particular, $\gamma(d)=m_i(\tau;d)$,
$q(0)=0$, $\gamma(0)=\gamma_i$, $\Gamma(0)=\Gamma_i$ and $z_t(0)=z_{it}$.
We use repeatedly that $\psi(v)$ is a subgradient of $\rho_\tau$ at $v$, so
that
\begin{equation}\label{eq:nb-sub}
 \rho_\tau(v')\geq\rho_\tau(v)+\psi(v)(v'-v),\qquad
 0\leq\rho_\tau(v-w)-\rho_\tau(v)+w\psi(v)\leq2|w|\ind(|v|\leq|w|),
\end{equation}
for all real $v$, $v'$ and $w$; the second bound holds because the middle
expression vanishes unless $v$ and $v-w$ lie on different sides of zero,
in which case $|v|\leq|w|$, and $\rho_\tau$ is $1$-Lipschitz.

\begin{lemma}[Population quantities]\label{lem:nb-pop}
\begin{enumerate}[label=(\roman*),leftmargin=2.4em]
\item For every $d$, $\mathbb M(d)=\min_\alpha\E\rho_\tau\{R_0(d)-\alpha\}>0$, and
$\mathbb M$ is differentiable with $\nabla \mathbb M(d)=-S(d)$.
\item There are $r_1\in(0,\min\{r_{\mathcal D}/2,1\}]$ and $c_1>0$ such that,
for $\norm d\leq r_1$, $q(d)$ is unique, $|q(d)|\leq C\norm d$, the
conditional density of $e_0(d)$ given $x_0$ is bounded by $C$, and
$|\Prb\{e_0(d)\leq v\}-\tau|\geq c_1\min(|v|,r_1)$ for all $v\in\R$. On
$\{\norm d<r_1\}$, $q$ and $S$ are differentiable, with
$\nabla q(d)=-\gamma(d)$ and Jacobian $-\Gamma(d)$ of $S$, and
$A(d)\geq c$, $\norm{\gamma(d)}+\norm{\Gamma(d)}\leq C$ and
$\norm{\gamma(d)-\gamma(0)}+\norm{\Gamma(d)-\Gamma(0)}\leq C\norm d$.
\item $S(0)=0$, $\mathbb M(0)\geq c$ and, for $\norm d\leq r_1$,
$\norm{S(d)}\leq C\norm d$ and
$|\mathbb M(d)-\mathbb M(0)-\tfrac12d\trans\Gamma(0)d|\leq C\norm d^3$.
\end{enumerate}
\end{lemma}

\begin{proof}
(i) Given $x_0$, $R_0(d)$ has the density $f(\cdot+x_0\trans d\mid x_0)$,
so $e_0(d)$ has a continuous distribution, $\E\psi\{e_0(d)\}=0$ and
$\mathbb M(d)>0$. The first inequality in \eqref{eq:nb-sub} with $v=e_0(d)$ and
$v'=R_0(d)-\alpha$ shows that $\mathbb M(d)$ is the minimum. With $v'=e_0(d')$ it
gives $\mathbb M(d')\geq \mathbb M(d)-(d'-d)\trans S(d)$, because
$e_0(d')-e_0(d)=-x_0\trans(d'-d)-\{q(d')-q(d)\}$ and
$\E\psi\{e_0(d)\}=0$. Conversely, by the second bound in \eqref{eq:nb-sub}
with $w=x_0\trans(d'-d)$,
\begin{align*}
 \mathbb M(d')&\leq\E\rho_\tau\{e_0(d)-x_0\trans(d'-d)\}\\
 &\leq \mathbb M(d)-(d'-d)\trans S(d)
 +2\norm{d'-d}\E[\norm{x_0}\ind\{|e_0(d)|\leq\norm{x_0}\norm{d'-d}\}],
\end{align*}
and the last expectation tends to zero as $d'\to d$ by dominated
convergence, since $\Prb\{e_0(d)=0\}=0$. Hence $\mathbb M$ is differentiable at $d$
with gradient $-S(d)$.

(ii) Let $A(d,v)=\E f(v+x_0\trans d\mid x_0)$ be the density of $R_0(d)$ at
$v$. Since $|f'|\leq C$, $|A(d,v)-a_i|\leq C'(|v|+\norm d)$, so that
$A(d,v)\geq c/2$ whenever $|v|+\norm d\leq2r_2$, with $r_2=c/(4C')$. Also
$|\Prb\{R_0(d)\leq0\}-\tau|=|\E\{F(x_0\trans d\mid x_0)-F(0\mid x_0)\}|
\leq C''\norm d$. Hence, for
$\norm d\leq r_1:=\min\{r_2/2,cr_2/(8C''),r_{\mathcal D}/2,1\}$, the
distribution function of $R_0(d)$ increases at rate at least $c/2$ on
$[-r_2,r_2]$, its $\tau$-quantile is unique and satisfies
$|q(d)|\leq2C''\norm d/c\leq r_2/4$, and the density of $e_0(d)$, namely
$A(d,q(d)+v)$, is at least $c/2$ for $|v|\leq r_1$; this gives the
lower bound on $|\Prb\{e_0(d)\leq v\}-\tau|$ with $c_1=c/2$. Given $x_0$,
the density of $e_0(d)$ is $f(\cdot+q(d)+x_0\trans d\mid x_0)\leq C$. The
map $(d,v)\mapsto\Prb\{R_0(d)\leq v\}=\E F(v+x_0\trans d\mid x_0)$ has the
continuous partial derivatives $\E\{x_0f(v+x_0\trans d\mid x_0)\}$ and
$A(d,v)$, by dominated convergence, and $A(d,q(d))\geq c/2$, so the
implicit function theorem and the uniqueness of $q(d)$ give
$\nabla q(d)=-\gamma(d)$. In the same way,
$S(d)=\tau\E x_0-\E[x_0F\{q(d)+x_0\trans d\mid x_0\}]$ has Jacobian
$-\E[f_d(x_0)x_0\{x_0-\gamma(d)\}\trans]=-\Gamma(d)$, because
$\E\{f_d(x_0)z_0(d)\}=0$ by the definition of $\gamma(d)$. Finally,
$A(d)=A(d,q(d))\geq c/2$, $\norm{\gamma(d)}\leq C\E\norm{x_0}/A(d)$ and
$\norm{\Gamma(d)}\leq C\E\norm{z_0(d)}^2$, and, since
$|f_d(x)-f_{d'}(x)|\leq C(|q(d)-q(d')|+\norm x\norm{d-d'})
\leq C(1+\norm x)\norm{d-d'}$, because $q$ is Lipschitz on
$\{\norm d\leq r_1\}$ with constant $\sup\norm\gamma\leq C$, and
$\E(1+\norm{x_0})^3\leq C$, the functions $A$, $\E\{x_0f_d(x_0)\}$ and
$\E\{x_0x_0\trans f_d(x_0)\}$, and hence $\gamma$ and $\Gamma$, are
Lipschitz on $\{\norm d\leq r_1\}$.

(iii) By \eqref{eq:model}, $S(0)=\E[x_0\{\tau-F(0\mid x_0)\}]=0$, and,
since the density of $u_0$ is at least $c/2$ on $[-r_1,r_1]$,
$\mathbb M(0)=\E\rho_\tau(u_0)\geq\min(\tau,1-\tau)\E\{|u_0|\ind(|u_0|\leq r_1)\}
\geq\min(\tau,1-\tau)cr_1^2/2$. By (ii), $\norm{S(d)}\leq C\norm d$, and,
by (i) and (ii),
$\mathbb M(d)-\mathbb M(0)=-\int_0^1d\trans S(sd)\,ds
=\int_0^1\int_0^1s\,d\trans\Gamma(ss'd)d\,ds'\,ds$, which differs from
$\tfrac12d\trans\Gamma(0)d$ by at most $C\norm d^3$.
\end{proof}

\begin{lemma}[Coupling]\label{lem:nb-coupling}
For $k<t$, $\norm d\leq r_1$ and $v\in\R$,
\[
 \norm{W_t-W_t^{*k}}_3
 +\Prb\{\ind(R_t(d)\leq v)\neq\ind(R_t^{*k}(d)\leq v)\}
 \leq C\theta_1(t-k).
\]
If $g:\R^p\to\R^{p'}$ satisfies $\norm{g(x)}\leq C(1+\norm x)$ and
$\norm{g(x)-g(x')}\leq C\norm{x-x'}$, then
$\xi_t=g(x_t)\psi\{R_t(d)-v\}$ satisfies
$\norm{\xi_t-\xi_t^{*k}}_3\leq C\theta_1(t-k)^{1/12}$.
\end{lemma}

\begin{proof}
By stationarity and (a)(G),
$\E\norm{W_t-W_t^{*k}}^\varsigma\leq C\varrho^{t-k}$, and
$\norm{W_t-W_t^{*k}}_\varkappa\leq2\norm{W_0}_\varkappa\leq C$, so the
interpolation inequality above gives
$\norm{W_t-W_t^{*k}}_3\leq C\varrho_3^{t-k}\leq C\theta_1(t-k)$. The
indicators differ only if $|R_t(d)-v|\leq\delta$ or
$|R_t(d)-R_t^{*k}(d)|>\delta$. The density of $R_t(d)$ is bounded by $C$,
and $|R_t(d)-R_t^{*k}(d)|\leq(1+\norm d)\norm{W_t-W_t^{*k}}$, so Markov's
inequality bounds the probability by
$C\delta+C\varrho_3^{3(t-k)}\delta^{-3}$, and
$\delta=\varrho_3^{3(t-k)/4}=\theta_1(t-k)$ gives the first claim. For the
second, $\xi_t-\xi_t^{*k}$ is the sum of
$\{g(x_t)-g(x_t^{*k})\}\psi\{R_t(d)-v\}$, whose $L^3$ norm is at most
$C\norm{W_t-W_t^{*k}}_3$, and $g(x_t^{*k})$ times the difference of the
indicators, whose $L^3$ norm is at most
$\norm{g(x_0)}_\varkappa\Prb\{\ind(R_t(d)\leq v)\neq\ind(R_t^{*k}(d)\leq v)\}^{1/3-1/\varkappa}$
by Hölder's inequality; since $1/3-1/\varkappa>1/12$, both are at most
$C\theta_1(t-k)^{1/12}$.
\end{proof}

\begin{lemma}[Weighted oscillation]\label{lem:nb-osc}
Let $T\geq2$, $\norm d\leq r_1$, $1\leq K\leq T^{1/2}$ and
$T^{-1}\leq\bar a\leq1$, and
let $g_K:\R^p\to[0,K]$ satisfy $|g_K(x)-g_K(x')|\leq\norm{x-x'}$ and
$g_K(x)\leq\norm x+C$. With
$\Delta_t(a)=g_K(x_t)\{\ind(e_t(d)\leq a)-\ind(e_t(d)\leq0)\}$,
\[
 \Bigl\|\sup_{|a|\leq\bar a}\Bigl|\frac1T\sum_{t=1}^T
   \{\Delta_t(a)-\E\Delta_t(a)\}\Bigr|\Bigr\|_2
 \leq C\Bigl\{\Bigl(\frac{\bar a}T\Bigr)^{1/2}\log T+\frac{K(\log T)^2}T\Bigr\}.
\]
\end{lemma}

\begin{proof}
The process $a\mapsto T^{-1}\sum_t\Delta_t(a)$ is non-decreasing, and, for
$a\leq a'$, $0\leq\E\{\Delta_t(a')-\Delta_t(a)\}\leq C(a'-a)\E g_K(x_0)\leq
C(a'-a)$, because the conditional density of $e_0(d)$ given $x_0$ is
bounded (Lemma~\ref{lem:nb-pop}(ii)). Let $\mathcal V$ be a grid of mesh
$T^{-3}$ in $[-\bar a,\bar a]$ that contains $\pm\bar a$, with at most $3T^3$ points. By
monotonicity, the supremum is at most the maximum over $\mathcal V$ plus
$CT^{-3}$; all suprema may be taken over rational $a$, by right
continuity. Fix $a\in\mathcal V$ and let
$\xi_t=\{\Delta_t(a)-\E\Delta_t(a)\}/(2K)$, a function of $W_t$ with
$|\xi_t|\leq1$ and $\E\xi_t=0$. Its variance satisfies
$\sigma^2\leq(2K)^{-2}\E[g_K(x_0)^2\Prb\{|e_0(d)|\leq\bar a\mid x_0\}]\leq
C\bar a/K^2$, and, by Lemma~\ref{lem:nb-coupling},
$\E|\xi_t-\xi_t^{*k}|\leq(2K)^{-1}\E\norm{x_t-x_t^{*k}}+2C\theta_1(t-k)\leq
C_\xi\theta_1(t-k)$ for a constant $C_\xi$. Since $s\mapsto s\lambda(s/C_\xi)$ is
non-decreasing and $\bar a/K^2\geq T^{-2}$, $\sigma^2\lambda(\sigma^2/C_\xi)\leq
C(\bar a/K^2)\log T$. Apply Lemma~\ref{lem:dep-moments}(ii) with
$x=C_2\{(T\bar a/K^2)^{1/2}\log T+m\log T\}$. The denominator of the
exponent is at most three times the largest of $C(T\bar a/K^2)\log T$, $C/T$ and
$Cmx$, and in each case the exponent is at least $cC_2\log T$, so the
exponential term is at most $T^{-10}$ for $C_2$ large; the last term is at
most $4C_\xi T\theta_1(m)\leq T^{-10}$. Outside an event of probability at
most $CT^{-7}$, therefore, the maximum over $\mathcal V$ is at most
$2Kx/T\leq C\{(\bar a/T)^{1/2}\log T+K(\log T)^2/T\}$, because $m\leq C\log T$.
On the exceptional event, the supremum is at most
$T^{-1}\sum_tg_K(x_t)+\E g_K(x_0)\leq2K$, which contributes at most
$2K(CT^{-7})^{1/2}\leq CT^{-3}$ to the $L^2$ norm.
\end{proof}

\begin{lemma}[Moments]\label{lem:nb-moments}
For $\norm d\leq r_1$, let $\widehat a(d)$ be the $(\ell+1)$th smallest of
$e_1(d),\ldots,e_T(d)$, let $\bar\psi_T=T^{-1}\sum_t\psi(u_t)$ and
$S_T^{\rm lin}(d)=T^{-1}\sum_tz_t(d)\psi\{e_t(d)\}$, and let
\[
 \epsilon_T^{\rm U}=T^{-11/16}\log T+T^{-1+2/\varkappa}(\log T)^2,
\]
which is $o(T^{-1/2})$ because $\varkappa>4$. Then, for $T\geq2$, with the convention
$0\cdot\lambda(0)=0$:
\begin{enumerate}[label=(\roman*),leftmargin=2.4em]
\item $\norm{\bar\psi_T}_2+\norm{\widehat a(d)}_2\leq CT^{-1/2}$,
$\E\min\{\widehat a(d)^4,1\}\leq CT^{-2}\log T$ and
$\Prb\{|\widehat a(d)|>T^{-3/8}\}\leq CT^{-10}$;
\item $\norm{S_T(d)-S_T^{\rm lin}(d)}_2\leq C\epsilon_T^{\rm U}$;
\item $\E S_T^{\rm lin}(d)=S(d)$ and
$\norm{S_T^{\rm lin}(d)-S_T^{\rm lin}(0)-S(d)}_2
\leq C\{\norm d\lambda(\norm d)/T\}^{1/2}$;
\item $S_T^{\rm lin}(0)=T^{-1}\sum_tz_t\psi(u_t)$, which has mean zero,
$\E\norm{T^{-1/2}\sum_tz_t\psi(u_t)}^3\leq C$, and
$V_T=\Var\{T^{-1/2}\sum_tz_t\psi(u_t)\}=\Lambda+O(T^{-1})$, where the series
$\Lambda=\sum_{h\in\mathbb Z}\E\{z_0z_h\trans\psi(u_0)\psi(u_h)\}$
converges absolutely and $\norm\Lambda\leq C$.
\end{enumerate}
\end{lemma}

\begin{proof}
(i) The variables $\psi(u_t)$ are bounded by one, centred by
\eqref{eq:model}, and satisfy the coupling condition of
Lemma~\ref{lem:dep-moments}(ii) by Lemma~\ref{lem:nb-coupling}, so
$\E\bar\psi_T^2\leq C/T$. For the other bounds, let $T$ be so large that
$T^{-3/8}\leq r_1$ and $c_1T^{5/8}\geq1$; for smaller $T$ they hold after
enlarging $C$, because $\norm{\widehat a(d)}_\varkappa\leq C$, as shown at
the end of this part. Write $\widehat a=\widehat a(d)$,
$N(v)=\sum_t\ind\{e_t(d)\leq v\}$ and $F_e(v)=\Prb\{e_0(d)\leq v\}$. For
$0<v\leq r_1$, Lemma~\ref{lem:nb-pop}(ii) gives
$TF_e(v)\geq\ell+c_1Tv$ and $TF_e(-v)\leq\ell+1-c_1Tv$, so that
\begin{gather*}
 \Prb(\widehat a>v)=\Prb\{N(v)\leq\ell\}
 \leq\Prb\{N(v)-TF_e(v)\leq-c_1Tv\},\\
 \Prb(\widehat a<-v)\leq\Prb\{N(-v)-TF_e(-v)\geq c_1Tv\}.
\end{gather*}
The variables $\ind\{e_t(d)\leq\pm v\}-F_e(\pm v)$ satisfy the conditions
of Lemma~\ref{lem:dep-moments}(ii) with a bounded $C_\xi$, by
Lemma~\ref{lem:nb-coupling}. Its fourth-moment bound gives
$\Prb(|\widehat a|>v)\leq CT^{-2}v^{-4}$, and its exponential bound with
$v=T^{-3/8}$ and $x=c_1T^{5/8}$, for which the exponent is at least
$cT^{1/4}$, gives
$\Prb(|\widehat a|>T^{-3/8})\leq8\exp(-cT^{1/4})+CT\theta_1(m)\leq CT^{-10}$.
Hence
$\E\min(\widehat a^4,1)\leq\int_0^14v^3\Prb(|\widehat a|>v)\,dv\leq
T^{-2}+\int_{T^{-1/2}}^{r_1}4Cv^{-1}T^{-2}\,dv+\Prb(|\widehat a|>r_1)\leq
CT^{-2}\log T$, and in the same way
$\E\{\widehat a^2\ind(|\widehat a|\leq r_1)\}\leq CT^{-1}$. Finally,
$|\widehat a|\leq|e_t(d)|$ for at least $T\min(\tau,1-\tau)$ indices $t$,
namely the $T-\ell$ largest if $\widehat a\geq0$ and the $\ell+1$ smallest
otherwise, so that
$|\widehat a|^\varkappa\leq\{T\min(\tau,1-\tau)\}^{-1}\sum_t|e_t(d)|^\varkappa$
and $\E|\widehat a|^\varkappa\leq C$. By Hölder's inequality,
$\E\{\widehat a^2\ind(|\widehat a|>r_1)\}\leq\norm{\widehat a}_\varkappa^2
\Prb(|\widehat a|>r_1)^{1-2/\varkappa}\leq CT^{-2(1-2/\varkappa)}\leq
CT^{-1}$, because $\varkappa>4$.

(ii) By Assumption~\ref{ass:affine}(c), the residuals are almost surely
distinct, so, by \eqref{eq:profile} and the rank weights of
Remark~\ref{rem:outline},
$\tau-w_t(d)=\psi\{e_t(d)-\widehat a\}+(1-r)\ind\{e_t(d)=\widehat a\}$,
with $r=T\tau-\ell$.
Since the weights sum to $T\tau$, $T^{-1}\sum_t\psi\{e_t(d)-\widehat a\}=-(1-r)/T$,
and therefore, with $t^*$ the index for which $e_{t^*}(d)=\widehat a$,
\[
 S_T(d)-S_T^{\rm lin}(d)=G_T(\widehat a)+J(\widehat a)+\frac{1-r}Tz_{t^*}(d),
\]
where, for $a\in\R$,
\begin{gather*}
 J(a)=\E[z_0(d)\{\psi(e_0(d)-a)-\psi(e_0(d))\}],\\
 G_T(a)=\frac1T\sum_{t=1}^T[z_t(d)\{\psi(e_t(d)-a)-\psi(e_t(d))\}]-J(a).
\end{gather*}
First,
$\norm{z_{t^*}(d)}\leq\max_{t\leq T}\norm{z_t(d)}$, whose $L^2$ norm is at
most $(T\E\norm{z_0(d)}^\varkappa)^{1/\varkappa}\leq CT^{1/\varkappa}$.
Second, given $x_0$, the density of $e_0(d)$ is
$f_e(s\mid x_0)=f\{s+q(d)+x_0\trans d\mid x_0\}$, so that
\[
 J(a)=-a\E\{z_0(d)f_d(x_0)\}
 -\E\Bigl[z_0(d)\int_0^a\{f_e(s\mid x_0)-f_e(0\mid x_0)\}\,ds\Bigr].
\]
The first term vanishes by the definition of $\gamma(d)$, and
$|f_e(s\mid x)-f_e(0\mid x)|\leq C|s|$, so $\norm{J(a)}\leq C\min(a^2,1)$
and, by (i), $\norm{J(\widehat a)}_2\leq CT^{-1}(\log T)^{1/2}$. Third, let
$\bar a_T=T^{-3/8}$ and $K=T^{2/\varkappa}\leq T^{1/2}$. For each coordinate $j$,
write $z_{tj}(d)=\zeta^+_{tj}-\zeta^-_{tj}+\varepsilon_{tj}$, where
$\zeta^\pm_{tj}$ are the positive and negative parts of
$\max[-K,\min\{K,z_{tj}(d)\}]$ and
$|\varepsilon_{tj}|\leq|z_{tj}(d)|\ind\{|z_{tj}(d)|>K\}$. The
$\zeta^\pm_{tj}=g_K^\pm(x_t)$ satisfy the conditions of
Lemma~\ref{lem:nb-osc}, and $\psi(e-a)-\psi(e)=-\{\ind(e\leq a)-\ind(e\leq0)\}$,
so their parts of $\sup_{|a|\leq\bar a_T}\norm{G_T(a)}$ have $L^2$ norm at most
$C\{(\bar a_T/T)^{1/2}\log T+K(\log T)^2/T\}$. The part of
$\varepsilon_{tj}$ is at most
$T^{-1}\sum_t|\varepsilon_{tj}|+\E|\varepsilon_{0j}|$, whose $L^2$ norm is at
most $2(\E|z_{0j}(d)|^\varkappa)^{1/2}K^{1-\varkappa/2}\leq CT^{-1+2/\varkappa}$.
Since $\norm{G_T(a)}\leq Y=2T^{-1}\sum_t\norm{z_t(d)}+2\E\norm{z_0(d)}$ for
every $a$, with $\norm Y_\varkappa\leq C$, Hölder's inequality and (i) give
$\E\{\norm{G_T(\widehat a)}^2\ind(|\widehat a|>\bar a_T)\}\leq
C(T^{-10})^{1-2/\varkappa}\leq CT^{-5}$. Collecting the terms,
$\norm{S_T(d)-S_T^{\rm lin}(d)}_2\leq C\{T^{-1+1/\varkappa}+T^{-1}(\log T)^{1/2}
+T^{-11/16}\log T+T^{-1+2/\varkappa}(\log T)^2\}\leq C\epsilon_T^{\rm U}$.

(iii) The mean is
$\E[x_0\psi\{e_0(d)\}]-\gamma(d)\E\psi\{e_0(d)\}=S(d)$. Let
$\xi_t(d)=z_t(d)\psi\{e_t(d)\}-S(d)$ and $\Delta\xi_t=\xi_t(d)-\xi_t(0)$, a
stationary, centred function of $W_t$. Since
$\Delta\xi_0=\{\gamma(0)-\gamma(d)\}\psi\{e_0(d)\}+z_0[\psi\{e_0(d)\}-\psi(u_0)]-S(d)$
and $|\psi\{e_0(d)\}-\psi(u_0)|$ has conditional mean
$|F\{q(d)+x_0\trans d\mid x_0\}-F(0\mid x_0)|\leq C(1+\norm{x_0})\norm d$
given $x_0$, Lemma~\ref{lem:nb-pop} and $\E\norm{z_0}^2(1+\norm{x_0})\leq C$
give $\norm{\Delta\xi_0}_2\leq C\norm d^{1/2}$. For $h\geq1$, $\Delta\xi_h^{*0}$ is
independent of $\Delta\xi_0$ and centred, so, by Lemma~\ref{lem:nb-coupling},
$|\E\Delta\xi_0\trans\Delta\xi_h|\leq\norm{\Delta\xi_0}_2\min\{\norm{\Delta\xi_h}_2,
\norm{\Delta\xi_h-\Delta\xi_h^{*0}}_2\}\leq C\min\{\norm d,\norm d^{1/2}\theta_1(h)^{1/12}\}$,
and the same holds for $-h$ by stationarity. Let $h_d$ be the least
$h\geq0$ with $\theta_1(h)^{1/12}\leq\norm d^{1/2}$, so that
$h_d\leq C\lambda(\norm d)$. Summing the first bound over $|h|<h_d$ and
the second over $|h|\geq h_d$ gives
$\sum_h|\E\Delta\xi_0\trans\Delta\xi_h|\leq C\norm d\lambda(\norm d)$, and hence
$\E\norm{\sum_t\Delta\xi_t}^2\leq T\sum_h|\E\Delta\xi_0\trans\Delta\xi_h|\leq
CT\norm d\lambda(\norm d)$.

(iv) The first claim holds because $e_t(0)=u_t$ and $\gamma(0)=\gamma_i$,
and the mean is zero because $\E\{\psi(u_0)\mid x_0\}=0$. Let
$\xi_t=z_t\psi(u_t)$ and let $\mathcal P_s$ be the projections in the proof
of Lemma~\ref{lem:dep-moments}(ii). Since $\xi_k^{*0}$ is independent of
$\mathcal F_0$ and centred, Jensen's inequality, stationarity and
Lemma~\ref{lem:nb-coupling} give
$\norm{\E(\xi_k\mid\mathcal F_0)}_3\leq\norm{\xi_k-\xi_k^{*0}}_3\leq
C\theta_1(k)^{1/12}$ for $k\geq1$, and likewise with $\mathcal F_{-1}$, so
that $\norm{\mathcal P_0\xi_k}_3\leq C\theta_1(k)^{1/12}$ for $k\geq0$ and
$\xi_t=\sum_{k\geq0}\mathcal P_{t-k}\xi_t$ in $L^3$. Burkholder's inequality
\citep{Burkholder73}, applied to each coordinate, and Minkowski's inequality
in $L^{3/2}$ give
$\norm{\sum_t\mathcal P_{t-k}\xi_t}_3\leq CT^{1/2}\norm{\mathcal P_0\xi_k}_3$,
and summation over $k$ gives $\norm{\sum_t\xi_t}_3\leq CT^{1/2}$. Next,
$\norm{\E\xi_0\xi_h\trans}\leq\norm{\xi_0}_2\norm{\xi_h-\xi_h^{*0}}_2\leq
C\theta_1(h)^{1/12}$ for $h\geq1$, so that
$\sum_h(1+|h|)\norm{\E\xi_0\xi_h\trans}<\infty$, uniformly in $i$, and
$V_T=\sum_{|h|<T}(1-|h|/T)\E\xi_0\xi_h\trans=\Lambda+O(T^{-1})$.
\end{proof}

\subsection{Proof of Proposition~\ref{prop:nobias}}\label{app:nobias-proof}

\emph{Part (i).} Fix $T$ and $d\in\mathcal D$, and let $\mathcal G_d$ be the
$\sigma$-field generated by $(R_s(d))_{s\in\mathbb Z}$. The rank weights
$w_t(d)$ are functions of $R_1(d),\ldots,R_T(d)$, and hence
$\mathcal G_d$-measurable. By \eqref{eq:profile},
$\mathbb M_T(d)=T^{-1}\sum_tR_t(d)\{\tau-w_t(d)\}$, and
$\sum_t\{\tau-w_t(d)\}=0$. Conditioning on $\mathcal G_d$ and using
\eqref{eq:affine} gives
\[
 \E\{S_T(d)\mid\mathcal G_d\}
 =\frac1T\sum_{t=1}^T\{\mu(d)+\varphi(d)R_t(d)\}\{\tau-w_t(d)\}
 =\varphi(d)\mathbb M_T(d),
\]
so that $\E S_T(d)=\varphi(d)\E\mathbb M_T(d)$. By \eqref{eq:affine} and the
tower property, $\E\{x_0\mid R_0(d)\}=\mu(d)+\varphi(d)R_0(d)$, and, since
$\E\psi\{e_0(d)\}=0$ and $v\psi(v)=\rho_\tau(v)$,
\[
 S(d)=\E[\{\mu(d)+\varphi(d)R_0(d)\}\psi\{e_0(d)\}]
 =\varphi(d)\E[e_0(d)\psi\{e_0(d)\}]=\varphi(d)\mathbb M(d).
\]
At $d=0$, $S(0)=0$ and $\mathbb M(0)>0$ give $\varphi(0)=0$, as claimed in
Appendix~\ref{sec:dep-nobias}.

Fix $d_1\in\mathcal D$, put $d_s=sd_1$ for $s\in[0,1]$, and let
$g_1(s)=\E\mathbb M_T(d_s)$ and $g_2(s)=\mathbb M(d_s)$. For each $s$, the
residuals at $d_s$ are almost surely distinct, so, almost surely, their
order, and hence the rank weights, do not change in a neighbourhood of
$d_s$, on which $\mathbb M_T$ is therefore affine with gradient
$-S_T(d_s)$. Since $|\mathbb M_T(d')-\mathbb M_T(d)|\leq\norm{d'-d}
T^{-1}\sum_t\norm{x_t}$, dominated convergence shows that $g_1$ is
differentiable, with
$g_1'(s)=-d_1\trans\E S_T(d_s)=-d_1\trans\varphi(d_s)g_1(s)$. By
Lemma~\ref{lem:nb-pop}(i),
$g_2'(s)=-d_1\trans S(d_s)=-d_1\trans\varphi(d_s)g_2(s)$, and $g_2>0$.
Hence $(g_1/g_2)'=0$ on $[0,1]$, and $g_1(1)/g_2(1)=g_1(0)/g_2(0)$,
that is, $\E\mathbb M_T(d_1)=k_T\mathbb M(d_1)$. Then
$\E S_T(d_1)=\varphi(d_1)k_T\mathbb M(d_1)=k_TS(d_1)$.

Since $q(0)=0$, $\mathbb M_T(0)\leq T^{-1}\sum_t\rho_\tau(u_t)$, whose mean
is $\mathbb M(0)$, so $k_T\leq1$. With $\widehat a=\widehat a(0)$, the minimising
intercept, the first inequality in \eqref{eq:nb-sub} gives
$\mathbb M_T(0)\geq T^{-1}\sum_t\rho_\tau(u_t)-\widehat a\bar\psi_T$, so that
$0\leq \mathbb M(0)-\E\mathbb M_T(0)\leq\norm{\widehat a}_2\norm{\bar\psi_T}_2\leq
C/T$ by Lemma~\ref{lem:nb-moments}(i), and $\mathbb M(0)\geq c$ gives
$k_T\geq1-C/T$. Since $\mathbb M_T(0)\geq0$, also $k_T\geq0$, so the
bound holds for every $T\geq1$ after enlarging $C$.

\emph{Part (ii).} Let $\theta\in\R^p$ be fixed, $d_\theta=\theta/\sqrt{nT}$,
and $n$ and $T$ so large that $\norm{d_\theta}\leq r_1$. The estimator
$\widehat\theta=\sqrt{nT}(\widehat\beta-\beta_0)$ minimises the convex
function
$H_{nT}(\theta)=T\sum_i\{\mathbb M_{iT}(d_\theta)-\mathbb M_{iT}(0)\}$. Put
\[
 Z_{nT}=\frac1{\sqrt{nT}}\sum_{i=1}^n\sum_{t=1}^Tz_{it}\psi(u_{it}),\qquad
 \Delta_{nT}=\sqrt{\frac Tn}\sum_{i=1}^nS_{iT}(0).
\]

\emph{Step 1: mean.} By part (i), Lemma~\ref{lem:nb-pop}(iii) and
$|k_{iT}-1|\leq C/T$,
\[
 \E H_{nT}(\theta)=T\sum_ik_{iT}\{\mathbb M_i(d_\theta)-\mathbb M_i(0)\}
 =\tfrac12\theta\trans\Gamma_n\theta
 +O\{\norm\theta^2/T+\norm\theta^3(nT)^{-1/2}\}.
\]

\emph{Step 2: the score at the true slope.} By part (i), $\E S_{iT}(0)=0$,
and $\E S^{\rm lin}_{iT}(0)=0$, so $\Delta_{nT}-Z_{nT}
=(T/n)^{1/2}\sum_i\{S_{iT}(0)-S^{\rm lin}_{iT}(0)\}$ is a sum of independent
centred vectors, and, by Lemma~\ref{lem:nb-moments}(ii),
$\E\norm{\Delta_{nT}-Z_{nT}}^2\leq CT(\epsilon_T^{\rm U})^2\to0$. Also
$\E\norm{Z_{nT}}^2=n^{-1}\sum_i\operatorname{tr}V_{iT}\leq C$.

\emph{Step 3: the local quadratic approximation.} Let
$D^c_i(d)=\mathbb M_{iT}(d)-\mathbb M_{iT}(0)+d\trans S_{iT}(0)$. The
integration argument in Step 1 of the proof of Lemma~\ref{lem:local}, which
uses only convexity and the subgradient inequality, gives
$D^c_i(d)=-\int_0^1d\trans\{S_{iT}(sd)-S_{iT}(0)\}\,ds$, and
$H_{nT}(\theta)=-\theta\trans\Delta_{nT}+T\sum_iD^c_i(d_\theta)$, so that
$\E H_{nT}(\theta)=T\sum_i\E D^c_i(d_\theta)$. Since
$\norm{S_{iT}(d)}\leq T^{-1}\sum_t\norm{x_{it}}$ for every $d$, which is
square integrable, Fubini's theorem allows expectations to be taken inside
the integral. For $s\in[0,1]$, write
$S_{iT}(sd)-S_{iT}(0)$ as
$[S_{iT}(sd)-S^{\rm lin}_{iT}(sd)]-[S_{iT}(0)-S^{\rm lin}_{iT}(0)]
+[S^{\rm lin}_{iT}(sd)-S^{\rm lin}_{iT}(0)]$. By
Lemma~\ref{lem:nb-moments}(ii) and (iii), and since centring does not
increase an $L^2$ norm, its centred version has $L^2$ norm at most
$C\{\epsilon_T^{\rm U}+(\norm d\lambda(\norm d)/T)^{1/2}\}$. By Minkowski's
inequality for integrals and independence across individuals,
\begin{align*}
 \Var\Bigl\{T\sum_iD^c_i(d_\theta)\Bigr\}
 &\leq CT^2n\norm{d_\theta}^2\{(\epsilon_T^{\rm U})^2
   +\norm{d_\theta}\lambda(\norm{d_\theta})/T\}\\
 &=C\norm\theta^2\{T(\epsilon_T^{\rm U})^2
   +\norm{d_\theta}\lambda(\norm{d_\theta})\}\longrightarrow0.
\end{align*}
With Steps 1 and 2,
$H_{nT}(\theta)=-\theta\trans Z_{nT}+\tfrac12\theta\trans\Gamma_n\theta+\op(1)$
for each $\theta$. Since $\Gamma_n$ is bounded by
Lemma~\ref{lem:nb-pop}(ii) and $\lambda_{\min}(\Gamma_n)\geq c$ by
Assumption~\ref{ass:affine}(d), Step 4 of the proof of
Lemma~\ref{lem:local}, applied to the convex function
$H_{nT}(\theta)+\theta\trans Z_{nT}$, makes the approximation uniform on
compact sets.

\emph{Step 4: expansion and Gaussian approximation.} Step 1 of the proof of
Theorem~\ref{thm:main}, with $Z_{nT}$ in place of $\Delta^c_{nT}$, gives
$\widehat\theta=\Gamma_n^{-1}Z_{nT}+\op(1)$, which is the expansion. In
$Z_{nT}=(nT)^{-1/2}\sum_is_i$, the sums $s_i=\sum_tz_{it}\psi(u_{it})$ are
independent and centred, their normalised covariances add up to
$n^{-1}\sum_iV_{iT}=\Lambda_n+O(T^{-1})$, whose eigenvalues lie in $[c,C]$
for large $n$ and $T$ by Lemma~\ref{lem:nb-moments}(iv) and
Assumption~\ref{ass:affine}(d), and
$\sum_i\E\norm{(nT)^{-1/2}s_i}^3\leq Cn^{-1/2}$ by
Lemma~\ref{lem:nb-moments}(iv). Step 3 of the proof of Theorem~\ref{thm:dep}
and Step 3 of the proof of Theorem~\ref{thm:main} complete the proof. No
step restricts the growth of $n$ relative to $T$. \qed

\subsection{Proof of Corollary~\ref{cor:gauss}}\label{app:gauss-proof}

Drop the index $i$, let $\eta_t=(G_t^{(0)},\ldots,G_t^{(p)})\trans$, and
note that the series defining $Z_t^{(k)}$ converge almost surely and in
$L^2$, with $\Var(Z_t^{(k)})=1$ and $\Cov(Z_t^{(k)},Z_{t+h}^{(k)})=r(h)$.

\emph{Assumption~\ref{ass:causal} and (a)(G).} The vector $W_t$ is a
measurable function of $\mathcal F_t=(\ldots,\eta_{t-1},\eta_t)$. Since
$Z_t^{(0)}$ is independent of $x_t$ and standard normal,
$\Prb(u_t\leq0\mid x_t)=\Prb(Z_t^{(0)}\leq q_\tau)=\tau$, so
\eqref{eq:model} holds. Replacing the innovations up to time zero changes
each $Z_t^{(k)}$, $t\geq1$, by $\sum_{j\geq t}c_j(G_{t-j}^{(k)}-G_{t-j}^{(k)\prime})$,
whose $L^2$ norm is $\{2\sum_{j\geq t}c_j^2\}^{1/2}\leq C\varrho^t$, so (a)(G)
holds with $\varsigma=1$.

\emph{Conditions (a)--(c).} The vector $W_0$ is Gaussian with bounded mean
and covariance, so (a) holds for every $\varkappa$. Given $x_0=x$, $u_0$
has the density $f(v\mid x)=\sigma_u^{-1}\phi(v/\sigma_u+q_\tau)$, which with its
derivative is bounded by $C$, and $a_i=\phi(q_\tau)/\sigma_u\geq c$, so (b) holds.
For fixed $d$, let $R^\circ_t=R_t(d)+\sigma_uq_\tau=\sigma_uZ_t^{(0)}-d\trans Q^{1/2}(Z_t^{(1)},\ldots,Z_t^{(p)})\trans$.
Since the coordinate processes are independent with the common
autocovariance $r$, $\Cov(R^\circ_t,R^\circ_{t+h})=\sigma(d)^2r(h)$, so that
$(R_t(d))_t$ has the law of $(\sigma(d)Z_t^{(0)}-\sigma_uq_\tau)_t$. For
$h\neq0$, the Cauchy--Schwarz inequality gives
$|r(h)|\leq(\sum_{j\geq|h|}c_j^2)^{1/2}\leq(1-c_0^2)^{1/2}<1$, so
$R_t(d)-R_s(d)$ is normal with positive variance for $t\neq s$, and the
residuals are almost surely distinct, which is (c).

\emph{Condition \eqref{eq:affine}.} Since
$\Cov(x_t,R^\circ_s)=-Qd\,r(t-s)$, the Gaussian vector
$x_t-\varphi(d)R^\circ_t$ has zero covariance with every $R^\circ_s$, so it is
independent of $\mathcal G_d=\sigma(R^\circ_s:s\in\mathbb Z)$, by the
independence of jointly Gaussian uncorrelated vectors and a monotone class
argument, and it has mean zero. Hence
$\E(x_t\mid\mathcal G_d)=\varphi(d)R^\circ_t=\mu(d)+\varphi(d)R_t(d)$ for every
$d\in\R^p$.

\emph{The formulas and condition (d).} By the equality in law of the
residual paths, and since $\rho_\tau$ is positively homogeneous,
\[
 \E\mathbb M_T(d)
 =\E\min_\alpha\frac1T\sum_{t=1}^T\rho_\tau\{\sigma(d)Z_t^{(0)}-\alpha\}
 =\sigma(d)\E\min_\alpha\frac1T\sum_{t=1}^T\rho_\tau(Z_t^{(0)}-\alpha),
\]
and
$\mathbb M(d)=\sigma(d)\E\rho_\tau(Z-q_\tau)=\sigma(d)\phi(q_\tau)$ for standard
normal $Z$. Since $f(0\mid x)=\phi(q_\tau)/\sigma_u$ does not depend on $x$,
$\gamma_i=\E x_0=0$ and $\Gamma_i=\phi(q_\tau)Q/\sigma_u$. By the independence of
the regressor and error processes,
$\E\{x_0x_h\trans\psi(u_0)\psi(u_h)\}=Qr(h)c_\tau\{r(h)\}$. It remains to
show that $rc_\tau(r)\geq0$ for $|r|\leq1$; the term with $h=0$ is
$Q\tau(1-\tau)=v_\tau Q$, and $\sum_h|r(h)|\leq(\sum_j|c_j|)^2\leq C$, so
that the eigenvalues of $\Lambda_n$ and $\Gamma_n$ lie in $[c,C]$. Let
$U,U_1,U_2$ be independent standard normal variables. For $r\in[0,1]$, let
$Z=\sqrt rU+\sqrt{1-r}U_1$ and $Z'=\sqrt rU+\sqrt{1-r}U_2$; with
$\pi(g)=\Prb(\sqrt rg+\sqrt{1-r}U_1\leq q_\tau)$, Jensen's inequality gives
$\Prb(Z\leq q_\tau,Z'\leq q_\tau)=\E\pi(U)^2\geq\{\E\pi(U)\}^2=\tau^2$. For
$r\in[-1,0)$, let $Z=\sqrt{-r}U+\sqrt{1+r}U_1$ and
$Z'=-\sqrt{-r}U+\sqrt{1+r}U_2$. Then
$\Prb(Z\leq q_\tau,Z'>q_\tau)=\E\{\pi_1(U)\pi_2(U)\}$, where
$\pi_1(g)=\Prb(\sqrt{-r}g+\sqrt{1+r}U_1\leq q_\tau)$ and
$\pi_2(g)=\Prb(-\sqrt{-r}g+\sqrt{1+r}U_2>q_\tau)$ are both non-increasing.
With $U'$ an independent copy of $U$,
$2\Cov\{\pi_1(U),\pi_2(U)\}=\E[\{\pi_1(U)-\pi_1(U')\}\{\pi_2(U)-\pi_2(U')\}]\geq0$,
so $\Prb(Z\leq q_\tau,Z'>q_\tau)\geq\tau(1-\tau)$ and
$\Prb(Z\leq q_\tau,Z'\leq q_\tau)\leq\tau-\tau(1-\tau)=\tau^2$. For
$0<|r|<1$, both inequalities are strict, because $\pi$, $\pi_1$ and $\pi_2$
are then strictly decreasing, so that $\pi(U)$ is non-degenerate and
$\{\pi_1(U)-\pi_1(U')\}\{\pi_2(U)-\pi_2(U')\}>0$ whenever $U\neq U'$; by
continuity, $rc_\tau(r)$ is bounded away from zero for $|r|$ in compact
subsets of $(0,1)$. \qed

\subsection{Proof of Proposition~\ref{prop:cusp}}\label{app:cusp}

Fix $i$ and drop it where no confusion arises, and let $\norm x_\infty$ be
the essential supremum of $|x_{it}|$. Choose $d_0>0$ and
$b_0>0$ so that $|s|+|d|\,\norm{x}_\infty\leq a_0/2$ whenever $|s|\leq b_0$
and $|d|\leq d_0$. At $d=0$, the errors coincide at adjacent times
precisely when $(B_t,B_{t+1})=(1,0)$, in which case
$u_t=u_{t+1}=\varepsilon_t$. Given $B_1,\ldots,B_T$, the residuals
$u_t-dx_t$ therefore form
$K_T=\sum_{t=1}^{T-1}B_t(1-B_{t+1})$ doubleton clusters
$(\varepsilon-dX_1,\varepsilon-dX_2)$, with a common centre and two
independent marks, and $T-2K_T$ singleton clusters $\varepsilon-dX$;
different clusters are independent, the centres have the law of
$\varepsilon_{it}$ and the marks that of $x_{it}$. Moreover,
$\E K_T=(T-1)/4$ and, by Hoeffding's inequality applied to the two
residue classes of adjacent pairs, $\Prb(|K_T-T/4|>cT)\leq Ce^{-c'T}$ for
small $c>0$. Write $p_s=1/2+f_0s$ and $\delta_d=f_0D_x|d|$.

\begin{lemma}[Central spacing for independent clusters]\label{lem:cusp-spacing}
Let ${T'}=K_1+2K_2$ be even, and form a sample of ${T'}$ residuals from
$K_1$ singleton and $K_2$ doubleton clusters as above, with
$K_1\geq c_*{T'}$ for a fixed $c_*>0$. If $Y_{(1)}(d)\leq\cdots\leq Y_{({T'})}(d)$ are the ordered
residuals, then, uniformly over such $K_1$, $K_2$ and $|d|\leq d_0$,
\[
 \E\{Y_{({T'}/2+1)}(d)-Y_{({T'}/2)}(d)\}=\frac1{{T'}f_0}
 +O\{{T'}^{-3/2}(\log {T'})^{1/2}\}.
\]
\end{lemma}

\begin{proof}
For $|s|\leq b_0$, the number of observations at most $s$ in a singleton
cluster is Bernoulli$(p_s)$, because $\E x_{it}=0$. In a doubleton
cluster, the two thresholds are $s+d\min(X_1,X_2)$ and
$s+d\max(X_1,X_2)$, and $\E\max(X_1,X_2)=D_x/2=-\E\min(X_1,X_2)$, so its
count $N^\circ$ has $\Prb(N^\circ=2)=p_s-\delta_d/2$, $\Prb(N^\circ=1)=\delta_d$ and
$\Prb(N^\circ=0)=1-p_s-\delta_d/2$. The total count $N_s$ therefore has
$\E N_s={T'}p_s$ and
$v_{T'}(s,d)=\Var(N_s)=(K_1+4K_2)(1/4-f_0^2s^2)-K_2\delta_d$, and $c{T'}\leq v_{T'}(s,d)\leq C{T'}$
after reducing $b_0$ and $d_0$ if necessary. The characteristic function
of the centred $N_s$ is
$\exp\{-v_{T'}(s,d)t^2/2+O({T'}|t|^3)\}$ on a fixed interval around zero,
uniformly in the parameters, because every count is bounded by two, and
the $K_1\geq c_*{T'}$ singleton factors give
$|\E e^{\mathrm itN_s}|\leq\exp\{-c{T'}\sin^2(t/2)\}$ for $|t|\leq\pi$. Fourier
inversion on the lattice therefore gives
\begin{equation}\label{eq:cusp-lclt}
 \sup_{k\in\mathbb Z}\left|\Prb(N_s=k)
 -\frac{\exp\{-(k-{T'}p_s)^2/(2v_{T'}(s,d))\}}{\{2\pi v_{T'}(s,d)\}^{1/2}}\right|
 \leq C{T'}\int_{\R}|t|^3e^{-c{T'}t^2}\,dt+Ce^{-c{T'}}\leq\frac C{T'}
\end{equation}
for $|s|\leq b_0$. Since $N_s={T'}/2$ if and only if
$Y_{({T'}/2)}(d)\leq s<Y_{({T'}/2+1)}(d)$,
$\E\{Y_{({T'}/2+1)}(d)-Y_{({T'}/2)}(d)\}=\int_{\R}\Prb(N_s={T'}/2)\,ds$, where the
residuals have bounded support. Hoeffding's inequality for the
independent cluster counts bounds the integrand by $2e^{-c{T'}s^2}$ for
$|s|\leq b_0$ and by $2e^{-c{T'}}$ otherwise. On
$|s|\leq w_{T'}=C_w\{\log({T'}+1)/{T'}\}^{1/2}$ with $C_w$ large, the error in
\eqref{eq:cusp-lclt} integrates to $O\{{T'}^{-3/2}(\log {T'})^{1/2}\}$, and the
rest of the line contributes less. Finally, $v_{T'}(s,d)=v_{T'}(0,d)-(K_1+4K_2)f_0^2s^2$,
and the substitution $z={T'}f_0s/v_{T'}(0,d)^{1/2}$, under which
$v_{T'}(s,d)/v_{T'}(0,d)=1-(K_1+4K_2)z^2/{T'}^2$ with $(K_1+4K_2)/{T'}^2=O({T'}^{-1})$, gives
\[
 \int_{-w_{T'}}^{w_{T'}}
 \frac{\exp\{-{T'}^2f_0^2s^2/(2v_{T'}(s,d))\}}{\{2\pi v_{T'}(s,d)\}^{1/2}}\,ds
 =\frac1{{T'}f_0}+O({T'}^{-2}).
\]
\end{proof}

\emph{Step 1: the mean score and $\kappa$.} Let $\ell=T/2$; the weights
of the residuals are one for the $\ell$ smallest and zero otherwise, so
that $S_T(d)=T^{-1}\sum_tx_t/2-T^{-1}\sum_tx_tw_t(d)$, where $w_t(d)$ is
the weight of $u_t-dx_t$ and $\E x_t=0$. For a singleton with mark $X$
and centre $\varepsilon$, let $q$ be the $\ell$th order statistic of the
other $T-1$ residuals; its weight is $\ind(\varepsilon<q+dX)$, and, since
$(\varepsilon,X)$ is independent of $q$ and $q+dX$ lies in the flat part
of the density outside an event of probability $Ce^{-cT}$,
$\E\{X\ind(\varepsilon<q+dX)\}=f_0\sigma_x^2d+O(e^{-cT})$. For a
doubleton, let $q_-\leq q_+$ be the $(\ell-1)$th and $\ell$th order
statistics of the other $T-2$ residuals. If $d>0$, the residual with the
larger mark is among the $\ell$ smallest when $\varepsilon<q_++d\max(X_1,X_2)$, and
the other when $\varepsilon<q_-+d\min(X_1,X_2)$; hence the expected
selected regressor sum of the pair is
$2f_0\sigma_x^2d+(f_0D_x/2)\E(q_+-q_-)+O(e^{-cT})$, and for $d<0$ the second
term changes sign. On the event $|K_T-T/4|\leq cT$, the other residuals
contain at least $c_*T$ singletons, and Lemma~\ref{lem:cusp-spacing} with
$T'=T-2$ gives $\E(q_+-q_-)=1/\{(T-2)f_0\}+O\{T^{-3/2}(\log T)^{1/2}\}$.
Adding the contributions and dividing by $T$,
\begin{align*}
 \E S_T(d)&=-\Gamma d-\operatorname{sgn}(d)\frac{D_x\E K_T}{2T(T-2)}
 +O\{T^{-3/2}(\log T)^{1/2}\}\\
 &=-\Gamma d-\operatorname{sgn}(d)\frac{D_x}{8T}+O\{T^{-3/2}(\log T)^{1/2}\}
\end{align*}
for $0<|d|\leq d_0$, and integration along the segment from zero gives
$\E\{\mathbb M_T(d)-\mathbb M_T(0)\}=\Gamma d^2/2+D_x|d|/(8T)+O\{|d|T^{-3/2}(\log T)^{1/2}\}$,
where $\mathbb M_T(d)$ is the profiled criterion of the individual. For
$\kappa$, the population residual median is zero and the density of
$u_0-dx_0$ at zero is $f_0$ for small $d$; residuals two or more periods
apart are independent, and the lag-one covariance is
$\Cov\{\psi(u_0-dx_0),\psi(u_1-dx_1)\}=1/16-f_0D_x|d|/8$, because with
probability $1/4$ the two residuals share their centre, in which case
both are at most zero with probability $1/2-f_0D_x|d|/2$, and otherwise
they are independent. Hence
$\Omega_\psi(d)=3/8-f_0D_x|d|/4$ and $\kappa(d)=3/(8f_0)-D_x|d|/4$.

\emph{Step 2: the local criterion and the limit.} At a deterministic
$|d|\leq d_0$, let $\xi_t(d)=x_t\psi(u_t-dx_t)$ and
$\mathcal R_T(d)=S_T(d)-T^{-1}\sum_t\xi_t(d)$. Within each parity class
of the periods the residuals are independent, so Hoeffding's inequality puts the
empirical residual median within $T^{-3/8}$ of zero except with
probability $Ce^{-cT^{1/4}}$. For $|v|\leq T^{-3/8}$, the population weighted
increment $\E[x_t\{\ind(u_t-dx_t\leq v)-\ind(u_t-dx_t\leq0)\}]=f_0v\E x_t$
vanishes; Bernstein's inequality on the two residue classes, on a grid of
mesh $T^{-3}$ in $[-T^{-3/8},T^{-3/8}]$, with monotonicity of the parts of the
weighted counts with positive and negative regressors between grid
points, bounds the $L^2$ norm of the supremum over $|v|\leq T^{-3/8}$ of the
centred weighted increment by
$C\{(T^{-11/8}\log T)^{1/2}+\log T/T\}$; and, at a deterministic slope, ties
involve at most two residuals, so the allocation of the weights adds at
most $2\norm x_\infty/T$. Hence
$\sup_{|d|\leq d_0}\norm{\mathcal R_T(d)}_2\leq CT^{-11/16}(\log T)^{1/2}$, and the
bounded density and 1-dependence give
$\norm{T^{-1/2}\sum_t[\xi_t(d)-\xi_t(0)-\E\{\xi_t(d)-\xi_t(0)\}]}_2\leq C|d|^{1/2}$.
With $\mathcal Q_{nT}(d)=n^{-1}\sum_i\mathbb M_{iT}(d)$ and
$Z_{nT}=(nT)^{-1/2}\sum_{i,t}x_{it}\psi(u_{it})$, these bounds, independence
across individuals, integration along deterministic segments and the
value expansion of Step 1 give, for each $\theta\in\R$,
\[
 nT\{\mathcal Q_{nT}(\theta/\sqrt{nT})-\mathcal Q_{nT}(0)\}
 =-\theta Z_{nT}+\tfrac12\Gamma\theta^2+\frac{D_x}8\Bigl(\frac nT\Bigr)^{1/2}|\theta|+\op(1),
\]
the error from Step 1 being $O\{|\theta|(n/T)^{1/2}(\log T/T)^{1/2}\}$. By
convexity, the expansion holds uniformly on compact sets. Given all
errors, the $x_{it}\psi(u_{it})$ are independent and distributed as
$x_{it}/2$, because the regressors are symmetric and independent of the
errors, so $Z_{nT}\Rightarrow N(0,\Lambda)$. With
$c_{nT}=(D_x/8)(n/T)^{1/2}\to c_\upsilon$, the approximating criterion
$-\theta Z_{nT}+\Gamma\theta^2/2+c_{nT}|\theta|$ is minimised at
$\theta_{nT}^*=\Gamma^{-1}\operatorname{sgn}(Z_{nT})(|Z_{nT}|-c_{nT})_+$ and
exceeds its minimum by at least $\Gamma(\theta-\theta_{nT}^*)^2/2$, so the
nearness-of-argmins lemma of \citet{HjortPollard93} shows that the
minimisers are $\theta_{nT}^*+\op(1)$, and the continuous mapping theorem
proves the weak convergence.

\emph{Step 3: the atom.} Let $J_{iT}$ indicate that the two central
residuals of individual $i$ at $d=0$, the $\ell$th and $(\ell+1)$th
order statistics, belong to the same doubleton, with marks
$X_{i1}^\circ,X_{i2}^\circ$, and put $D_{iT}=J_{iT}|X_{i1}^\circ-X_{i2}^\circ|$.
For $d$ slightly above zero, the residual of that pair with the larger
mark becomes the $\ell$th smallest, and for $d$ slightly below zero the
one with the smaller mark does; all other weights are unchanged. The
one-sided scores therefore satisfy $S_{iT}(0\pm)=S^\circ_{iT}\mp D_{iT}/(2T)$,
where $S^\circ_{iT}=T^{-1}\sum_tx_{it}\{1/2-w_{it}(0)\}$ with the pair, if any,
given weight $1/2$ each. Given the errors, $S^\circ_{iT}$ is a sum of
independent regressors with coefficients $\pm1/(2T)$, or $0$ for the pair,
so that $\E S^\circ_{iT}=0$, $T\E(S^\circ_{iT})^2\to\sigma_x^2/4=\Lambda$ and
$\sqrt{nT}\,\overline S^\circ_{nT}\Rightarrow N(0,\Lambda)$, with
$\overline S^\circ_{nT}=n^{-1}\sum_iS^\circ_{iT}$. Given the other clusters, a doubleton
straddles the central ranks precisely when its centre lies between their
two central order statistics; by the flat density and
Lemma~\ref{lem:cusp-spacing}, $\Prb(J_{iT}=1)=\E K_T/(T-2)
+O\{T^{-1/2}(\log T)^{1/2}\}\to1/4$, and, the marks being independent of
$J_{iT}$, $\overline D_{nT}=n^{-1}\sum_iD_{iT}\to_pD_x/4$. By convexity,
zero is a minimiser if and only if
$n^{-1}\sum_iS_{iT}(0+)\leq0\leq n^{-1}\sum_iS_{iT}(0-)$, that is,
$|\sqrt{nT}\,\overline S^\circ_{nT}|\leq\frac12(n/T)^{1/2}\overline D_{nT}$, and it
is the unique minimiser under strict inequality. If $n/T\to\upsilon<\infty$,
the probability of equality tends to zero, so
$\Prb(\widehat\beta=\beta_0)\to\Prb(|Z|\leq\upsilon^{1/2}D_x/8)
=2\Phi\{D_x\upsilon^{1/2}/(4\sigma_x)\}-1$. If $n/T\to\infty$, then,
because the $S^\circ_{iT}$ are independent across $i$ and centred,
$nT\E(\overline S^\circ_{nT})^2=T\E(S^\circ_{iT})^2\leq\Lambda$, so the
left-hand side of the inequality is $\Op(1)$, whereas the right-hand side
tends to infinity in probability, because $\overline D_{nT}\to_pD_x/4>0$.
The strict inequality therefore holds with probability tending to one, and
$\Prb(\widehat\beta=\beta_0)\to1$. This argument uses neither Step 2 nor any
upper bound on the growth of $n$. \qed

\newpage
\bibliographystyle{ecta}
\bibliography{biblio}

\end{document}